\documentclass{aa}  

\usepackage{graphicx}
\usepackage{txfonts}
\usepackage[dvipsnames]{xcolor}
\usepackage{aalongtable}
\usepackage{lscape}
\usepackage{amsmath}
\usepackage{caption}
\usepackage[skip=0.5ex]{subcaption}
\usepackage{supertabular}
\usepackage{array}
\usepackage{placeins}
\usepackage{float}
\usepackage{footmisc}

\usepackage[colorlinks, citecolor=blue, linkcolor=blue]{hyperref}
\hypersetup{colorlinks,breaklinks, linkcolor=blue,urlcolor=magenta, anchorcolor=blue,citecolor=blue}
\bibpunct{(}{)}{;}{a}{}{,} 

\newcommand{\be}{\begin{equation}}
\newcommand{\ee}{\end{equation}}

\usepackage{natbib}
\bibpunct{(}{)}{;}{a}{}{,} 

\begin{document}

   \title{Planetary nebulae seen with TESS: New and revisited short-period binary central star candidates from Cycles 1 to 4}
   \titlerunning{Planetary nebulae seen with TESS Cycles 1-4}
     \authorrunning{A. Aller, J. Lillo-Box \& D. Jones}

   \author{Alba Aller
          \inst{1},
          Jorge Lillo-Box\inst{2}, 
           \and
          David Jones\inst{3,4,5}
          }

\institute{Observatorio Astron\'omico Nacional (OAN), Alfonso XII 3, 28014, Madrid, Spain\\
\email{a.aller@oan.es}
\and 
Centro de Astrobiolog\'ia (CAB), CSIC-INTA, Camino Bajo del Castillo s/n, 28692, Villanueva de la Ca\~nada (Madrid), Spain 
\and 
Instituto Astrof\'isico de Canarias, E-38205, La Laguna, Spain
\and
Departamento de Astrof\'isica, Universidad de la Laguna, E-38206 La Laguna, Tenerife, Spain
\and
Nordic Optical Telescope, Rambla Jos\'e Ana Fern\'andez P\'erez 7, 38711, Bre\~na Baja, Spain
             }

 
  \abstract
   {High-precision and high-cadence photometric surveys such as Kepler or TESS are making huge progress not only in the detection of new extrasolar planets but also in the study of a great number of variable stars. This is the case for central stars of planetary nebulae (PNe), which have similarly benefited from the capabilities of these missions, increasing the number of known binary central stars and helping us to constrain the relationship between binarity and the complex morphologies of their host PNe.}
   {In this paper, we analyse the TESS light curves of a large sample of central stars of PNe with the aim of detecting signs of variability that may hint at the presence of short-period binary nuclei. This will have important implications in understanding PN formation  evolution as well as the common envelope phase.}
   {We analysed 62 central stars of true, likely, or possible PNe and modelled the detected variability through an MCMC approach accounting for three effects: reflection, ellipsoidal modulations due to tidal forces, and the so-called Doppler beaming. 
   Among the 62 central stars, only 38 are amenable for this study. The remaining 24 show large contamination from nearby sources preventing an optimal analysis. Also, eight targets are already known binary central stars, which we revisit here with the new high precision of the TESS data.} 
   {In addition to recovering the eight already known binaries in our sample, we find that 18 further central stars show clear signs of periodic variability in the TESS data, probably resulting from different physical effects compatible with the binary scenario. We propose them as new candidate  binary central stars. We also discuss the origin of the detected variability in each particular case by using the \texttt{TESS\_localize} algorithm. Finally, 12 targets show no or only weak evidence of variability at the sensitivity of TESS.}
  {Our study demonstrates the power of space-based photometric surveys in searching for close binary companions of central stars of PNe. Although our detections can only be catalogued as candidate binaries, we find a large percentage of possible stellar pairs associated with PNe, supporting the hypothesis that binarity plays a key role in shaping these celestial structures.}


   \maketitle
%

\section{Introduction}

Planetary nebulae (PNe) are one of the most fascinating astronomical objects in the Universe. They are formed when low-to-intermediate mass stars (0.8-8 $M_\odot$) reach the final stage of their lives, ejecting their outer envelopes into the interstellar medium. As a result, a great diversity of PNe are observed in our Galaxy, and there are no two PNe that are exactly alike. But one thing seems clear: spherical PNe are almost the exception and they represent only one-fifth of all the PNe discovered in the Milky Way. On the contrary, the remaining 80\% show a wide range of complex morphologies \citep{Parker2006}.

Explaining the precise processes that lead to the formation of these kinds of structures is a difficult task, although some progress has been made in the last decades with central star binarity emerging as the most likely culprit \citep[see][and references therein]{Boffin-Jones2019}.
Binary central stars of PNe (bCSPNe) appear to be key not only to explain the variety of shapes in PNe but also to respond to many other open questions in stellar evolution, such as the mass-loss rate and history, the little-understood common envelope phase, or the estimation of the binary population in general.

Thus, many of the efforts in the PN community today are devoted to the search for, and characterisation of, new binary central stars in the heart of PNe (with both short and long orbital periods), with the aim of shedding light on the formation of the striking morphologies they present. Dedicated ground-based observations of CSPNe are very costly in terms of time, so very good candidates among the large sample of PNe (more than 3000 in our Galaxy) are required to guide targeted observational campaigns. Within this context, large photometric surveys such as OGLE (Optical Gravitational Lensing Experiment, \citealt{Udalski1992}), \textit{Kepler} \citep{Borucki2003}, and TESS (Transiting Exoplanets Survey Satellite, \citealt{Ricker2015}) have enormously contributed to the discovery of a large fraction of (candidate) binary systems in the nuclei of PNe. Numerous papers that have been published recently focus on this topic with encouraging results, significantly increasing the binary population  \citep[see, e.g.][]{Miszalski2009a, DeMarco2015, Jacoby2021, Aller2020}. However, the fraction of known binary central stars in the whole population of PNe remains very low,  representing less than 1\% of the whole sample of known Galactic PNe, while unbiased surveys indicate that the true fraction should be $\sim$20\% or more \citep{Jones-Boffin2017,Jacoby2021}. The need to continue this search for binary central stars is, therefore, essential in order to ultimately obtain a statistically significant and unbiased sample with which to definitively understand the role of binarity in the formation and evolution of PNe (as well as related phenomena).

In this paper, we continue the analysis carried out in \cite{Aller2020}, in which we investigated the variability of the eight central stars of PNe observed in Cycle 1 of TESS for indications of binarity. Here, we extend this study to cover the first four years of TESS operations (Cycles 1 to 4) in order to find new short-period binary central star candidates in the whole sample of galactic PNe. In this search, as well as some already known binary central stars, we also identified new binary candidates that present modulations in their TESS light curves compatible with the presence of companion stars. These modulations are likely due to irradiation effects (i.e. the reflection of the light of the central star on the companion's surface), ellipsoidal modulations (consequence of tidal forces with the companion), and Doppler beaming (relativistic effects due to the orbital motion of a binary system with respect to the observer). Previous works have used these effects to detect binary companions \citep[see, for instance,][]{Santander-Garcia2015,DeMarco2015, maxted02,faigler11} and even planets (e.g. \citealt{shporer11, mazeh12,lillo-box2014,Millholland17}). All these physical effects open the door to the detection of close binaries without requiring the detection of eclipses or the need for radial velocity monitoring. Also, TESS represents a valuable opportunity to revisit close binary CSPNe already known by improving the determination of their properties, especially in those eclipsing systems.

The paper is organised as follows. In Section\,\ref{sec:observations}, we present our target sample and briefly introduce the TESS datasets. The light curve modelling process is explained in Section\,\ref{Sect:modeling} and  Section\,\ref{Sect:analysis} is dedicated to the analysis of the results, discussing the detected variability in the light curves and the periodicities found, as well as distinguishing whether they are already known binaries, new binary candidates or stars with weak evidence or no variability at all. In Section\,\ref{sec:discussion}, we discuss the origin of the variability in each particular case and, finally, we close the paper in Section\,\ref{sec:conclusions} with some short final remarks and main conclusions.

\section{Target selection and observations}
\label{sec:observations}
\subsection{The PN sample in TESS}

TESS was launched in 2018 and was initially designed as a two-year mission, although it is now conducting its second-extension. During the first year of observations, TESS completed Cycle 1 in a total of 13 sectors and provided two-minute cadence light curves with a baseline of $\sim$ 27 days for eight central stars of PNe which were analysed in detail in \cite{Aller2020}. 
In July 2019, TESS started its Cycle 2, observing the northern ecliptic hemisphere in another 13 additional sectors. Since then, every year TESS completes a new cycle. In this paper, we have analysed all the CSPNe that have been observed with two-minute cadence until September 2022, that is, Cycles 1, 2, 3, and 4. They are listed in Table\,\ref{table:main_table}, along with their most relevant information as PNG designation, common name, TESS input catalogue (TIC) designation, equatorial coordinates, TESS magnitude, the sector or sectors of the observation and the PN status found in the HASH database\footnote{\url{http://202.189.117.101:8999/gpne/index.php}}
\citep{Bojicic2017}. In total, we analysed the available two-minute cadence light curves for 62 central stars of true (T), possible (P) or likely (L) PNe, including the eight previously presented in \cite{Aller2020} from Cycle\,1.
We note that almost half of the targets do not have the TIC designation linked to SIMBAD, so we have carefully inspected all the target pixel files (TPFs) and targets, to ensure that the light curves correspond to the target itself.

 \begin{figure*}
     \includegraphics[width=0.335\textwidth]{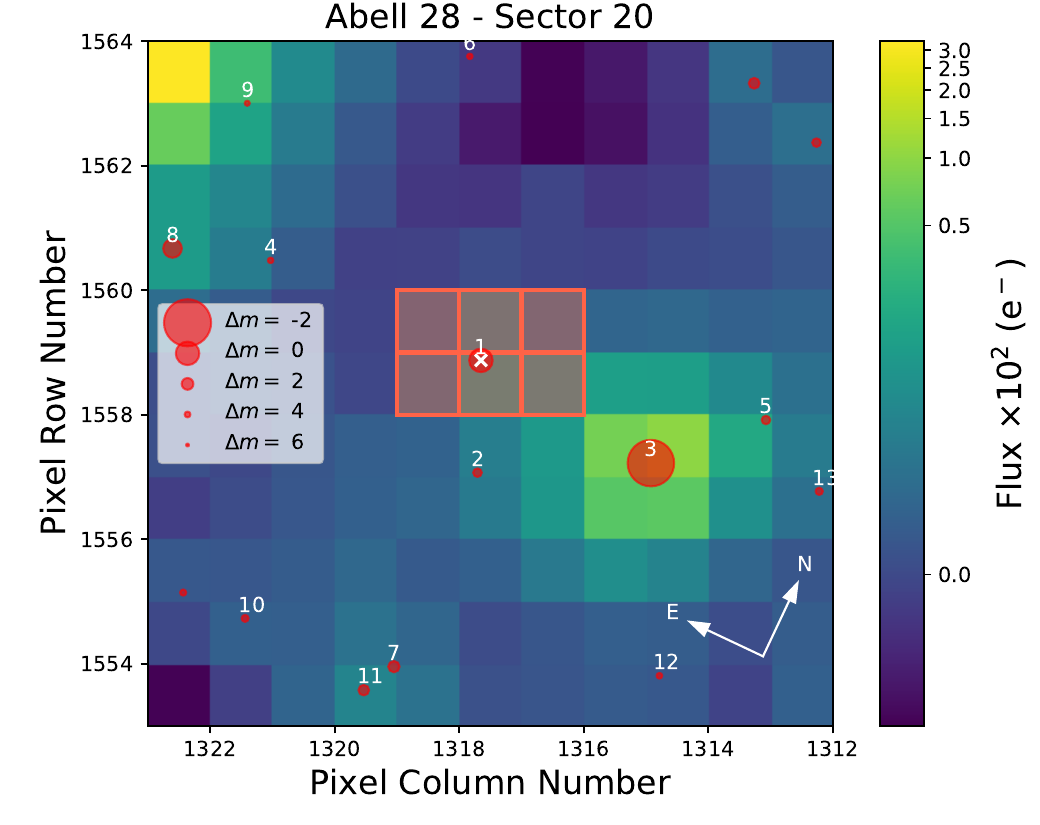}
     \includegraphics[width=0.335\textwidth]{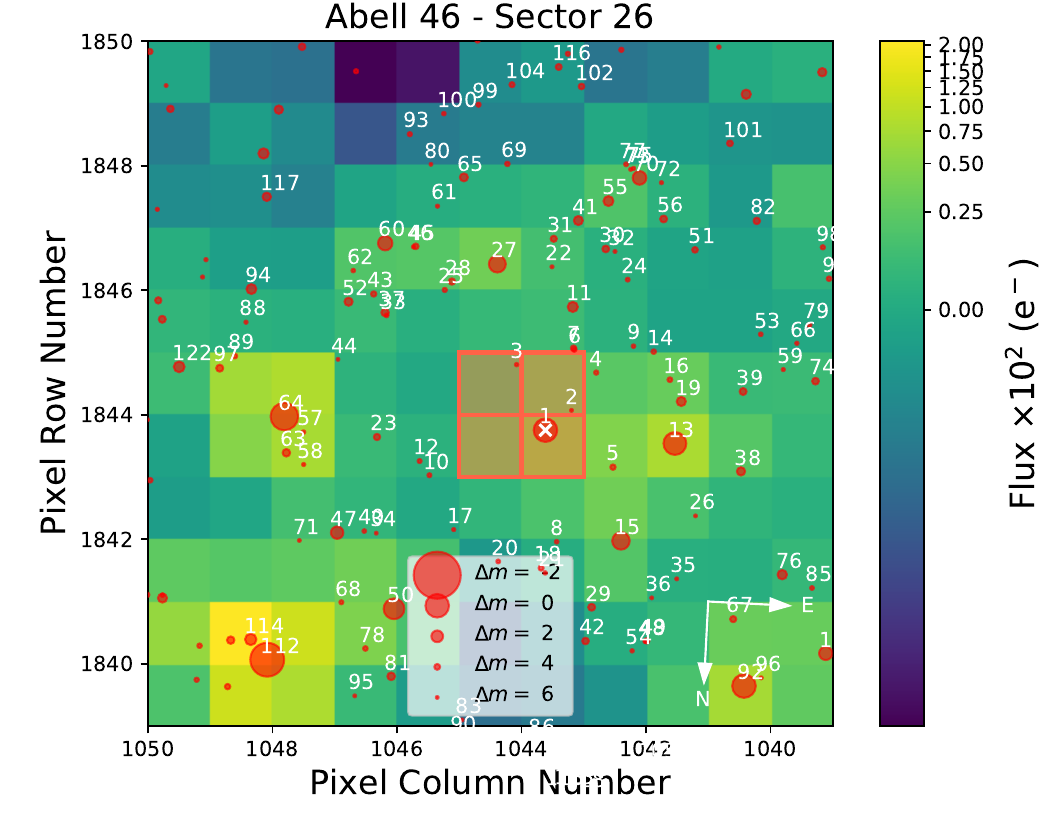}     
     \includegraphics[width=0.335\textwidth]{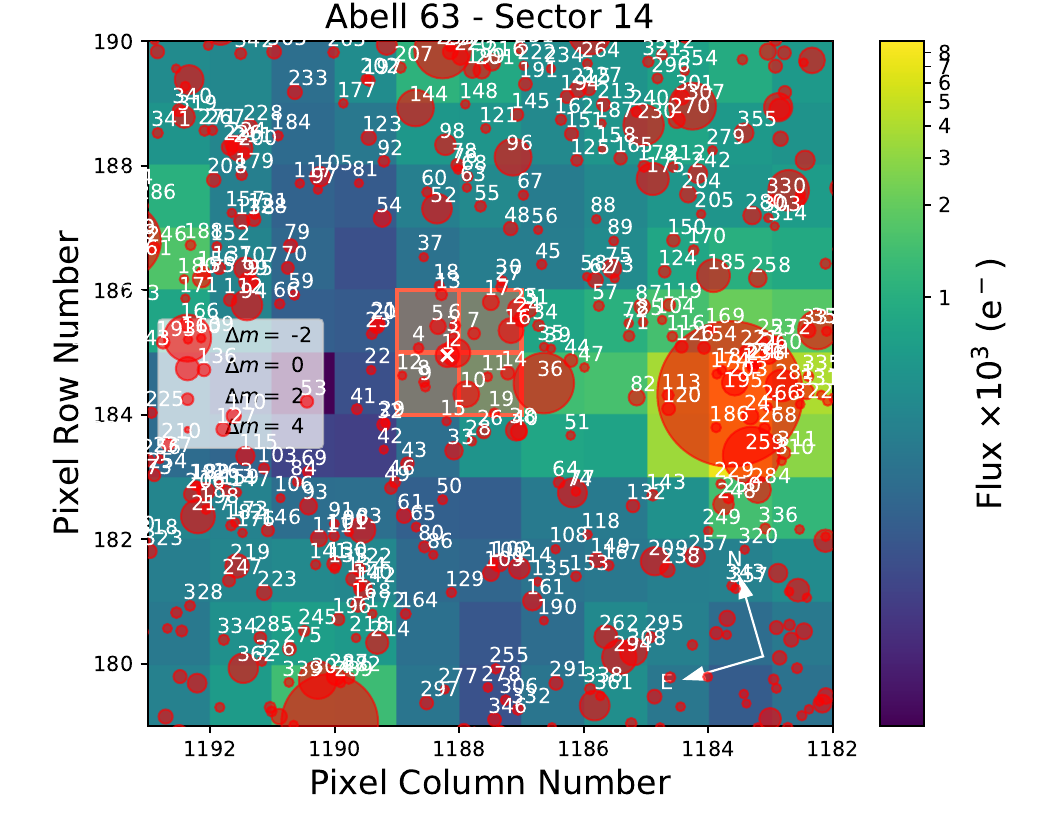}
\caption{Target pixel files (TPFs) of three central stars in the sample (marked with white crosses) obtained with {\sc tpfplotter}. The red circles are sources from the \textit{Gaia} DR3 catalogue in the field with scaled magnitudes (see legend). The aperture mask used by the pipeline to extract the photometry is also marked. The pixel scale is 21 arcsec\,pixel$^{-1}$. Left, middle and right panels represent examples of fields with negligible, minimal or severe contamination, respectively (see Sect.\,\ref{Sect:description_TESS_data} for more details).}
\label{fig:example_TPFs}
\end{figure*}

\subsection{TESS photometric data}
 \label{Sect:description_TESS_data}

We retrieved the light curves provided by the Science Processing Operations Center (SPOC) pipeline \citep{Jenkins2016} from the Mikulski Archive for Space Telescopes MAST\footnote{\url{https://mast.stsci.edu/portal/Mashup/Clients/Mast/Portal.html}}). We used the PDCSAP (Pre-search Data Conditioning Simple Aperture Photometry) flux for the analysis, only removing those data points with a non-zero 'quality' flag. For a more detailed description of the TESS data, we point the interested reader to \cite{Aller2020}. When more than one sector was available in the MAST archive for a particular target, we combined all the individual light curves to produce a single and continuous multi-sector light curve. This is done with the \texttt{stitch} method in the \texttt{Lightkurve} package, which concatenates and normalises all the sectors. 

In order to check for possible blends and contaminating sources in the apertures, we plotted the TPF of each central star in the sample using \texttt{tpfplotter}\footnote{\url{https://github.com/jlillo/tpfplotter}} \citep{Aller2020}. This check of the TPFs is especially important in the most crowded fields, since the relatively large pixel size of the TESS CCDs (21 arcseconds per pixel) can lead to strong photometric contamination from nearby sources. 
Figure\,\ref{fig:example_TPFs} shows an example TPF for three different degrees of contamination. In all cases, the central star is marked with a white cross and the red circles represent \textit{Gaia} sources from the DR3 catalogue \citep{GaiaCollaboration2021}, scaled by magnitude contrast against the target source. We have over-plotted all the sources with a magnitude contrast up to $\Delta$m = 6 (i.e. six magnitudes fainter than our target, which corresponds to $\sim$ 0.4\% of contamination if entirely inside the aperture). The aperture mask used by the pipeline to extract the photometry has also been plotted over the TPF. 

As shown in the figure, we find very crowded fields with severe contamination (larger than 10\%), fields with little contamination and cases with negligible contamination where no other star within a contrast magnitude of $\Delta$m = 6 is identified by \textit{Gaia} inside the TESS aperture. We note that this is only a first selection criterion that allows us to avoid critically contaminated cases where the attribution of the true source of the variability will not be possible with the current dataset and tools. For targets passing this criterion, we perform an analysis using the \texttt{TESS\_localize} code to unveil the origin of the photometric variations and account for the possible contamination of stars outside the aperture. Figure B.1\footnote{Available at \url{https://zenodo.org/records/13370453}.\label{zenodo}} shows the TPFs of the first sector observed for each target. 

For each individual central star and sector, we have calculated the contribution of the field stars to the total flux in the aperture mask. Out of the 62 targets, we found severe contamination for 28 stars. We decided to exclude these targets from further analysis, since the probability that any potential variability can come from any of the nearby stars is too high. Only for specific contaminated objects, we carried out additional studies. In particular, Abell\,63 (because it is a well-known eclipsing binary central star in an extremely crowded field), and also three objects that we already analysed in \cite{Aller2020} from Cycle 1: NGC\,246, NGC\,2867, and NGC\,5189 (see Sect.\,\ref{Sect:analysis}). In addition, we found 16 cases with minimal ($\le$10\% but not null) contamination and 18 central stars where there was no appreciable contamination from nearby sources in the aperture mask. The level of contamination (negligible, minimal or severe) is marked in Table\,\ref{table:main_table} with a, b or c, respectively. We subsequently discuss these cases in detail in Section~\ref{Sect:analysis}. 

Once this contamination cleaning step is done, each single multi-sector light curve was inspected by performing a Lomb-Scargle analysis in order to identify possible periodic variability in the data as a first look.

 At this point, it is important to note that apart from the possible contamination of other stars in the field, we sometimes have bright PNe around our targets of interest, whose light might dilute the amplitude of the detected variability and hence, have an impact on the derived physical parameters. Quantifying this PN contribution is out of the scope of this study, which is to identify new binary candidates in the sample of central stars of PNe.

\section{Methodology: Data analysis and photometric modelling}
 \label{Sect:modeling}
 
\subsection{Parametric approach}   
 \label{Sect:parametric_model}

In a first approach, similar to the analysis performed in our previous work in Cycle 1 \citep{Aller2020} and other works using the same detection technique (e.g. \citealt{Millholland17,Lillo-Box2016}), we model the TESS light curves by using simple sinusoidal functions for the three effects (ellipsoidal, Doppler beaming and reflection). This corresponds to

\begin{equation}
	\left(\frac{\Delta F}{F}\right)_{\rm ellip} = - A_{\rm ellip} \cos{(2\theta)},
\end{equation}
	
\begin{equation}
	\left(\frac{\Delta F}{F}\right)_{\rm beam} =  A_{\rm beam} \sin{(\theta)}, {\rm and}
\end{equation} 

\begin{equation}
	\left(\frac{\Delta F}{F}\right)_{\rm ref} =  -A_{\rm ref} \cos{(\theta)},
\end{equation} 

\noindent with $\theta = 2\pi (t-T_0)/P$, where $T_0$ is the time of inferior conjunction of the companion star, $P$ is the orbital period, and $A_{\rm ell}$, $A_{\rm beam}$, and $A_{\rm ref}$ are the amplitudes of the ellipsoidal, Doppler beaming, and reflection effects, respectively. The simple model thus corresponds to:

\begin{equation}
\label{eq:3E}
\begin{split}
	\left(\frac{\Delta F}{F}\right)_{\rm 3E} & =  Z_0  + \left(\frac{\Delta F}{F}\right)_{\rm ellip} + \left(\frac{\Delta F}{F}\right)_{\rm beam} + \left(\frac{\Delta F}{F}\right)_{\rm ref} = \\
	 & = Z_0 - A_{\rm ellip} \cos{(2\theta)} + A_{\rm beam} \sin{(\theta)} -A_{\rm ref} \cos{(\theta)}.
\end{split} 
\end{equation} 

As in most cases only one of the effects might be present, we also test a simpler model with only one sinusoidal function, namely:

\begin{equation}
\label{eq:1E}
	\left(\frac{\Delta F}{F}\right)_{\rm 1E}  =  Z_0  + \left(\frac{\Delta F}{F}\right)_{\rm 1E} = Z_0 + A_{\rm 1E} \cos{(\theta)}. 
\end{equation} 

Based on these equations, we have seven parameters in the case of the 3-effects (3E, Eq.~\ref{eq:3E}), namely the photometric level ($Z_0$), the three amplitudes ($A_{\rm ellip}$, $A_{\rm ref}$, $A_{\rm beam}$), the time of inferior conjunction ($T_0$), the orbital period $P$, and the photometric jitter (a term that we add in quadrature to the photometric uncertainties to account for other unknown systematics). For the 1-effect model (1E, Eq.~\ref{eq:1E}), we only have four parameters as only one amplitude is needed ($A_{\rm 1E}$).

We acknowledge the fact that this purely sinusoidal model is only an approximation to a physically motivated model. For instance, gravity darkening and geometrical effects produce non-sinusoidal ellipsoidal modulations. However, this approximation is sufficient for the exploratory goals of this work.

 We explore the parameter space through a Monte Carlo Markov Chain sampler using the \texttt{emcee} algorithm \citep{Foreman-mackey2013}. In particular, we set uninformative (uniform) priors for the orbital period between 0.1-10 days and for the time of inferior conjunction between the first date of observation and 10 days afterwards. In general terms, we allow the amplitudes to vary from 0 to twice the maximum peak-to-peak value of each light curve. The photometric level is also left with a uniform prior $\mathcal{U}(0.9,1.1)$, as well as the jitter term that we sample in the range $\mathcal{U}(0,100)$ in ppm.

In total, three models are tested for each of the targets studied in this paper: a flat model (FL) with only the photometric level and jitter as parameters, a 1-effect model (1E), and the 3-effects model (3E). For each of them, we sample the posterior distribution of the parameters using a number of walkers equal to four times the number of parameters of the model, and a total of 100\,000 steps per walker. We check the convergence of the chains by ensuring that the length of the chain is at least 30 times the autocorrelation time (\citealt{Foreman-mackey2013}). We estimate the Bayesian evidence ($\ln{\mathcal{Z}_i}$) of each model using the \texttt{perrakis}\footnote{\url{https://github.com/exord/bayev}. A python implementation by R. D\'iaz of the formalism explained in \cite{Perrakis2014}.} code \citep{Diaz2016}. This metric is then used to select the simplest model that best represents the data. For a complex model to be selected over a simpler one (i.e. the 1E over the FL or the 3E over the 1E), we require the Bayesian evidence of the more complex model to be larger than the evidence of the simpler model by 6 in logarithmic space (i.e. $\mathcal{B} = \Delta \ln{\mathcal{Z}}> 6$, \citealt{Trotta2008}), corresponding to strong evidence in favour of the more complex model. Otherwise, the simpler model is selected.

\subsection{Identification of the variability source}   
 \label{Sect:tess_localize_newsect}

As mentioned in Sect.\,\ref{Sect:description_TESS_data}, the large plate scale of TESS (21 arcsec\,pixel$^{-1}$) makes it particularly difficult to undoubtedly assign the source of the variability to one specific target (e.g. \citealt{lillo-box24}). Essentially, the high frequency resolution of TESS is at the expense of its low spatial resolution. Even in those stars in which the aperture mask has little contamination from nearby sources, the analysis is delicate, since other stars outside the aperture can contributed to the measured flux. Recently, \cite{Higgins-Bell2023} developed a method to localise the origin of variability on the sky to better than one fifth of a pixel given a measured frequency (or frequencies). Basically, the method can resolve the variable source in frequency space for each pixel. The authors showed that even stars more than three pixels outside the aperture, can produce significant contamination in the extracted light curves. 

Once the MCMC analysis presented in Sect.~\ref{Sect:parametric_model} is performed and the periodicities in the TESS light curves are measured, we can apply this methodology. To this end we use the implementation presented in the open-source Python package \texttt{TESS\_localize}\footnote{\url{https:/github.com/Higgins00/TESS-Localize}}. We apply this algorithm to our sample (both the already known binaries and those with newly detected variability) in order to identify the origin (i.e. location on the sky) of the periodicities found in this work. This analysis is presented in Sec.\,\ref{sec:tess_localize}.

  \begin{table*}
\centering 
  \caption{Results from the modelling process for the 38 stars analysed with minimal or negligible contamination in the TESS aperture.}
     \label{table:results_fit} 
\begin{tabular}{cccccccc}
\hline
Name & TIC & Morphology & Known binary?&$\mathcal{B}_{10}$ & $\mathcal{B}_{30}$ & $\mathcal{B}_{31}$ & BestModel \\
\hline\hline
Abell\,63 			& 	1842385646	 & B   & yes &&  & & - \\
Abell\,46 			& 423311936 &  E  & yes & &  &  & PHOEBE \\
DS\,1 				& 120596335 & I  & yes &  &  &  & PHOEBE \\
Abell\,30 			& 331986754 & R& yes &$>100$ & $>100$ & -0.2 & 1E \\
LoTr\,5 			& 357307129 & E  &yes &$>100$ & $>100$ & $>100$ & 1E \\
PG\,1520+525 		& 193990614 & R  & no &13.5 & 7.5 & -5.9 & 1E \\
PNG\,136.7+61.9	& 471013481 & na  & no &34.4 & 1.9 & -32.5 & 1E \\
Pa\,165 			& 367508287 &I  &  no &23.9 & 20.2 & -3.6 & 1E \\
NGC\,2867 		& 387196603 & E  & no &13.3 & 16.7 & 3.4 & 1E \\
WPS\,28 			& 229966905 & na  & no &41.0 & 24.3 & -16.7 & 1E \\
Fr1-4 			& 401795856 & na 
& no &23.6 & 2.6 & -21.0 & 1E \\
NGC\,7094 		& 387371712 & E   & no &$>100$ & $>100$ & -2.1 & 1E \\
RWT\,152 			& 65145453 & S  & no&34.9 & 17.8 & -17.1 & 1E \\
NGC\,2371 		& 446005482 & B   & no &25.5 & 20.8 & -4.7 & 1E \\
NGC\,7293 		& 69813909 & B   & no &74.4 & 70.2 & -4.2 & 1E \\
Hen\,3-1863 		& 290274851 & S  &no &$>100$ & $>100$ & $<-100$ & 1E \\
WPS\,54 			& 453444667 & na  & no& $>100$ & $>100$ & -20.8 & 1E \\
NGC\,5189 		& 341689253 & B   & yes &$>100$ & $>100$ & $>100$ & 3E \\
NGC\,246 			& 3905338 & E  & yes &22.7 & 31.3 & 8.5 & 3E \\
NGC\,2392 		& 60947424 & E   & yes &$>100$ & $>100$ & $>100$ & 3E \\
IC\,2149 			& 440340055 & B  & no &$>100$ & $>100$ & $>100$ & 3E \\
NGC\,1501 		& 84306468 & E  & no &$>100$ & $>100$ & 95.6 & 3E \\
PG\,1034+001 		& 124598476 & na  & no &$>100$ & $>100$ & $>100$ & 3E \\
IC\,4593 			& 119955971 & E  & no &$>100$ & $>100$ & $>100$ & 3E \\
K\,1-16 			& 233689607 & B  & no &24.6 & 36.8 & 12.2 & 3E \\
HDW\, 7 			& 760317391 & R  & no &-1.3 & -8.0 & -6.7 & FL \\
JnEr\,1 			& 741726060 & B   & no &-2.8 & -8.3 & -5.5 & FL \\
Abell\,28 			& 471015303 & R  & no & -0.9 & -7.1 & -6.2 & FL \\
EC\,13290-1933 		& 32145732 & na & no & -2.9 & -7.9 & -5.0 & FL \\
Abell\,31 			& 444036737 & E   & no&-1.4 & -6.6 & -5.2 & FL \\
Abell\,7 			& 169305167 & E  & no&-2.4 & -10.1 & -7.8 & FL \\
StDr\,56*			& 63505607 & na  & no &-0.3 & -5.6 & -5.2 & FL \\
Fr\,2-21*		& 352568802 & na  & no &4.5 & 5.0 & 0.5 & FL \\
Sh\,2-216 			& 391260920 & R  & no &-4.0 & -15.7 & -11.6 & FL \\
Fr\,2-46 			& 92709449 & na  & no&-4.0 & -10.4 & -6.4 & FL \\
Lo\,1 				& 146711168 & B  & no &-1.9 & -8.6 & -6.6 & FL \\
IC\,5148/50 		& 2027562015 & R    & no &-3.3 & -5.1 & -1.8 & FL \\
AMU\,1** 			& 63373833 & B   & yes &-0.6 & -9.1 & -8.5 & FL \\
\hline
\end{tabular}
 \tablefoot{
\tablefoottext{*}{Although both StDr\,56 and Fr\,2-21 converge to a flat model, they have significant signals in their periodograms so we think they deserve a more detailed analysis (see Section\,\ref{sect.novar}). {**} AMU\,1 has a confirmed binary system in the nucleus, detected with Kepler data.
 }}
\end{table*}

  
 \section{Results in the TESS dataset}
\label{Sect:analysis}

The results of the variability analysis described in Section\,\ref{Sect:modeling} are summarised in Table~\ref{table:results_fit}. The column labelled as `Best model' shows the preferred solution of the modelling process based on the Occam's razor quantified by the comparison of the log of the Bayesian evidence between two models ($\mathcal{B}_{ij}$, where "i" and "j" represent each of the two models). This is, we compare the evidence of the 1E (1-effect) model against the flat (FL) model ($\mathcal{B}_{10}$), the evidence of the 3E model against the flat model ($\mathcal{B}_{30}$) and the evidence of the 3E model against the 1E model ($\mathcal{B}_{31}$). Out of 38 targets, a total of 26 show modulations compatible with one or more effects. These central stars are described in Sections.\,\ref{sect.known_binaries} and \ref{sect.new_candidates}, depending on whether they are already known binaries or new binary candidates, respectively. For the remaining 12 central stars, the flat model is the most plausible, that is, there are no statistically significant modulations in the light curve down to the sensitivity of the data. These central stars are briefly discussed in Sect.\,\ref{sect.novar}.

\begin{figure}
  \centering
\includegraphics[width=0.5\textwidth]{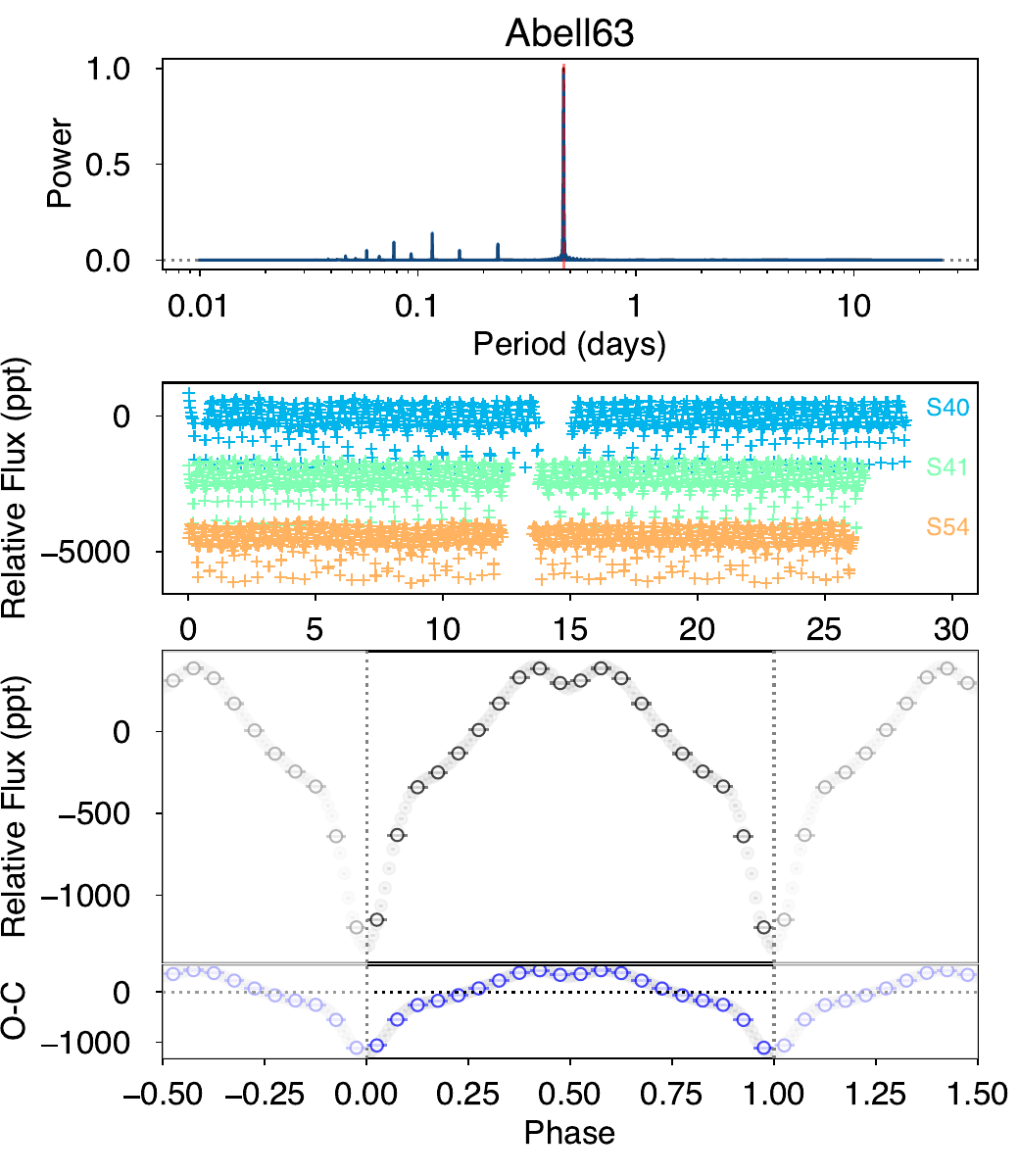}
       \caption{Periodogram ($\textit{upper panel}$), TESS photometric time series for sectors 40, 41, and 54 ($\textit{middle panel}$), and phase-folded combined light curve ($\textit{bottom panel}$) of Abell\,63. Two different bin sizes are shown with black bin size of 0.01 in phase) and grey (bin size of 0.05 in phase) circles.}
\label{fig:Abell63LC}
\end{figure}

\begin{figure*}
          \includegraphics[width=1\textwidth]{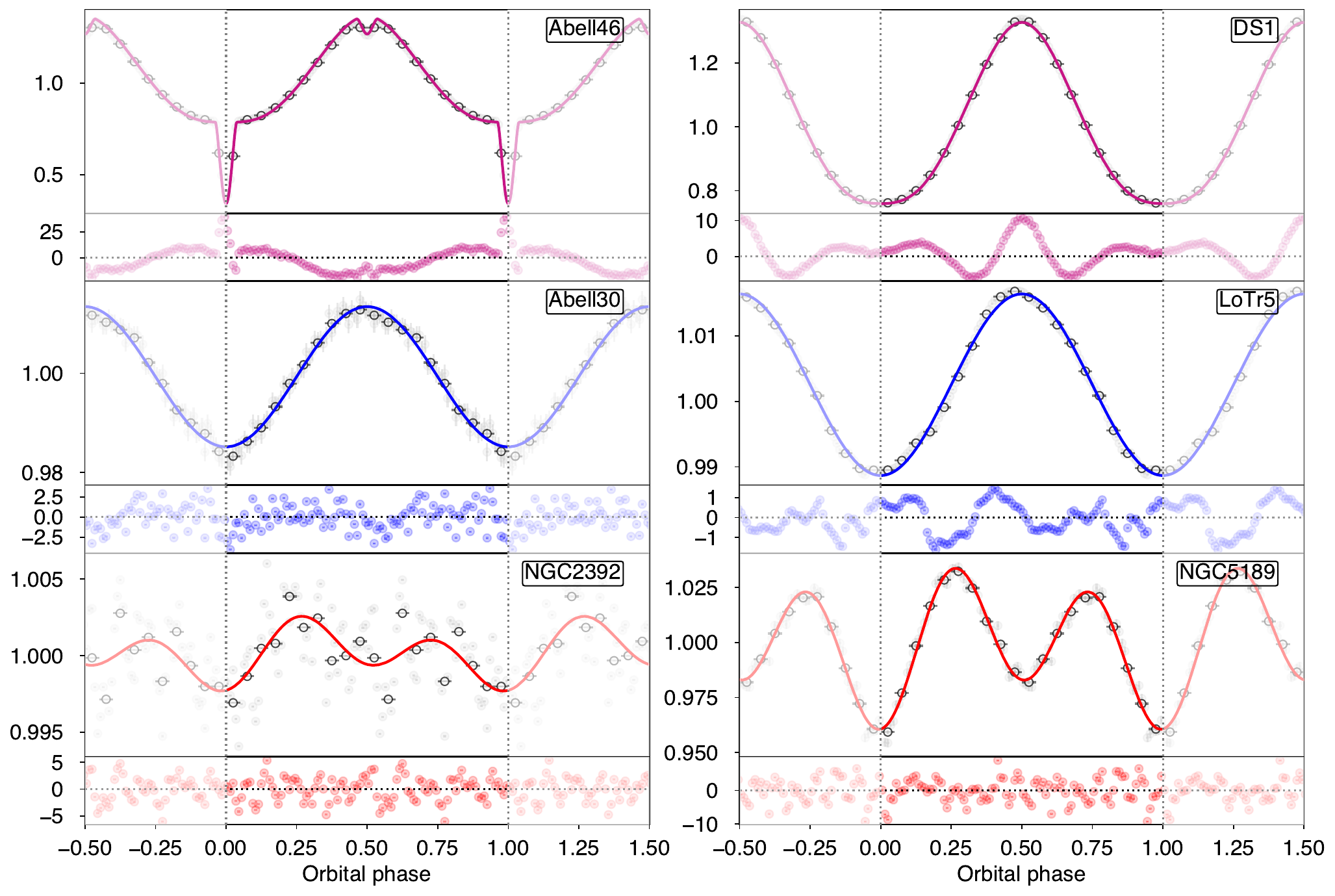}
       \caption{Phase-folded light curves of the already known binaries, with the best-fitting model overlaid. Results from the PHOEBE2 analyses are overplotted in pink and from the MCMC analyses in blue (1E solution) and red (3E solution). Grey symbols correspond to a bin size of 0.01 in phase (i.e. 100 datapoints) while the black symbols correspond to bin sizes of 0.05 in phase (i.e. 20 datapoints). The period derived from the fitting can be found in Table\,\ref{table:results_fit_known} and the corresponding periodograms are shown in Fig.\,B.2\footref{zenodo}). We note that the period of NGC\,5189 does not correspond with the orbital period published in the literature for this binary.}
\label{fig:results_known_binaries}
\end{figure*}
   
    \begin{table*}
\centering
   \caption{Results from the light curve analysis of those already known binary systems analysed with the parametric approach described in  Sect.\,\ref{Sect:parametric_model}. }
   \label{table:results_fit_known} 
   \setlength\extrarowheight{3pt}
\begin{tabular}{ccccccc}
  \hline    
Name  & BestModel & Period (days) & $T_0$ (2458300) & A$_{\rm ref}$/A$_{\rm 1E}$ (ppt) & A$_{\rm beam}$ (ppt) & A$_{\rm ellip}$ (ppt)\\
  \hline \hline    
Abell30 			 & 1E & $1.06115^{+0.00013}_{-0.00013}$ & $55.06^{+0.15}_{-0.15}$ & $14.14^{+0.25}_{-0.25}$ & - & - \\
LoTr5 			 & 1E\tablefootmark{1} & $5.8090049^{+0.0000036}_{-0.0000036}$ & $71.35280^{+0.00064}_{-0.00064}$ & $13.8657^{+0.0033}_{-0.0033}$ & - &  -\\
AMU1 	 & 1E\tablefootmark{2} &  $2.9726^{+0.0013}_{-0.0014}$& $57.94^{+0.55}_{-0.54}$ & $0.424^{+0.077}_{-0.077}$ & - & -\\
NGC5189 		& 3E & $1.7151718^{+0.0000065}_{-0.0000065}$ & $66.3145^{+0.0031}_{-0.0031}$ & $11.21^{+0.28}_{-0.28}$ & $5.38^{+0.28}_{-0.28}$ & $27.95^{+0.28}_{-0.28}$ \\
NGC2392 		 & 3E & $2.13769^{+0.00016}_{-0.00016}$ & $73.949^{+0.091}_{-0.091}$ & $1.310^{+0.011}_{-0.011}$ & $0.913^{+0.013}_{-0.013}$ & $0.7787^{+0.0097}_{-0.0097}$ \\
\hline
\end{tabular}
  \tablefoot{
   \tablefoottext{1}{This periodicity in LoTr\,5 corresponds to the rotation period.}
\tablefoottext{2}{Although the evidence for the flat model is larger in our analysis of AMU\,1, we can recover the 1-effect model with similar evidence to the flat model by discarding the data with a high dispersion. The fit then converges to the known orbital period but still with insufficient significance (see Section\,\ref{sect.known_binaries}).
 }
 }
\end{table*}

 \subsection{Already reported binaries from the literature}
   \label{sect.known_binaries}
   
Eight of the targets in the sample are either already known binary systems (confirmed through radial velocity observations) or candidates reported in the literature by using other techniques that still await confirmation. We do not include here our candidates from Cycle 1 of TESS presented in \cite{Aller2020}, which will be discussed in Sect.\,\ref{sect.new_candidates} with the rest of the sample. The eight already known binaries are: Abell\,63, Abell\,46, AMU\,1, DS\,1, Abell\,30, NGC\,2392, NGC\,5189, and LoTr\,5. The first seven are short-period binaries and we are able to recover the expected variability in the TESS data. In all the cases except one (Abell\,63), the contamination in the aperture mask from nearby stars is less than 10\%. The remaining system is the long-period binary LoTr\,5. In this case, we detect the variability associated with the rotation of one of the stars in the binary nucleus but obviously not the variability associated to the orbital period of the binary \citep[we note that the central star of LoTr\,5 has one of the longest orbital periods -- P$\sim$2700 days -- of all the known central stars of PN,][]{Jones2017}. In the following, we revisit and briefly discuss each system, and present our new analyses. The results are summarised in Table~\ref{table:results_fit_known}.

 \paragraph{{\bf Abell\,63 and Abell\,46.}} The eclipsing binary nuclei of PNe Abell\,63 and Abell\,46 have been extensively studied by several authors in the past years. Their nuclei, named UU\,Sge and V477\,Lyr, respectively, consist of a primary hot white dwarf or subdwarf and a low-mass star. Their orbital periods are 0.465 days for UU\,Sge \citep{Bell1994} and 0.472  days for V477\,Lyr \citep{Pollaco-Bell1994}. 
 
 Both systems have been the subject of detailed modelling in the literature \citep[e.g.][]{Afsar-Ibanoglu2008}, in order to constrain the physical properties of the binary components (masses, temperatures and radii). In the case of Abell\,46, for example, this has resulted in a potentially post-RGB scenario for the central star based on the low derived luminosity for the primary component.  However, the derived temperature and surface gravity do not lie on post-RGB evolutionary tracks for the derived mass \citep[instead being consistent with a much lower mass;][]{Jones2022}. Our photometric modelling does not take into account eclipses (see, for instance the TESS light curve from Abell\,63 in Fig.~\ref{fig:Abell63LC}) and, therefore, the analysis of these light curves can not be performed by using the methodology explained in Sect.\,\ref{Sect:modeling}.
 
We, therefore, decided to model the system using a version of the PHOEBE2 code \citep{Prsa16,Horvat2018,Jones2020,Conroy2020}, adapted to incorporate model atmospheres calculated by \cite{Reindl2016, Reindl2023} using the T\"ubingen Model Atmosphere Package \citep[TMAP;][]{rauch03,werner03, tmap2012}, and previously used to model the binary central star of the PN Ou~5 \citep{Jones2022}.  This new approach also has the benefit of using more appropriate TMAP atmospheres for the primary star, where previous models either assumed \citet[more suitable for main sequence stars]{Kurucz1993} or black body atmospheres - both of which can lead to important differences in the derived temperature (potentially leading to the aforementioned mismatch with evolutionary tracks).

 Preliminary explorations of the parameter space, fixing the stellar masses to those from previous modelling efforts (as they are based primarily on the observed radial velocities), indicated that a solution consistent with evolutionary tracks should be possible. As such, we ran a MCMC sampling of the binary parameters, allowing the primary temperature to vary but forcing its radius to lie on the appropriate evolutionary track of \citet{Miller-Bertolami2016}, while leaving the temperature and radius of the companion free.  The resulting best fit is shown in Fig.\,\ref{fig:results_known_binaries}, and the best-fitting parameters are listed in Table \ref{tab:A46fit}. Ultimately, a more detailed fit is required (taking into account multi-band ground-based photometry as well as directly fitting the observed radial velocities - both of which are beyond the scope of this work) before claiming that the discrepancy between parameters from evolutionary tracks and binary modelling has been resolved or that the new parameters (and their associated uncertainties) are more reliable than those in the literature. Nevertheless, the fit presented here highlights that the high quality data from TESS combined with more advanced modelling tools has the potential to improve the derived parameters of post-CE binary central stars of PNe.

  \begin{table}
  \setlength\extrarowheight{3pt}
  \centering
   \caption{PHOEBE2 model parameters for V477\,Lyr, the central star of Abell\,46.}
   \label{tab:A46fit} 
\begin{tabular}{rl}
  \hline    
Parameter & Value\\
  \hline \hline    
Primary temperature ($T_1$) & 89.90$\pm$0.21~kK\\
Primary radius ($R_1$)\tablefootmark{$\dagger$} & 0.2010$\pm$0.0011~R$_\odot$\\
Primary mass ($M_1$)\tablefootmark{*} & 0.51~M$_\odot$\\
Secondary temperature ($T_2$) & 7.26$\pm$0.13 ~kK\\
Secondary radius ($R_2$) & 0.4317$\pm$0.0010~R$_\odot$\\
Secondary mass ($M_2$)\tablefootmark{*} & 0.15~M$_\odot$\\
Binary inclination ($i$)\tablefootmark{*} & 80.5$^\circ$\\
\hline
\end{tabular}
\tablefoot{
    \tablefoottext{$\dagger$}{Derived from the evolutionary tracks of \citet{Miller-Bertolami2016}.}
   \tablefoottext{*}{Fixed in the model to match previous modelling efforts \citep{pollacco93,Afsar-Ibanoglu2008}.}
}
\end{table}

 \paragraph{{\bf DS\,1.}} In the nucleus of this PN there is a well-known double-lined spectroscopic binary (named KV Vel or LSS 2018) with an orbital period of $\sim$0.357 days \citep{Drilling1985, Kilkenny1988}. It is a post-common envelope system with an unusual and extremely strong reflection effect, with an amplitude of 0.55 mag in V  \citep{Hilditch1996}. The TESS light curve clearly shows the same variability with a prominent signal at the same periodicity as the published orbital period. Preliminary fitting of the light curve using PHOEBE2 (following the scheme outlined above for the central star of Abell~46) indicates that the data are entirely consistent with the parameters derived by \cite{Hilditch1996}, albeit with a slightly lower albedo for the secondary (0.55$\pm$0.01 c.f.\ 0.6).

 \paragraph{{\bf Abell\,30.}} \cite{Jacoby2020} reported the presence of light curve brightness variations in the K2 mission \citep{howell14} data of the central star of this born again PN. Although it has not been confirmed through radial velocity observations, the authors concluded that these variations in the light curve were highly suggestive of a binary central system with a period of $\sim$1.060 days. After analysing other possible physical processes, \cite{Jacoby2020} proposed the irradiation of a cooler companion as the most likely origin for the observed photometric variability in this system, and discarded other possible effects as Doppler beaming and ellipsoidal modulations. We find the same variability in the three sectors of the TESS light curves. Figure\,B.2\footref{zenodo} shows the Lomb-Scargle periodogram, where the False Alarm Probabilities (FAPs) at 10\%, 1\% and 0.1\% are also indicated with grey horizontal dotted lines. The phase-folded light curve with the period derived in the fitting is plotted in the corresponding panel of Fig.\,\ref{fig:results_known_binaries}. The one sinusoidal model appears as the preferred solution, showing a large amplitude of 14 ppt (see Table\,\ref{table:results_fit_known}). However, we note that the evidence of a 3-effects model is almost the same as the simple model ($\mathcal{B}_{31}=-0.2$), with an ellipsoidal amplitude of 0.95$\pm{0.25}$ ppt. Indeed, these ellipsoidal modulations can be easily recognised in the residuals of the one sinusoidal model in  Fig\,\ref{fig:results_known_binaries}. However, although the time span covered by the three sectors of TESS (44, 45 and 46) is similar to that from the K2 mission, the precision of the TESS data is clearly worse than that from its predecessor and, therefore, the TESS light curves do not provide new information about the properties of the system.

 \begin{figure}
   \centering
   \includegraphics[width=0.5\textwidth]{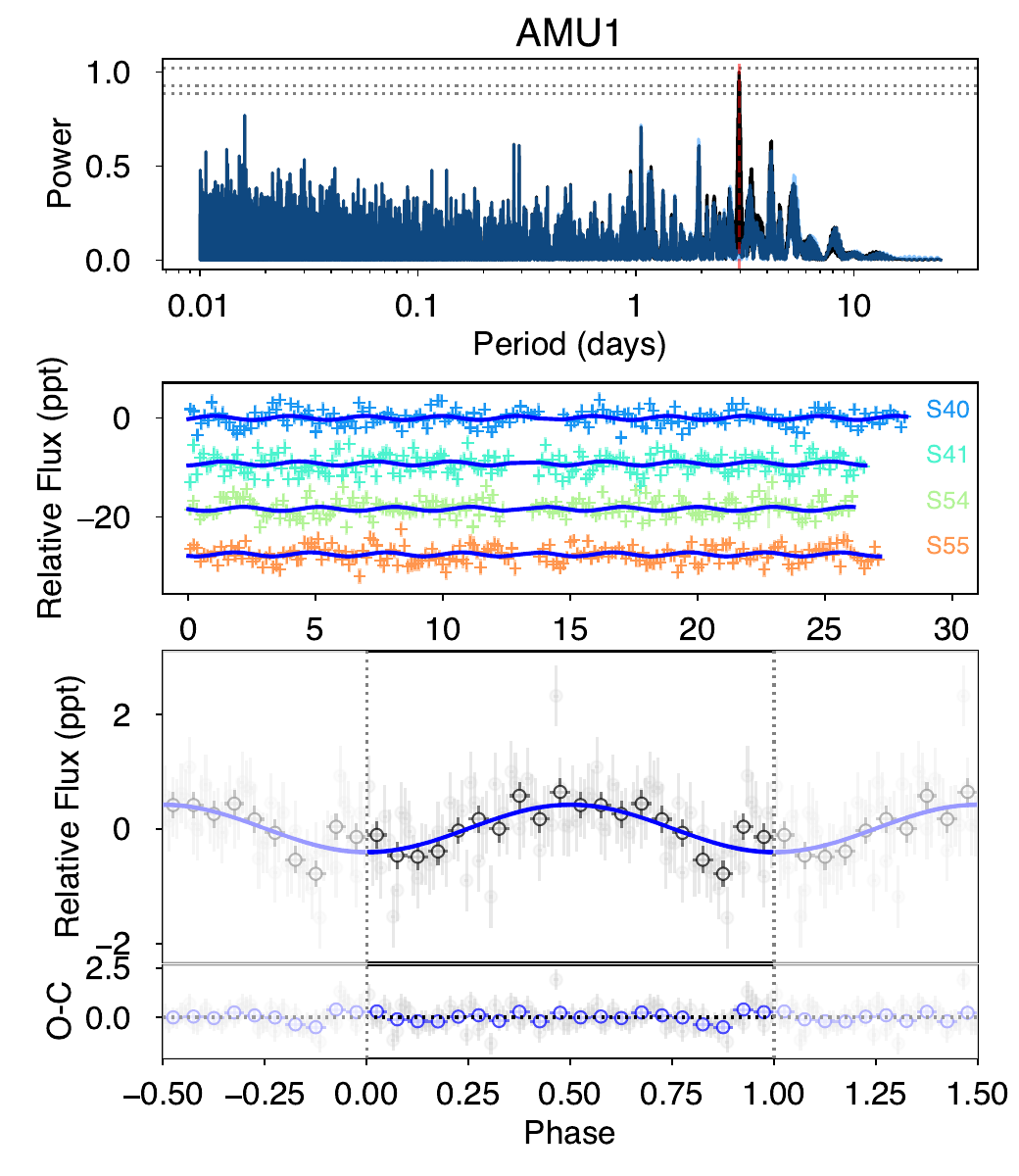}
       \caption{Periodogram ($\textit{upper panel}$), time-flux light curves for several sectors ($\textit{middle panel}$), and phase-folded single, multi-sector light curve ($\textit{bottom panel}$) of AMU\,1. Two different bin sizes are shown with black (bin size of 0.01 in phase) and grey (bin size of 0.05 in phase) symbols. The period derived from the fitting can be found in Table\,\ref{table:results_fit_known}.}
\label{fig:AMU1}
\end{figure}

 \paragraph{{\bf AMU\,1.}} The central star of this multipolar PN \citep{Aller2013} was found to be a binary with a short period of $\sim$2.928 days by \cite{DeMarco2015}. The Kepler light curve showed a photometric sinusoidal variability with a very low amplitude (0.73 ppt), consistent with relativistic beaming effects according to those authors. In the TESS light curves, we also detect a signal at the same periodicity, although the confidence level is much worse (approaching the False Alarm Probability at 10\,\% in some sectors) because of the lower photometric precision and shorter baseline of TESS compared to Kepler. Though the results from the TESS light curve analysis show that the evidence for the flat model is larger, we can recover the 1-effect model with similar evidence discarding the sector\,15, which has a large dispersion in the data. In this way, the MCMC chains converge to the known orbital period, although still with insufficient significance ($\mathcal{B}_{10}=-0.6$, see Figure\,\ref{fig:AMU1}). In summary, the TESS data do not provide any new information beyond what Kepler already showed. 

      \begin{figure}
        \centering
     \includegraphics[width=0.4\textwidth]{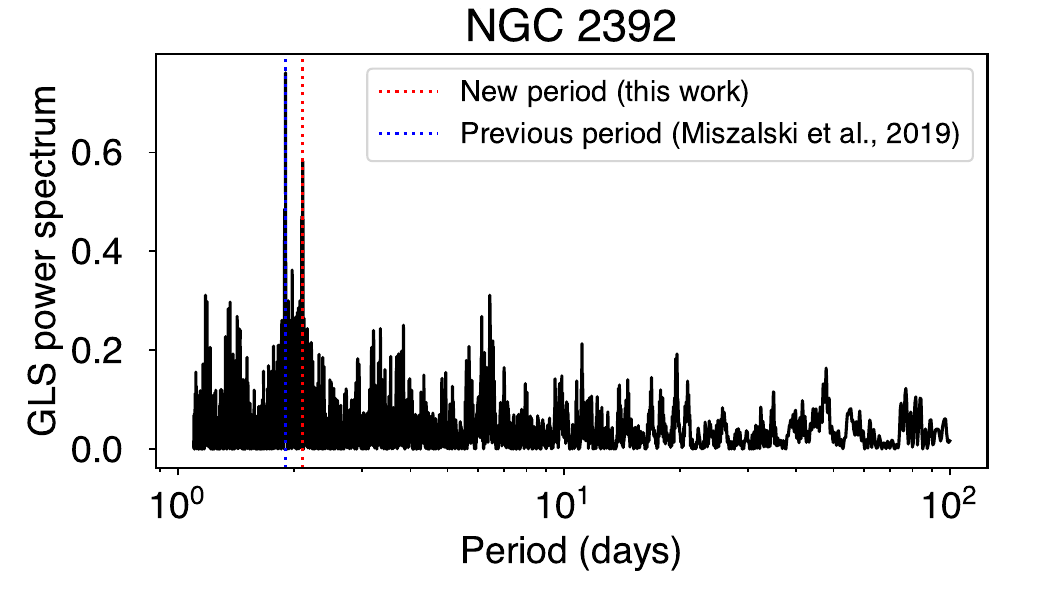}
          \includegraphics[width=0.4\textwidth]{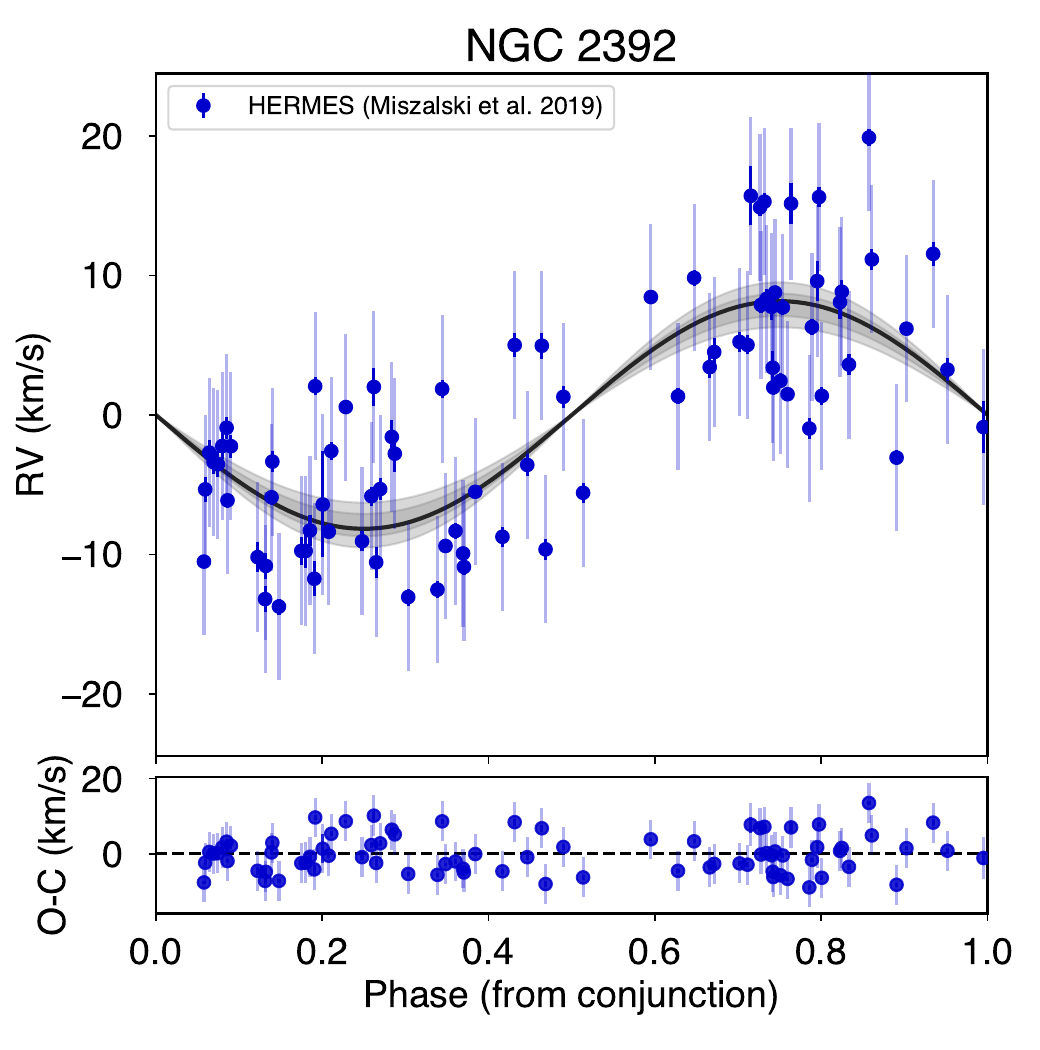}
       \caption{Lomb-scargle periodogram (upper panel) and radial velocity analysis (bottom panel) of NGC\,5189, based on the data presented in \cite{Manick2015}.}
\label{fig:NGC2392_RV}
\end{figure}

 \paragraph{{\bf NGC\,2392.}} The central star of this well-known PN was discovered to be a single-lined spectroscopic binary by \cite{Miszalski2019} after a radial velocity monitoring campaign. According to the authors, the binary system appears to be a double degenerate system with an orbital period of $\sim$1.9 days. The periodogram of the TESS light curve shows a forest of significant periodicities above the 0.1\% FAP (see Figure\,B.2\footref{zenodo}). Among all these signals, we can identify two pairs of peaks more prominent at around $\sim$4.2 and $\sim$2.1 days, but none at the exact periodicity of 1.9\,d found by \cite{Miszalski2019}. We note, however, that the peak at 2.1d is also present in the data by \cite{Miszalski2019}, although the authors chose to analyse the 1.9d because it was the highest in their RV periodogram. For that reason, we decided to re-analyse the radial velocity data presented in \cite{Miszalski2019}. Our independent analysis using the same MCMC principles explained for the analysis of the TESS light curves but now applied to the radial velocity model, shows that the data converges to an orbital period of 2.10842273 $\pm$ 0.000085 days, which is actually significant in the RV periodogram (see upper panel of Fig.~\ref{fig:NGC2392_RV}) and slightly longer than that derived by the authors with the same data. This result is in strong agreement with what we observe in the TESS light curve, and allows us to conclude that the 4.2 days periodicity seen in the light curve is, actually, an alias of the real period. This is a clear example of the importance of confirming the orbital period by more than one method. Taking this result into account, we forced the MCMC analysis to converge to the 2.1\,days periodicity and obtained that the best solution is a 3E model.

  \paragraph{{\bf NGC\,5189.}} The central star of this complex, quadrupolar PN was already analysed in our previous work \citep{Aller2020}. Now, we re-analyse it taking into account new TESS data as well as the previous data from Cycle 1, drawing similar conclusions. After a radial velocity monitoring campaign, \cite{Manick2015} identified the periodicity found at 4.04 days as the orbital period of the binary. The periodogram of the TESS light curve (sectors\,11 and 38) show several prominent peaks above the 0.1 $\%$ FAP (the most significant at 0.125, 0.858 and 1.71 d, see Figure\,B.2\footref{zenodo}) but none of them around the 4.04 days period previously reported. The result of MCMC analysis is, therefore, very similar to those presented in \citep{Aller2020}. A 3E model is the most plausible for the 1.71 d periodicity (the peak at 0.858\,d is certainly an alias of this one) and the amplitudes are listed in Table\,\ref{table:results_fit_known}. However, the resulted Doppler beaming amplitude is quite large and clearly not physically feasible for this type of object. This is a case with a lot of contamination from nearby stars in the photometric aperture so we cannot reach any firm conclusions from the TESS light curve. In fact, it is very likely that the origin of the periodicities we see in the TESS light curves does not correspond to our central star (see Section.\,\ref{sec:tess_localize} for more detail).

   \begin{figure}
     \includegraphics[width=0.5\textwidth]{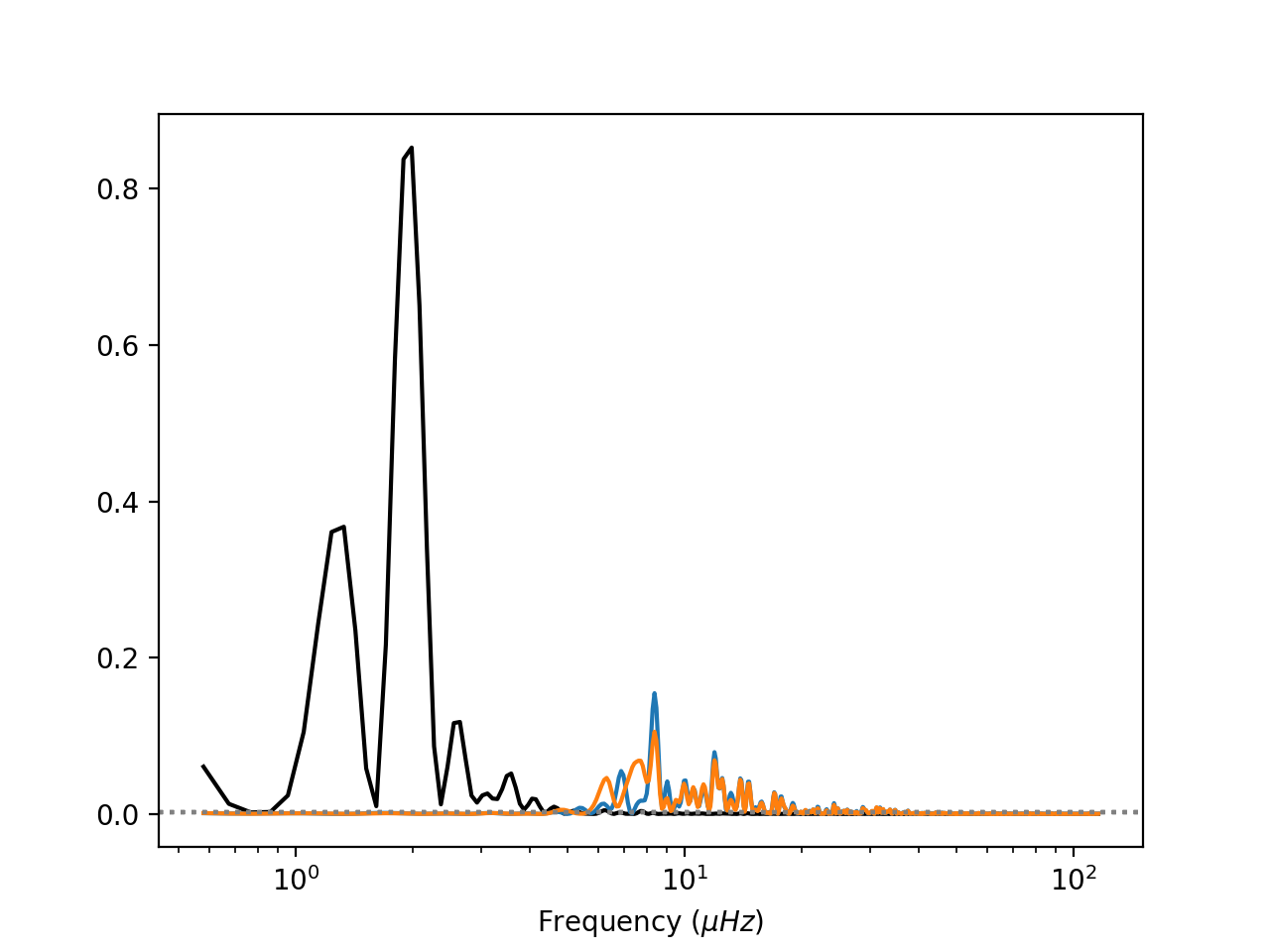}
\caption{Periodogram of LoTr\,5 after subtracting the rotation period with the {\sc wotan} package (in blue and orange, with two different window sizes in the \texttt{flatten} module) and the original periodogram (in black).}
\label{fig:wotan_LoTr5}
\end{figure}

 \paragraph{{\bf LoTr\,5.}} This PN is known for having one of the longest period binary central star \citep[P$\sim$2717 days;][]{VanWinckel2014, Jones2017} so far. The system consists of a hot star and a barium star with a rotation period of 5.95 days. A possible third component in the system was discussed by \cite{Aller2018} on the basis of new radial velocity observations, although without firm conclusions. The rotation period is clearly recovered in the TESS dataset (see Fig.\,\ref{fig:results_known_binaries}). In the MCMC analysis, the reflection and the beaming effects are both degenerate, and the ellipsoidal variations go clearly to zero. Therefore, the 1E model is the only one possible, although there are clearly more effects in the light curve (as, for example, oscillation signals). These other effects certainly show up when subtracting the rotation signal from the light curve. We made this by using the detrending methods implemented in the {\sc wotan} package \citep{Hippke2019}. Figure\,\ref{fig:wotan_LoTr5} shows the resulting periodogram of LoTr\,5 after this detrending process. A more detailed analysis of these frequencies would be desirable to obtain precise information on the physical parameters of the pair, but this is out of the scope of this paper.

 \begin{table*}
 \setlength\extrarowheight{3pt}
\centering 
   \caption{Results from the light curve analysis for those central stars with the 1-effect model as the best solution.}
   \label{table:results_fit_candidates_1E} 
\begin{tabular}{ccccc}
  \hline    
Name & BestModel & Period& $T_0$-2458300 & A$_{\rm 1E}$ \\
&&      (days)  &  (days) & (ppt)\\
  \hline \hline     
NGC7293* 		 & 1E & $2.79151^{+0.00021}_{-0.00021}$ & $68.592^{+0.046}_{-0.046}$ & $1.48^{+0.11}_{-0.11}$  \\
NGC2867* 		 & 1E & $4.5762^{+0.0055}_{-0.028}$ & $58378.41^{+1.1}_{-0.42}$ & $0.316^{+0.042}_{-0.042}$\\
RWT152* 			& 1E & $6.0090^{+0.0015}_{-0.0015}$ & $75.59^{+0.17}_{-0.17}$ & $0.721^{+0.071}_{-0.071}$  \\
PG1520+525		& 1E & $5.1992^{+0.0036}_{-0.0035}$ & $67.22^{+0.87}_{-0.91}$ & $4.13^{+0.66}_{-0.65}$ \\
PNG136.7+61.9 	 & 1E & $8.477^{+0.062}_{-0.058}$ & $68.2^{+8.7}_{-9.2}$ & $25.3^{+2.7}_{-2.7}$  \\
Pa165 			 & 1E & $3.836^{+0.019}_{-0.023}$ & $64.6^{+8.7}_{-7.1}$ & $0.639^{+0.075}_{-0.076}$  \\
WPS28 			 & 1E & $2.82171^{+0.00035}_{-0.00035}$ & $62.27^{+0.15}_{-0.15}$ & $1.26^{+0.12}_{-0.12}$ \\
Fr1-4 			 & 1E & $6.4446^{+0.0048}_{-0.0049}$ & $78.20^{+0.64}_{-0.63}$ & $5.14^{+0.60}_{-0.60}$ \\
NGC7094 		 & 1E & $4.263^{+0.012}_{-0.017}$ & $59.9^{+5.7}_{-4.0}$ & $2.45^{+0.14}_{-0.14}$ \\
NGC2371 		 & 1E & $2.30726^{+0.00034}_{-0.00034}$ & $70.34^{+0.16}_{-0.16}$ & $0.828^{+0.089}_{-0.089}$ \\
Hen3-1863 		 & 1E & $5.19725^{+0.00038}_{-0.00038}$ & $63.979^{+0.051}_{-0.050}$ & $31.804^{+0.019}_{-0.019}$  \\
WPS54 			 & 1E & $1.72669^{+0.00003}_{-0.00003}$ & $71.2247^{+0.0214}_{-0.0212}$ & $30.819^{+0.74}_{-0.74}$  \\
 			  & 1E & $3.45297^{+0.00005}_{-0.00005}$ & $69.6921^{+0.0165}_{-0.0164}$ & $74.794^{+0.74}_{-0.75}$  \\
\hline
\end{tabular}
 \tablefoot{
\tablefoottext{*}{Targets already analysed in Cycle\,1 \citep{Aller2020}.
 }}
\end{table*}


\begin{table*}
\setlength\extrarowheight{3pt}
 \centering
   \caption{Results from the light curve study for those central stars with negligible or minimal contamination analysed with models including reflection, ellipsoidal, and Doppler beaming.}
   \label{table:results_fit_candidates_3E} 
\begin{tabular}{ccccccc}
  \hline    
Name &  BestModel & Period & $T_0$-2458300 & A$_{\rm ref}$ & A$_{\rm beam}$ & A$_{\rm ellip}$\\
&&      (days)  &  (days) & (ppt)& (ppt)& (ppt)\\
  \hline \hline     
PG1034+001* 		 & 3E & $1.856796^{+0.000012}_{-0.000012}$ & $55.2250^{+0.0061}_{-0.0062}$ & $0.0091^{+0.015}_{-0.0068}$ & $4.202^{+0.063}_{-0.064}$ & $3.051^{+0.063}_{-0.062}$ \\
NGC\,246* 			 & 3E & $4.99574^{+0.00072}_{-0.00073}$ & $61.030^{+0.096}_{-0.093}$ & $0.238^{+0.035}_{-0.035}$ & $0.103^{+0.037}_{-0.037}$ & $0.256^{+0.034}_{-0.035}$ \\
IC\,2149 & 3E & $1.31274^{+0.00066}_{-0.00065}$ & $80.300^{+0.227}_{-0.230}$ & $7.545^{+0.126}_{-0.127}$ & $1.030^{+0.249}_{-0.255}$ & $2.091^{+0.136}_{-0.133}$ \\
NGC\,1501 		& 3E & $3.2830^{+0.0053}_{-0.0043}$ & $80.05^{+0.59}_{-0.72}$ & $1.74^{+0.12}_{-0.12}$ & $0.50^{+0.13}_{-0.13}$ & $2.03^{+0.12}_{-0.12}$ \\
IC\,4593 			 & 3E & $2.50369^{+0.00061}_{-0.00064}$ & $63.70^{+0.34}_{-0.33}$ & $12.772^{+0.051}_{-0.053}$ & $0.0049^{+0.0087}_{-0.0037}$ & $2.683^{+0.058}_{-0.055}$ \\
K\,1-16 			 & 3E & $3.86624^{+0.00046}_{-0.00043}$ & $80.69^{+0.14}_{-0.16}$ & $0.56^{+0.15}_{-0.15}$ & $0.99^{+0.13}_{-0.13}$ & $0.68^{+0.12}_{-0.12}$ \\
\hline
\end{tabular}
 \tablefoot{
\tablefoottext{*}{Targets already analysed in Cycle\,1 \citep{Aller2020}.
 }}
\end{table*}


 \subsection{New binary candidates from TESS}
  \label{sect.new_candidates}

In this section, we describe the results from the analysis of those central stars with negligible or minimal contamination in the TPFs and showing variability in the light curves compatible with the presence of (previously unknown) binary systems. In addition, we also revisit those binary candidates already analysed in Cycle\,1 \citep{Aller2020}, even if they are found in severely contaminated fields. In total, we analyse in this section 18 central stars. The results from the analysis are summarised in Tables~\ref{table:results_fit_candidates_1E} and ~\ref{table:results_fit_candidates_3E}, depending on whether the best fit in the MCMC analysis is a 1E or a 3E solution, respectively.

We find that the variability of 12 central stars is better explained based on the current data with a simple sinusoidal model (see Table~\ref{table:results_fit_candidates_1E}). Three of them (NGC\,7293, NGC\,2867 and RWT\,152) are indeed binary candidates from Cycle\,1 \citep{Aller2020} that we have re-analysed by taking into account the new observations from Cycles 2, 3 and 4. For two of these targets (NGC\,2867 and NGC\,7293), the new results are compatible to those presented in \cite{Aller2020}. In the case of NGC\,2867, and after analysing sectors\,9, 10, 35 and 37 together, we detect a significant periodic variability at $\sim$ 4.6 days with an amplitude of 0.3 ppt, similar to those from our previous paper and thus reinforcing the binary status of this central star. It is also important to highlight that in this case the 3E model has a larger evidence than the 1E model with a log-evidence of $\mathcal{B}_{31}=+3.4$ and the same periodicity. As shown in the corresponding panel of Fig.~\ref{fig:phase_new_candidates}, two bumps are appreciated at around phases 0.25 and 0.75, which is the reason for the 3E model having such a large evidence. However, the current dataset is still insufficient to prefer the more complex 3E model. It is also important to emphasise that the photometric aperture of NGC\,2867 is quite crowded with other stars, so the origin of the variability is not entirely clear (see also Section\,\ref{sec:tess_localize}). The results from the analysis of NGC\,7293 (the well-known Helix Nebula) adding the new sectors\,28 and 42, are also quite similar to those obtained in \cite{Aller2020}. A clear photometric variability is identified with a periodicity of $\sim$2.8 days and amplitude of 1.5 ppt. Although, in this case, there is no contamination from nearby sources in the aperture, the origin of the variability may be also questionable (see Section\,\ref{sec:tess_localize}). Only for the case of RWT\,152, the new dataset from sector\,34 leads to different conclusions than in Cycle\,1. Now, apart from the forest of significant periodicities that we already saw in sector\,7, a new clear signal at $\sim$ 6 days shows up in the periodogram (see corresponding panel in Fig.\,B.3), and the MCMC converges to that period by using a single sinusoidal model. As explained in \cite{Aller2020}, some of the found periodic variabilities can be caused by the presence of spots. In any case, dedicated observations are required to confirm or discard the presence of a companion star.  

Another nine central stars in the sample present variability compatible with a 1E model. They are: PG\,1520+525, PNG\,136.7+61.9, Pa\,165, WPS\,28, WPS\,54, Fr\,1-4, NGC\,7094, NGC\,2371 and Hen\,3-1863. Fig.\,\ref{fig:phase_new_candidates} shows the phase-folded light curves with the periods derived in the fitting (see also Table\,\ref{table:results_fit_candidates_1E}). Objects with 1E model solution are plotted in blue. We briefly discuss here some key aspects of them as follows:

$PG\,1520+525$: This is the only "classical" round \citep[with a ring appearance,][]{Jacoby-VanDeSteene1995} PN in the sample showing variability in the TESS light curve. We find a periodicity of $\sim$5.2 d with an amplitude of $\sim$ 4.13 ppt, compatible with irradiation on a close companion. Although there are no other stars identified by \textit{Gaia} inside the TESS aperture within $\Delta$m = 6 magnitudes, the location of the variability is not clearly associated with the central star (see Sec.\,\ref{sec:tess_localize}), so dedicated follow-up is desirable to confirm the origin of the detected variability.

$PNG\,136.7+61.9$: This extremely faint PN\footnote{\url{https://telescopius.com/spa/pictures/view/93299/deep_sky/by-boris_us5wu}} with elliptical appearance, has a DAO (hydrogen and helium) white dwarf central star, suggested as a post-RGB candidate by \cite{Reindl2023}. The authors did not find significant light curve variations in the TESS data. We find a 25 ppt variability amplitude in the only sector available in TESS that we could attribute to the irradiation of a cool companion.

$Pa\,165$: Although the 1E solution is the most likely for this central star with an irregular PN, the 3E model has also large evidence ({$\mathcal{B}_{31}=-3.6$). The low amplitude of the variability (0.65 ppt) is compatible with either irradiation or relativistic effects. 

$WPS\,28$: The light curve of this white dwarf (with a possible PN around it according to the HASH PN database) exhibits a clear sinusoidal variability with a period of $\sim$ 2.8 days. The large temporal coverage of the TESS data, with a total of 25 sectors from Cycles 1 to 4, provides an exquisite precision. 
Interestingly, the phase-folded light curve with the period corresponding to this variability also shows a small dimming at the minimum of the relative flux (see corresponding panel in Fig.~\ref{fig:phase_new_candidates}). Although we note that this dimming is not yet statistically significant, this is the expected location (within the uncertainties), of an eclipse when the variability is due to the reflection effect and the orbital inclination is such that the companion passes between the stellar disk and our line of sight. In order to further test the source of the variability and this tentative eclipse signal, we use a modified version of the TESS positional probability software (\texttt{tpp}\footnote{\url{https://github.com/ahadjigeorghiou/TESSPositionalProbability}}) described in \cite{tpp} and \cite{lillo-box24} to compute the probability of each \textit{Gaia}-detected source in the field around the target as being the true source of the transiting signal. By doing so, we find that the star with the largest probability of being the transit host is actually $WPS\,28$, although still with a 32\% probability, which makes highly interesting the follow-up of this target. However, their TPFs show severe contamination from other stars in all the sectors except one (sector 21, which has minimal contamination). Additional observations are thus critical to ensure the source of the variability (but see also Sect.~\ref{sec:tess_localize}). 

$WPS\,54$: This star was classified as a RS CVn type star by \cite{Drake2014} based on their Catalina Sky Survey light curve. The RS CVn variable type consists of binary systems with high chromospheric activity FGK companions. After analysing the Catalina light curve, \cite{Werner2019} found a variability with a periodicity of 3.45 days, that the authors interpreted as the white dwarf's rotation period. The TESS light curve actually shows this periodicity in the periodogram (see Fig.\,\ref{fig:WPS54_fits}). Although we acknowledge that the variations in the light curve may be due to multiple spots on the star, we still try our modelling process. The 1E model on this periodicity produces large residuals, suggestive of a wrong model (see Fig.\,\ref{fig:WPS54_fits}, upper panel). The 3E model on this periodicity still produce large residuals with the presence of a phase-shifted signal at half of the periodicity (1.7d; see also Fig.\,\ref{fig:WPS54_fits}, upper panel). We then decided to simultaneously model a one sinusoidal model for the 3.45d signal (assuming it as the rotation period of the white dwarf) and 1E or 3E model to the 1.7d periodicity. By doing so, we find that the residuals of both 1E and 3E model are largely reduced (see Fig.~\ref{fig:WPS54_fits}, bottom panel), with the 1E model being preferred against the 3E model. Still, this result may point to either irradiation with 1.7d orbital period or perfectly symmetric ellipsoidal effect with a 3.45d period. Both scenarios are thus very interesting because any of them suggest a binary companion with a coupled orbital period commensurable with the rotation period of the star, suggestive of a coupling between the orbital and rotation frequencies.

$Fr\,1-4$: The central star of this possible (according to the HASH database) PN  is the hot subdwarf JL\,102. The single TESS light curve of this target (produced by combining sectors 27 and 39) shows a clear variability with $\sim$ 6.4 d period. The corresponding signal in the periodogram is very strong and our analysis reveals a 1E model with $\sim$ 5 ppt amplitude, which is compatible with irradiation on a stellar companion. 

$NGC\,7094$: This object was classified as a PN by \cite{Kohoutek1963}. NGC\,7094 has a filamentary structure, with a shell deviating  sphericity \citep{Rauch1999}, and a PG1159-type central star \citep{Lobling2019}. The variability in the TESS light curve is very clear, with a strong peak at $\sim$ 4.3 d in the periodogram, which may be perfectly compatible with irradiation of a companion star. The contamination from other field stars inside the TESS aperture is negligible and NGC\,7094 is one of the most promising candidates to reside in a binary system (see Sect.~\ref{sec:tess_localize}).

$NGC\,2371$:  This complex PN has been extensively studied in the literature \citep[see, for example, ][]{Gomez-Gonzalez2020, Ramos-Larios2012b}. It exhibits a high-excitation, multipolar morphology with a Wolf-Rayet-type as a central star. This central star is a well-known pulsator \citep{Ciardullo-Bond1996}, whose  TESS light curve was recently analysed by \cite{Corsico2021}. The authors detected several pulsation periods for this star between 878.5 s and 1032.6 s \citep[shorter than the pulsation frquencies previously reported by][from the ground]{Ciardullo-Bond1996}, but no evidence of binary signatures. On the contrary, we find several periodicities above the 0.1\% FAP in the TESS periodogram, that populate a wide region at lower frequencies (most of them between 0.3 and 4 days, see Fig.\,B.3\footref{zenodo}). Our analysis converges to the one at $\sim$ 2.3 days, showing a low-amplitude ($\sim$ 0.8 ppt) sinusoidal variability.

$Hen\,3-1863$: The light curve of this central star is highly dominated by a strong variability, which appears to have its origin in the central star (see Sect.~\ref{sec:tess_localize}). The periodogram shows a strong a broad peak at $\sim$ 4.3 d (see Fig.\,B.3) and our MCMC analysis converges to that solution. The phase-folded light curve with that periodicity shows a structure at phase $\phi=0.5$ and large amplitude high-frequency variations that are not explained with our sinusoidal model (see corresponding panel in Fig.\,\ref{fig:phase_new_candidates}), indicating that additional effects are present in the data. The nature of these effects are out of the scope of this paper. However, the large scale variability at 4.3 days is clear and well represented by our model.

On the other hand, we found 6 targets showing a 3E solution as the most likely one. The phase-folded light curves with the periods derived in the fitting of each of them are shown in Fig.\,\ref{fig:phase_new_candidates}, with the 3E model solution overplotted in red.

Among these, two (NGC\,246 and PG\,1034+001) are binary candidates from Cycle\,1 \citep{Aller2020} that  we re-analyse by adding the new data from Cycles 2-4. The results for PG\,1034+001 are practically the same as in \cite{Aller2020}, with a strong periodicity at $\sim$ 1.86 days and evident ellipsoidal modulations. We invite the reader to revise our previous paper for more details. The case of NGC\,246 is quite different. The central star of this elliptical PN is a known hierarchical triple stellar system \citep{Adam2014}. In \cite{Aller2020}, we found a simple sinusoidal model with a period of $\sim$ 6.8 days as the best solution. With the new data (sector\,30), the MCMC analysis converges to 3E solution at $\sim$ 5 days. However, we note that both signals may have the same origin, since the peak is quite broad in the periodogram (see Fig.\,B.3\footref{zenodo}), which is dominated by the low frequencies (in the range of 500-700 $\mu$Hz) associated with pulsations \citep{Ciardullo-Bond1996}. Also, it is important to remember that this is one of the targets with substantial photometric contamination by other nearby stars in the TPF (see corresponding panel in Fig.\,B.1\footref{zenodo}), which makes it extremely complicated to draw firm conclusions on the possible binary scenario.

Other 4 objects in the sample present variability compatible with a 3E modulation. They are briefly discussed as follows: 
 
$NGC\,1501$: The central star of this PN is a pulsating hydrogen-deficient, pre-white dwarf. Its TESS light curve was already analysed by \cite{Corsico2021} in order to search for pulsation frequencies and binary signatures. Although evidences for the latter scenario were not found, the authors presented asteroseismological analysis for the central star. We find several significant peaks in the periodogram, apart from the low-frequency signal analysed by \cite{Corsico2021}, which may be compatible with the presence of another star in the system. Our analysis converges to the one at $\sim$ 3.3 days by using a 3E solution, showing a low ($\sim$ 0.5 ppt) beaming amplitude that is compatible with relativistic effects in the system.

$K\,1-16$: As NGC\,1501, this is also a pulsating hydrogen-deficient, pre-white dwarfs whose TESS light curves were already analysed by \cite{Corsico2021}. Binary signatures were not found and their asteroseismological analysis was quite limited in this case because of the dramatic changes in the pulsations of this star. The periodogram shows a forest of significant periodicities (between 1 and 11 days, approximately, see corresponding panel in Fig.\,B.3\footref{zenodo}). The MCMC analysis found the second most prominent peak (corresponding to $\sim$ 3.9 days) as the most probable, with low amplitudes for the three effects (irradiation, ellipsoidal modulations, and Doppler beaming), although the beaming amplitude still is unrealistically large. Therefore, the origin of this variability is hard to interpret due to the large amount of different signals. Radial velocity observations are crucial to confirm the nature of this variability.

$IC\,2149$: The central star of this elongated PN shows a high degree of variability in the TESS light curve, resulting in a large number of periodicities above the 0.1\% FAP in the periodogram (see Fig.\,B.3\footref{zenodo}). Given the large amplitude and high-frequency of these variations and the fact that we are interested in longer periodicities, we included in the modelling an extra jitter noise term to account for these short-period signals in a proper way. After doing so, the preferred solution for the MCMC fit is a 3E model with a period of 1.31 days. The derived Doppler beaming amplitude (A$_{\rm beam}$ = $1.01 \pm 0.25$~ppt) is indeed too large to be realistic and might still be driven partly by the high-frequency variability. However, it is important to note that its uncertainty is still large and might be compatible with 0 at 4$\sigma$.

$IC\,4593$: As the case of the previous two PNe, the periodogram of this central star is dominated by several peaks above the 0.1\% FAP. However, there are three of them (between 2 and 4 days) that predominate above the rest. The analysis converges to the one at 2.6 days, with a large reflection amplitude (A$_{\rm ref}$ $\sim$ 12.8 ppt), a moderate A$_{\rm beam}$ of $\sim$ 2.7 ppt and no Doppler beaming.

 \begin{figure}
\includegraphics[width=0.5\textwidth]{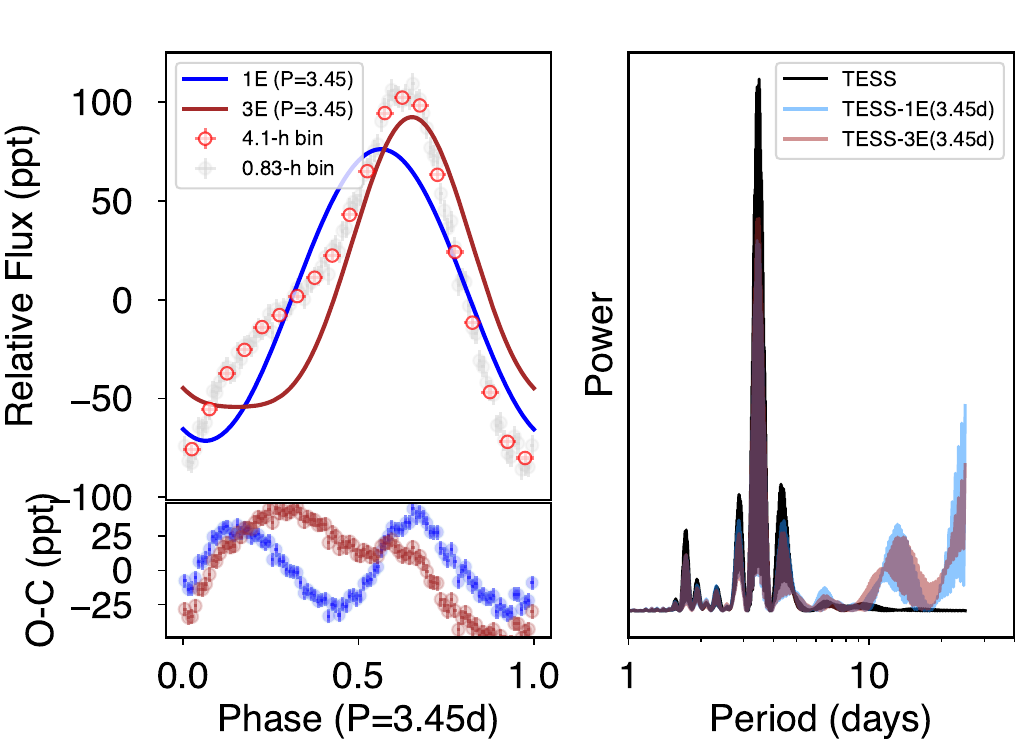}
\includegraphics[width=0.5\textwidth]{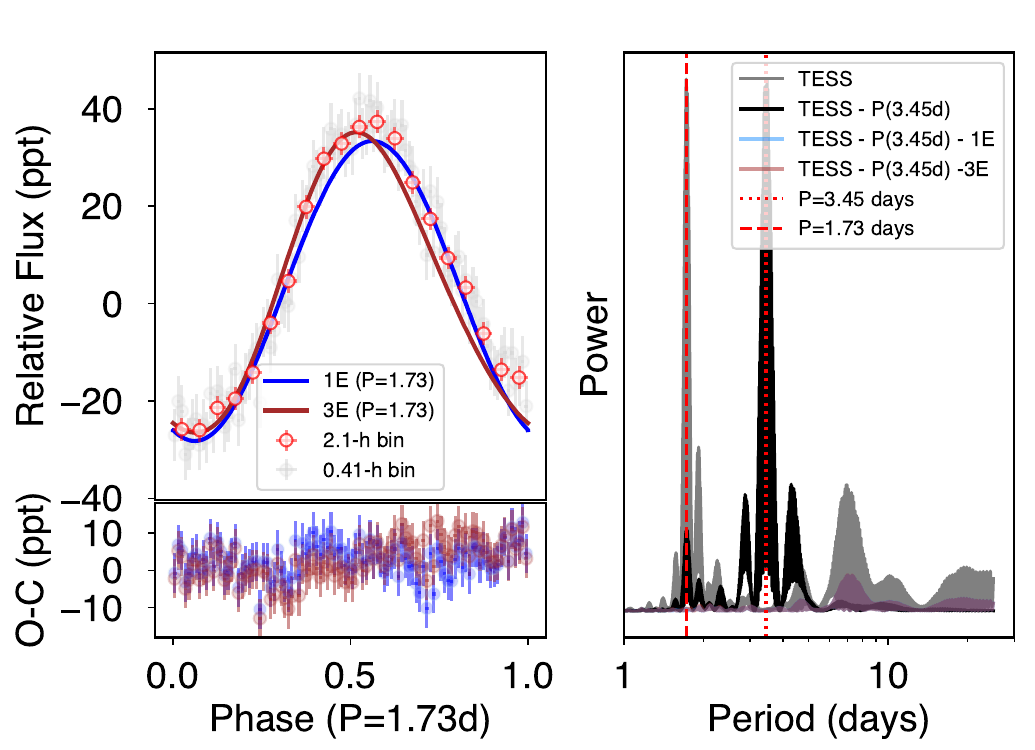}
 \caption{Modelling of the WPS\,54 TESS light curve. Top panels: Phase-folded light curve (top-left) with the periodicity corresponding to the largest peak in the periodogram (top-right panel), including the 1E (blue) and 3E (red) models and their corresponding residuals. The model is not satisfactory due to the presence of additional signals. Bottom panels: Phase-folded light curve (bottom-left) with the 1.73d periodicity after removing the sinusiondal patter corresponding to the 3.45d (potentially the rotation period). The 1E (blue) and 3E (red) models are shown, as well as their corresponding residuals.}
  \label{fig:WPS54_fits}
\end{figure}


 \subsection{Targets with no variability or with weak evidence}
 \label{sect.novar}

We find 12 central stars (leaving aside AMU\,1, already discussed in Section\,\ref{sect.known_binaries}) in which the best solution is the flat model (see Table\,\ref{table:results_fit}). In general terms, we can discard sinusoidal variabilities down to the TESS sensitivity for these targets. Although in some few cases like AMU\,1 and Fr\,2-21 the MCMC finds compatible solutions with sinusoidal patterns, the Bayesian evidence still favours the simpler flat model. For all these non-detections, we can use the posteriors of the MCMC analysis to provide an upper limit to the amplitude of potential variability of a hypothetical companion. We use the marginalised posterior distribution of the 1E model ($A_{\rm 1E}$) and use its 95\% percentile as the upper limit. This is shown in Fig.\~B.5\footref{zenodo}, where we display the corner plots corresponding to the dependency between the period and  $A_{\rm 1E}$ as well as the marginalised distributions and the 95\% upper limit. 

At this point, there are some cases that deserve special attention. The case of Abell\,7 is intriguing since, in Cycle\,1 (sector\,5), we identified two clear significant (and independent) peaks in the periodogram at $\sim2.6$ and $\sim 3$ days \citep{Aller2020} with unclear origins. Interestingly, these peaks no longer appear in the new sector 32 data (see Fig.~B.4\footref{zenodo}), so that the possible binary scenario remains questionable. As a possibility, the detected variability in sector\,5 may be due to other non-binary scenarios as, for example, starspots.

We would also like to mention the case of StDr\,56\footnote{This PN was recently discovered by amateur astronomers Marcel Drechsler and Xavier Strottner, who named it the Goblet of Fire Nebula {\url{(https://www.astrobin.com/s9vmx1/)}}.}. It is a faint and filamentary nebula, in which central region Drechsler and Strottner identified two possible white dwarfs, either of which could be the central star of the nebula. It is worthy to note that, although the flat model is the best solution in the MCMC fitting, the periodogram of the TESS light curve shows a very clear significant peak at $\sim$ 7 days (see Fig.\,\ref{fig:StDr56}, top panels). However, the uncertainty of the data is still large compared to the amplitude of the candidate signal and so the flat model is preferred. This is a promising candidate for subsequent follow-up.

Another good binary candidate is the central star of Fr\,2-21. Although, as in the case of StDr\,56, the best solution for this target is the flat model (see Table\,\ref{table:results_fit}), the 1E and 3E models have comparably large evidences compared to the flat model ($\mathcal{B}_{10}=4.5$, $\mathcal{B}_{30}=5.0$). The 1E solution is shown in Fig.~\ref{fig:Fr2-21}, where we present the TESS time-flux light curve, the periodogram and the phase-folded light curve at the strongest peak ($\sim$ 2.5 days). For this analysis, several photometric points present just before and after the TESS data downlink have been removed, since they are typically the result of instrumental effects. It is also worth mentioning that a possible eclipse located at the expected minimum of the phase folded light curve is also hinted. However, the current data do not allow its confirmation.

Finally, the central stars of PNe JnEr\,1 and Abell\,28 were found to have infrared excess compatible with a possible companion spectral type later than M4 by \cite{Douchin2015}. The TESS light curves of these objects do not show any variability suggestive of a close binary scenario. Only in the case of Abell\,28 do we find a significant signal at 0.11 days above the 0.5 \% FAP in the periodogram (see Fig.\,B.4\footref{zenodo}), but the MCMC analysis does not converge to that peak. Similarly, \cite{DeMarco2013} detected infrared excess in Abell\,31 suggesting an M4V companion, in agreement with the works of \cite{Frew2008} and \cite{Ciardullo1999}. The high-precision photometry of the four sectors (34, 44, 45, 46) of TESS show no evidence of photometric variability at least in the time span covered by the light curves (see Fig.\,B.4). This is in accordance with the wide companion proposed by \cite{Ciardullo1999}.    

The rest of the targets which have no significant photometric variability in their light curves are HDW\,7, Lo\,1, EC\,13290-1933, Sh\,2-216, IC\,5148/50 and Fr\,2-46. Their periodograms, which do not show any peak above the FAPs, are also presented in Fig.\,B.4. Only for the case of Sh\,2-216, the most prominent peak (at $\sim$ 0.2 days) could be a significant signal with more data.
\\

 \begin{figure}
    \centering
\includegraphics[width=0.45\textwidth]{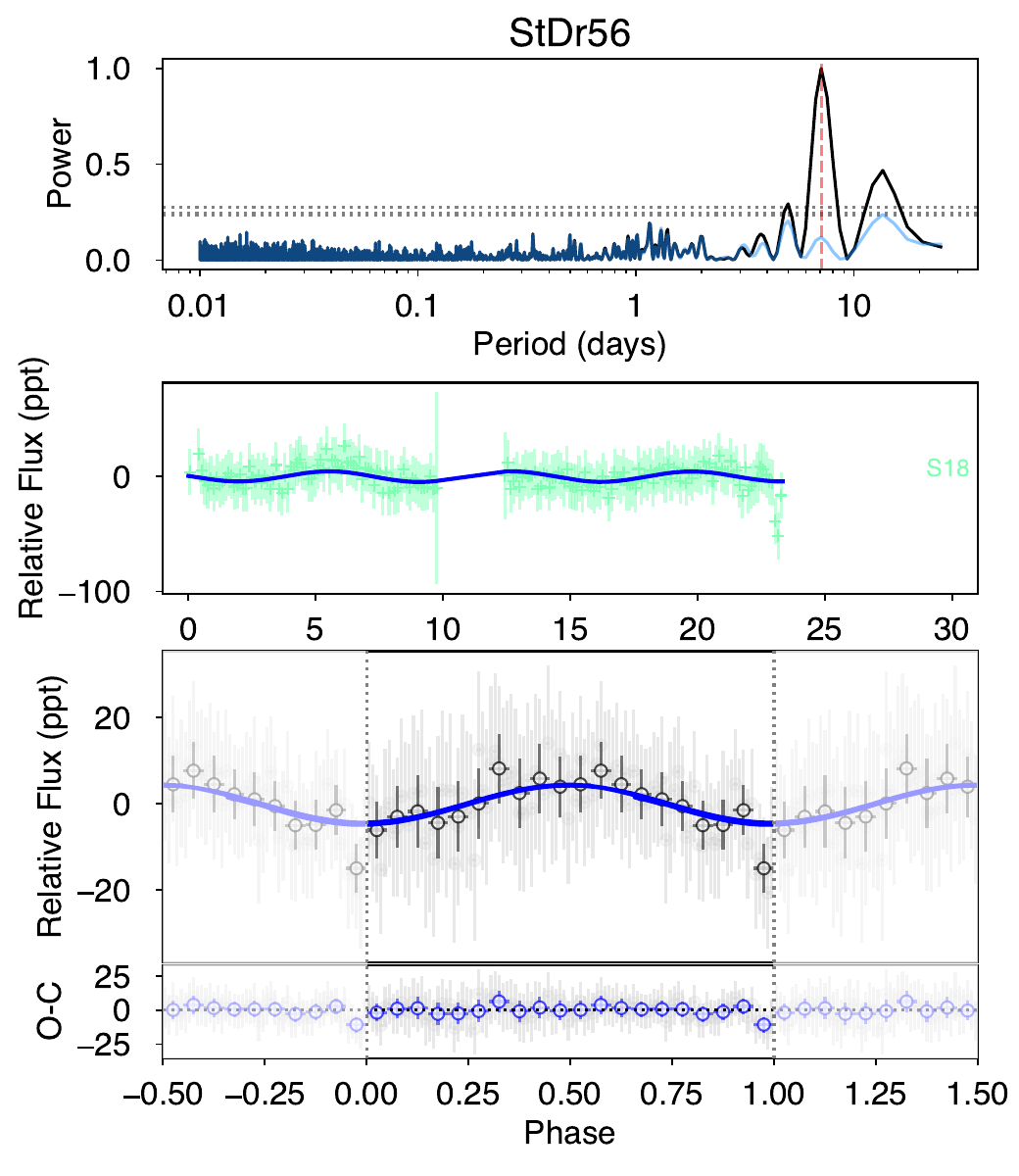}
    \caption{Periodogram ($\textit{upper panel}$), time-flux light curve ($\textit{middle panel}$), and phase-folded light curve ($\textit{bottom panel}$) of StDr\,56. Two different bin sizes are shown with black (bin size of 0.01 in phase) and grey (bin size of 0.05 in phase) symbols. The period derived from the fitting can be found in Table\,\ref{table:stdr_Fr2-21_values}.}
    \label{fig:StDr56}
\end{figure}

\begin{figure}
    \centering
\includegraphics[width=0.45\textwidth]{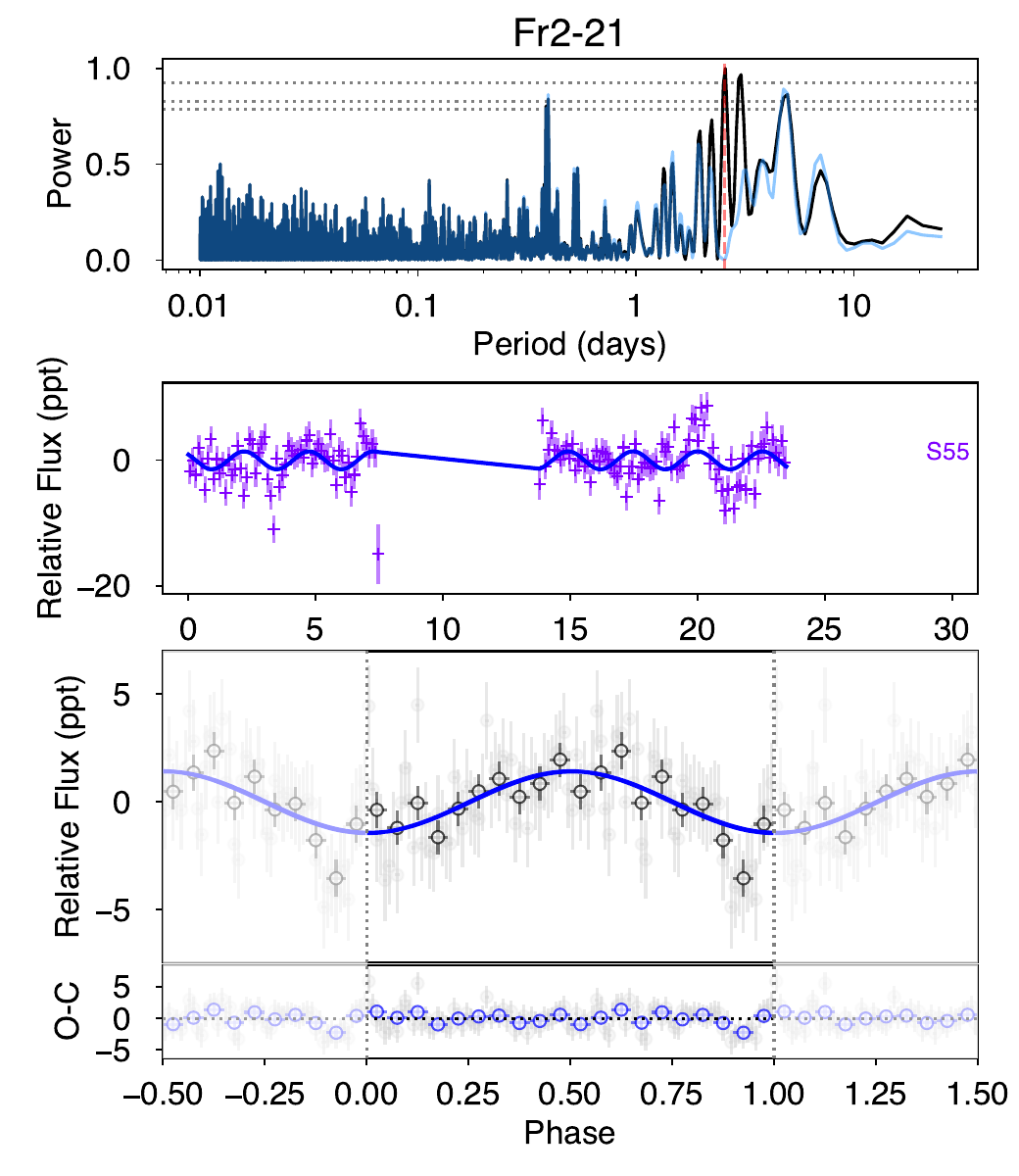}
    \caption{Periodogram ($\textit{upper panel}$), time-flux light curve ($\textit{middle panel}$), and phase-folded light curve ($\textit{bottom panel}$) of Fr\,2-21.  Two different bin sizes are shown with black (bin size of 0.01 in phase) and grey (bin size of 0.05 in phase) symbols.The period derived from the fitting can be found in Table\,\ref{table:stdr_Fr2-21_values}.}
    \label{fig:Fr2-21}
\end{figure}

 \begin{table}
\setlength{\extrarowheight}{3pt}
 \centering
  \caption{Results from the light curve study for the two central stars with weak evidence: StDr\,56 and Fr\,2-21.}
  \label{table:stdr_Fr2-21_values}
    \begin{tabular}{lccc}

  \hline    
Object   &      Period  &  $T_0$ - 2458300 & Amplitude \\
   &      (days)  &  (days) & (ppt)\\
  \hline \hline
StDr\,56                &       7.1135$^{+0.5924}_{-0.5924}$ &  67.4266$^{+9.1473}_{-9.1224}$        &       5.5325$^{+2.5579}_{-3.0553}$ \\
Fr\,2-21       &       2.5462$^{+0.0119}_{-0.0155}$ &  63.8790$^{+8.8033}_{-6.7630}$        &         1.4423$^{+0.2619}_{-0.2640}$ \\
\hline
\end{tabular}
\label{Table:results_LCs}
\end{table}


 \section{Discussion of the variability}
 \label{sec:discussion}

\subsection{Physical reliability of the Doppler beaming amplitudes}
\label{sec:beaming_amplitudes}

It is important to note that, as the modelling process does not put meaningful constraints on the physical parameters, the solution from the MCMC analysis has to be checked for unfeasible solutions. Particularly, we discuss the plausibility of the relativistic effects (Doppler beaming) as responsible for the sinusoidal variability in both 1E and 3E solutions. This effect produces low-amplitude sinusoidal variations \citep{Zucker2007} in the light curve, so in most of the cases we can directly reject this scenario by just checking for large amplitudes inferred from the MCMC analysis. 

In the same fashion as in \cite{Aller2020}, following the equations (2) and (3) from \cite{vanKerkwijk2010}, we can estimate the radial velocity amplitude derived from the Doppler beaming amplitude obtained in the analysis (A$_{\rm beam}$ in the 3E solution and A$_{\rm 1E}$ in the simple 1E solution under the assumption that the sinusoidal variability is due only to the Doppler beaming effect), and, therefore, estimate the binary mass function ($f$), defined as (assuming circular orbit): 

  \begin{eqnarray}
  \label{eq:mass_function}
f = \frac{P_{orb}\varv{_r}^3}{2 \pi G} = \frac{M_2^3 sin^3 i}{(M_1+M_2)^2}, 
   \end{eqnarray} 

\noindent where M$_1$ and M$_2$ are the masses of the two components, $i$ is the inclination of the system, $P_{\rm orb}$ the orbital period and $\varv{_r}$ the corresponding radial velocity semi-amplitude calculated from the measured Doppler beaming amplitude in the TESS light curve \citep[Eq.~2 in][]{Aller2020}. By definition, the binary mass function will be always lower than M$_2$.

We derive this mass function for all the targets with 1E or 3E solutions after the analysis of their light curves. To simplify, and as there is no available information on the effective temperatures of all the central stars in the sample, we have used for the calculation two reasonable values for this parameter that approximately cover the whole possible range of it in this type of stars, that is T$_{eff}$ = 40 000K, for the lower limit, and 150\,000K for the upper one). 

Figure\,\ref{fig:massfunc} shows the calculated mass functions for 0.5, 1, 5 and 10 M$_{\odot}$ in the Amplitude vs P$_{\rm orb}$ plane. For visualisation purposes, cases with A$_{\rm beam}$ (in golden) or A$_{\rm 1E}$ (in purple) greater than 6 are not shown in the plot (IC\,4593, with A$_{\rm beam}$ $\sim$ 0.005 ppt, is also not included). As expected, large Doppler beaming amplitudes imply unrealistic masses for the hypothetical secondary component (assuming that the primary has a typical mass of a white dwarf or similar). Only for very short orbital periods (a few hours) would the solution be reasonable but we do not have orbital periods in that region. Thus, only a handful of central stars have amplitudes (A$_{\rm beam}$ or A$_{\rm 1E}$) compatible with relativistic effects. Subsequent follow-up efforts must take these unphysical scenarios into consideration.

 \begin{figure}
  \includegraphics[width=0.45\textwidth]{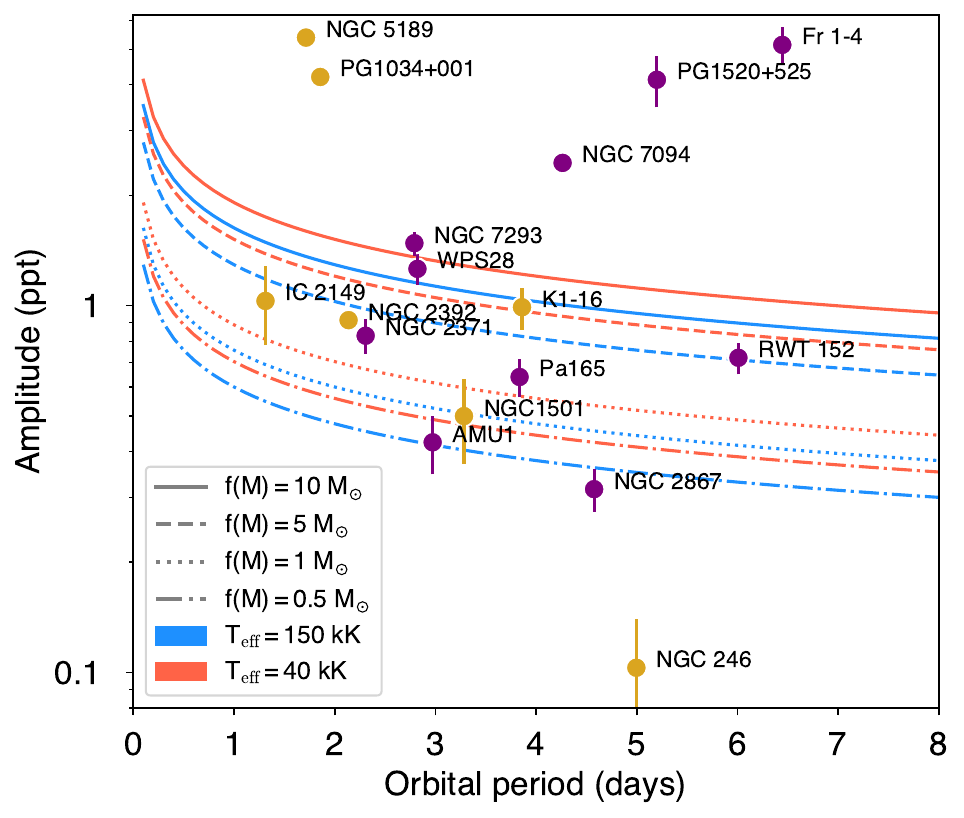}
       \caption{Amplitudes versus orbital periods measured in the TESS light curves. Doppler beaming amplitudes (from the 3E solution) are plotted in gold colour, while A$_{\rm 1E}$ (that could be due to irradiation either Doppler beaming), are in purple. Four different mass functions (0.5, 1, 5 and 10 $M_\odot$) for two different effective temperatures (T$_{eff}$ = 40 and 150kK) have been overplotted to see the position of the stars according to their corresponding effective temperature.}
\label{fig:massfunc}
\end{figure}

\subsection{Identification of the variability source}
\label{sec:tess_localize}

As explained in Sect.\,\ref{Sect:description_TESS_data}, we applied the \texttt{TESS\_localize} method to our sample in order to identify the origin of the detected variability. The results are summarised in Table\,\ref{table:results_tesslocalize} where we list the name of the target in our sample, its TIC designation, whether or not the variability is attributed to the corresponding target (according to $\texttt{TESS\_localize}$) and the most likely \textit{Gaia} source which would be responsible for the variability, together with its probability.

In some cases, the result from the \texttt{TESS\_localize} is not reliable according to its standards (see \citealt{Higgins-Bell2023} for further information), and we classify such targets as inconclusive. In these cases, the fit to the location of the source of the variability is not significant based on the $height$ parameter of the code. According to the documentation of the algorithm, a proper fit is achieved if the relative uncertainty in the height is smaller than 20\%, which corresponds to 5$\sigma$ significance. In brief, only cases where the signal with the indicated frequency can be properly fit, can provide a good location for the variability. Otherwise, the solution is unreliable. We applied the algorithm to each individual sector of each target\footnote{Unfortunately, the code is not prepared to be used with multi-sector light curves.}, by using the optimal number of PCA components that \texttt{TESS\_localize} calculates. For simplicity, Table\,\ref{table:results_tesslocalize} shows a mean of the likelihood calculated only with those sectors in which the height is $\le$ 20\%. If none of the sectors has a height $\le$ 20\%, the solution is unreliable.

The already known binary central stars in our sample are the best targets to test the method. The results from \texttt{TESS\_localize} for these stars reveal that, as expected, the source of the periodicity found in their extracted TESS light curves match the corresponding positions on the sky of each specific central star. However, for NGC\,5189, this is not the case. According to the \texttt{TESS\_localize} results, the two periodicities detected in the TESS light curve (none of them coinciding with the published orbital period, detected through radial velocities, see above) have origins different from the central star. The photometric variability in TESS associated with the RV frequency (measured by \citealt{Manick2015}) is however too shallow to apply the \texttt{TESS\_localize} test. Also, in the case of AMU\,1, the signal of the variability is too weak to provide a reliable result, so \texttt{TESS\_localize} does not recover the position of the central star as the origin of such variability.

The case of NGC\,2392 is also of interest, since the variability at period 2.1 d (the one detected in the TESS light curve) matches with a 100\% probability the position of the central star. Conversely, when asking the algorithm about the origin of the 1.9 days period previously reported by \cite{Miszalski2019}, the probability that the origin of such periodicity being  the central star is negligible. This result reinforces the 2.1 days periodicity as the true orbital period of this binary central star.

In addition, we find that for seven of our new binary candidates, the origin of the variability appears to coincide with the position of the central star (see Table\,\ref{table:results_tesslocalize}). Thus, these targets have a very high probability to reside in binary systems and radial velocity follow-up would be desirable to confirm their binary nature and derive orbital and physical parameters. For one central star (NGC\,7293), the detected variability is significantly attributed to another star in the field (\textit{Gaia} DR3 6628875408232926336). Although it is a target with negligible contamination in the aperture mask, the probability that the measured variability corresponds to the mentioned field star is basically 100\%, according to \texttt{TESS\_localize}. This star is outside the aperture (more than 5 pixels far away from the central star) which highlights the importance of doing this kind of analysis. Thus, we can discard NGC\,7293 as a potential binary candidate. 

We would also like to mention the case of WPS\,28. Out of 25 available sectors for this target, only three provide a reliable fit (with the relative uncertainty of the "height" parameter in these cases close to the 20\%), so we think this result should be taken with caution. 

For the remaining targets (8), the result of \texttt{TESS\_localize} is unreliable because of the low significance of the detection. Thus, we cannot unequivocally confirm or discard the binary scenario for these targets and new dedicated observations are required.

\begin{table*}
\setlength\extrarowheight{1pt}
\centering 
  \caption{Results after applying the \texttt{TESS\_localize} method for those variable stars analysed with minimal or negligible contamination in the aperture mask.}
     \label{table:results_tesslocalize} 
\begin{tabular}{cccccc}
\hline
Name & TIC & Period (d) & Attributed & Likelihood & \textit{Gaia} ID source variability \\
\hline\hline
 \multicolumn{5}{l}{\textit{Already known binaries}}\\
 \hline
Abell\,30 			& 331986754 & 1.06115 & yes & 91.0 & 660071056749861888  \\
Abell\,46 			& 423311936 & 0.47173& yes & 100.0 & 4585381817643702528  \\
Abell\,63 			& 1842385646	     & 0.46499& yes & 100.0 & 4585381817643702528  \\
LoTr\,5 			& 357307129 &5.85705 & yes & 100.0 & 3958428334589607552  \\
DS\,1 				& 120596335 & 0.35711 & yes & 51.5 & 5362804330246457344  \\
NGC\,2392 		& 60947424 & 2.10842 & yes & 100 & 865037169677723904  \\
AMU\,1 			& 63373833 & 2.99685& inconclusive & 
& \\
NGC\,5189 		& 341689253 & 1.71517 & no & 90.5 & 5863702868275418240  \\
NGC\,5189 		& 341689253 & 0.125& no & 95.2 & 5863726301619914496  \\
\hline
 \multicolumn{5}{l}{\textit{Binary candidates after TESS analysis (this work)}}\\
 \hline
NGC\,1501 		& 84306468 & 3.2830& yes & 100 & 473712872456844544 \\
PG\,1034+001 		& 124598476 &1.856796 & yes & 100.0 & 3806885288337214848  \\
IC\,4593 			& 119955971 & 2.50369 & yes &  100.0 & 4457218245479455744  \\
NGC\,7094 		& 387371712 & 4.263 & yes & 99.8 & 1770058865674512896  \\
Hen\,3-1863 		& 290274851 & 5.19725& yes & 99.9 & 6680314483486887552  \\
IC\,2149 			& 440340055 & 2.63877& yes &  99.9 & 196996680850087936  \\
WPS\,54 			& 453444667 & 3.45 / 1.73& yes & 100.0 & 1020641700210987520  \\
PG\,1520+525 		& 193990614 & 5.1992& inconclusive &  &   \\
PNG\,136.7+61.9	& 950518812 & 8.477& inconclusive &  &   \\
Pa\,165 			& 367508287 & 3.836& inconclusive &  &   \\
NGC\,2867 		& 387196603 & 4.5762& inconclusive &  &   \\
RWT\,152 			& 65145453 & 6.0090& inconclusive &  &   \\
NGC\,2371 		& 446005482 & 2.30726& inconclusive &  &   \\
NGC\,246 			& 3905338 & 4.99574& inconclusive &  &  \\
K\,1-16 			& 233689607 & 3.86624& inconclusive &  &   \\
WPS\,28 			& 229966905 & 2.82171& no & 98.6 &  2239709656843139072\\
Fr\,1-4 			& 401795856 & 6.4446& no & 100  &  6343834891409105664\\
NGC\,7293 		& 69813909 & 2.79151& no & 100 & 6628875408232926336  \\

\hline
\end{tabular}
\end{table*}

   
\section{Final remarks and conclusions}
\label{sec:conclusions}

\begin{figure*}
\centering
\includegraphics[width=0.99\textwidth]{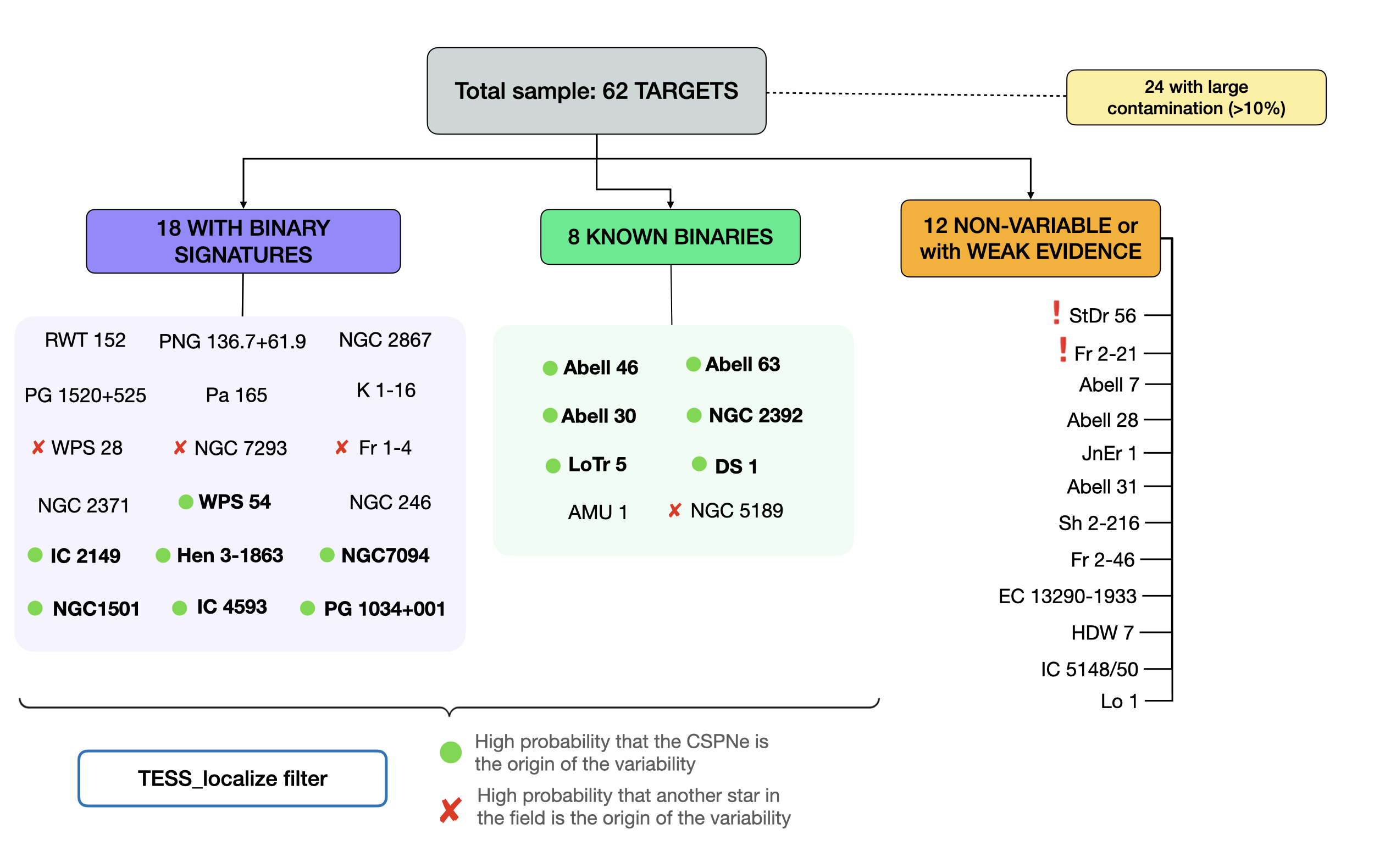}
\caption{Results after the MCMC variability analysis of the total sample of central stars of planetary nebulae. Green dots indicate those targets with a good location of the variability in the central source after applying \texttt{TESS\_localize} and red crosses indicate those targets with high probability that the variability is  due to other star in the field. Red X marks indicate those objects with weak evidence which may become binary candidates with more data.}
\label{fig:diagram_TESS}
\end{figure*}  

As a continuation of the work presented in \cite{Aller2020}, we analyse here the 2-minute cadence TESS light curves available in the MAST archive from the first four years of data from this mission (Cycles 1 to 4) of a sample of planetary nebulae nuclei in our Galaxy. In total, we analyse 62 central stars of (true, possible or likely) PNe in order to identify new binary candidates.

Figure\,\ref{fig:diagram_TESS} shows a summary of our results. Among the 62 central stars, 24 were set aside since they were in crowded fields, with large photometric contamination from nearby stars and, therefore, the origin of a hypothetically detected variability in such cases would be difficult to assess. The large size of the TESS pixel scale (21 arcsec\,pixel$^{-1}$) makes this first exploration essential. We then performed MCMC analyses to the remaining central star light curves (i.e. those with negligible or minimal contamination) in order to study the binary scenario, accounting for different physical effects: irradiation, ellipsoidal modulations and Doppler beaming. 

Among the modelled sample, eight central stars are already known binary systems; another 18 present modulations in the light curves that may be compatible with binary signatures; and 12 central stars show weak or no significant variability. 

For the already known binary systems, we analysed the new high-precision data from TESS, also using the radial velocity observations available in the literature for a handful of them. This analysis allows us to provide new information in some of the systems, for example, refining the orbital parameters in the case of NGC\,2392. 

Additionally, we propose as binary candidates the 18 central stars showing sinusoidal variability in their light curves (four of them already analysed in Cycle 1). For all of them, we derived short orbital periods ranging from $\sim$ 1.3 to 6.5 days. With the goal of confirming the source of the variability, we used the code \texttt{TESS\_localize}. This algorithm has been designed to identify, with an estimated likelihood, the location of the origin of a given frequency in the TESS target pixel file. Thus, after applying this method, we find that 7 central stars seem to be the origin of the detected periodicities, being our best candidates, while in three central stars the cause of the variability is likely another star in the field. For the remaining targets (8), \texttt{TESS\_localize} provides unreliable fits so the results are inconclusive.

It is worth highlighting that the periods of the detected candidates do not match with the current period distribution of the known close binary central stars \citep[see, for example,][]{Jacoby2021}. The candidates detected in our work have larger periods than the median of the distribution of published binaries. However, it is also true that most of them have been discovered  through ground-based observations, that are biased towards short periods and large amplitudes. With the recent space surveys like Kepler, K2 and TESS, we are populating a region with lower amplitudes and longer periods, changing the period distribution we know so far. In any case, it is remarkable that in our sample, we do not find binary candidates with periods below $\sim$ 1 day. A priori, there is not any selection effect in the sample which may affect this finding, and it is true that other effects like stellar spots may be the answer to some of the variability encountered. To resolve this, more detailed analysis has to be done in each case to confirm the potential binarity and to constrain the parameters of the systems.

We note that among the binary systems (both already known and new candidates) only two PNe are classified as round in the HASH PN database, which represents less than 10\%. They are Abell\,30 and PG\,1520+525. The first one, certainly has a spherical shell but with other structures like dots and cometary tails in the inner regions. On the other hand, PG\,1520+525 is a more classical and perfectly round PNe. The rest, are PNe with morphologies that deviate from sphericity, such as bipolar, elliptical, and/or with other complex structures, that can be in many cases perfectly explained by the presence of a binary system in their nuclei.

On the contrary, if we attend to the morphology of the sample of the 12 central stars that do not show variability (i.e, they converge to a flat model in our MCMC analysis), the results are quite different. About $\sim$33\% have round PNe and another $\sim$33\% have non-defined morphological type according the HASH PN database. The remaining $\sim$33\% have non-spherical shapes, such as elliptical or bipolar. The non-detected variability in such cases is by no means a confirmation that they are indeed single stars but instead, we provide a detection limit for a possible companion.

Definitively, TESS is providing a huge amount of high-precision photometric data that are enormously contributing to the study of a broad range of variable stars, including binary central stars of planetary nebula. The discovery and characterisation of more of these systems is crucial to further our limited understanding on the formation of the beautiful, varied and complex sample of PNe. The PLATO mission \citep{rauer14} in the near future will also help in this way by providing long-term high-precision photometry for a large region of the sky.

\begin{acknowledgements}

We thank Nicole Reindl and Mikkel Nørup Lund for useful discussions on the pulsations of LoTr\,5, and Nicole also for her help in the selection of the targets and providing feedback on the manuscript. AA acknowledges support from Government of Comunidad Aut\'onoma de Madrid (Spain) through postdoctoral grant `Atracci\'on de Talento Investigador' 2018-T2/TIC-11697. 
J.L.-B. was partly funded by grants LCF/BQ/PI20/11760023,  Ram\'on y Cajal fellowship with code RYC2021-031640-I, by the  Spanish MICIU/AEI/10.13039/501100011033 grant PID2019-107061GB-C61, and by the MICIU/AEI/10.13039/501100011033 and NextGenerationEU/PRTR grant CNS2023-144309. DJ acknowledges support from the Agencia Estatal de Investigaci\'on del Ministerio de Ciencia, Innovaci\'on y Universidades (MCIU/AEI) and the European Regional Development Fund (ERDF) with reference PID-2022-136653NA-I00 (DOI:10.13039/501100011033). DJ also acknowledges support from the Agencia Estatal de Investigaci\'on del Ministerio de Ciencia, Innovaci\'on y Universidades (MCIU/AEI) and the European Union NextGenerationEU/PRTR with reference CNS2023-143910 (DOI:10.13039/501100011033). This research is part of the I+D+i project PID2019-105203GB-C21 founded by Spanish AEI (MICIU) grant 10.13039/501100011033.

This research has made use of the SIMBAD database, operated at the CDS, Strasbourg (France), Aladin, NASA's Astrophysics Data System Bibliographic Services, and the Spanish Virtual Observatory (http://svo.cab.inta-csic.es) supported by the Spanish Ministry of Science and Innovation/State Agency of Research MCIN/AEI/10.13039/501100011033 through grant PID2020-112949GB-I00. This paper includes data collected with the TESS mission, obtained from the MAST data archive at the Space Telescope Science Institute (STScI). Funding for the TESS mission is provided by the NASA Explorer Program. STScI is operated by the Association of Universities for Research in Astronomy, Inc., under NASA contract NAS 5-26555.
We acknowledge the use of pipelines at the TESS Science Office and at the TESS Science
Processing Operations Center.

\end{acknowledgements}


\bibliographystyle{aa} 
\bibliography{Bibliography.bib} 

\Online

\begin{appendix} 

\appendix

\section{Additional material}

{\fontsize{9}{12}\selectfont 
\begin{longtable}{lccccccc}
\caption{Sample of CSPNe studied in this work. The PN\,G designations, common names, TESS input catalogue (TIC), equatorial coordinates, TESS magnitude, and the sector and camera of the observation of each target are listed.}    
\label{table:main_table}      
\\
\hline\hline       
PNG	& 	Common name 	& 	TIC&		$\alpha$(2000.0) & $\delta$(2000.0) & TESS$_{mag}$ & Sector  & PN status \\ 
\hline
\endfirsthead
\caption{continued.}\\
\hline\hline
PNG	& 	Common name 	& 	TIC&		$\alpha$(2000.0) & $\delta$(2000.0) & TESS$_{mag}$ & Sector  & PN status \\ 
\hline
\endhead
\hline
\endfoot
   \hline
 \multicolumn{8}{l}{\textit{Targets where at least one sector has negligible contamination}}\\
 \hline
   PN\,G\,255.3-59.6		&Lo\,1 		&  146711168	&  02$^h$\,56$^m$\,58$\fs$4		&  -44$^{\circ}$\,10$'$\,18$''$		&15.66 & 30\tablefootmark{$a$}	&	T	 \\
 PN\,G\,197.8+17.3 	&NGC\,2392		&  60947424      & 07$^h$\,29$^m$\,10$\fs$8         &  +20$^{\circ}$\,54$'$\,42$''$  & 10.91 &  44\tablefootmark{$a$}, 45\tablefootmark{$a$}, 46\tablefootmark{$a$}	&  T   \\
    PN\,G\,164.8+31.1  	&JnEr\,1			& 741726060 	& 07$^h$\,57$^m$\,51$\fs$6         &  +53$^{\circ}$\,25$'$\,17$''$  & 17.58	&  47\tablefootmark{$a$}	&  T \\
  PN\,G\,158.8+37.1		& Abell\,28 		&  471015303 	&  08$^h$\,41$^m$\,35$\fs$6 	& +58$^{\circ}$\,13$'$\,48$''$   &15.91& 20\tablefootmark{$a$}, 47\tablefootmark{$a$}	& L	\\
 PN\,G\,219.1+31.2  	&Abell\,31	& 444036737 	&   08$^h$\,54$^m$\,13$\fs$2		&  +08$^{\circ}$\,53$'$\,53$''$		 &15.90& 34\tablefootmark{$b$}, 44\tablefootmark{$b$}, 45\tablefootmark{$a$}, 46\tablefootmark{$b$}	&  T	 \\
    PN\,G\,162.1+47.9		&WPS\,54  		&  453444667	& 09$^h$\,51$^m$\,25$\fs$9    & +53$^{\circ}$\,09$'$\,31$''$    &15.73 & 21\tablefootmark{$a$}, 48\tablefootmark{$a$}	& P	\\
    PN\,G\,136.7+61.9 	&PN\,G\,136.7+61.9	&  471013481\tablefootmark{$*$} 	& 12$^h$\,07$^m$\,28$\fs$4         &  +54$^{\circ}$\,01$'$\,29$''$  & 16.43	&  48\tablefootmark{$a$}	& T  \\
    PN\,G\,339.9+88.4		& LoTr\,5  		&  357307129	& 12$^h$\,55$^m$\,33$\fs$7    & +25$^{\circ}$\,53$'$\,31$''$    &8.09 & 23\tablefootmark{$a$}, 49\tablefootmark{$a$}	& T	\\
PN\,G\,315.7+42.0		&EC\,13290-1933 	& 32145732  	&    13$^h$\,31$^m$\,46$\fs$3    & -19$^{\circ}$\,48$'$\,26$''$    & 	14.64& 37\tablefootmark{$a$}	&  P	\\ 
PN\,G\,085.3+52.3  	&PG\,1520+525			& 193990614 	&     15$^h$\,21$^m$\,46$\fs$6     & +52$^{\circ}$\,22$'$\,04$''$ 	& 16.06	& 16\tablefootmark{$a$}, 49\tablefootmark{$a$}, 51\tablefootmark{$b$}	&  T  \\
  PN\,G\,094.0+27.4		& K\,1-16\tablefootmark{$1$} 		&  233689607	&  18$^h$\,21$^m$\,52$\fs$1 	& +64$^{\circ}$\,21$'$\,53$''$   & 15.47 & (14--17, 19, 20, 22--26)\tablefootmark{$a,b$} & T \\
     	& 	&  	&   	&    &  &  (40, 47, 49, 50, 52--56)\tablefootmark{$a,b$} 	 \\
  PN\,G\,359.2-33.5		&Hen\,3-1863 	& 290274851 	&   20$^h$\,19$^m$\,28$\fs$7  		&  -41$^{\circ}$\,31$'$\,27$''$ 		&10.95 & 27\tablefootmark{$a$}	&  T		\\
   PN\,G\,011.7-41.3  		&Fr\,2-46 		& 92709449 	&   21$^h$\,05$^m$\,47$\fs$6		&  -32$^{\circ}$\,49$'$\,52$''$    		&13.77 & 28\tablefootmark{$a$}	& 	P \\
  PN\,G\,066.7-28.2  		&NGC\,7094&  387371712      &  21$^h$\,36$^m$\,53$\fs$0       &  +12$^{\circ}$\,47$'$\,19$''$  & 13.90	&  55\tablefootmark{$a$}		&  T \\
 \hline
 \multicolumn{8}{l}{\textit{Targets where at least one sector has minimal contamination}}\\
 \hline
PN\,G\,141.8-29.9		&StDr\,56 		&  63505607	& 02$^h$\,07$^m$\,17$\fs$4    & +30$^{\circ}$\,05$'$\,12$''$    & 15.42 & 18\tablefootmark{$b$}	&New cand.	\\
  PN\,G\,144.1+06.1		& NGC\,1501 	&  84306468	&  04$^h$\,06$^m$\,59$\fs$4 	& +60$^{\circ}$\,55$'$\,14$''$   & 13.99 & 19\tablefootmark{$b$}	& T \\
  PN\,G\,158.5+00.7		&Sh\,2-216 		&  391260920	&  04$^h$\,43$^m$\,21$\fs$3 	& +46$^{\circ}$\,42$'$\,06$''$   &13.05 & 19\tablefootmark{$b$}	& T	\\ 
 PN\,G\,166.1+10.4		& IC\,2149 		&  440340055	&  05$^h$\,56$^m$\,23$\fs$9 	& +46$^{\circ}$\,06$'$\,17$''$   &11.27 & 19\tablefootmark{$b$} 	& T	\\
   PN\,G\,189.1+19.8 	&NGC\,2371	& 446005482	& 07$^h$\,25$^m$\,34$\fs$7    & +29$^{\circ}$\,29$'$\,26$''$   &15.19 & 20\tablefootmark{$b$}, 44\tablefootmark{$b$}, 45\tablefootmark{$b$}, 46\tablefootmark{$b$}, 47\tablefootmark{$b$}& T	 \\
 PN\,G\,211.4+18.4 	&HDW\,7		& 760317391	&  07$^h$\,55$^m$\,11$\fs$3         & +09$^{\circ}$\,33$'$\,09$''$   & 	18.38&  45\tablefootmark{$b$}, 46\tablefootmark{$b$}	& T  \\
    PN\,G\,208.5+33.2  	&Abell\,30 			& 331986754 	&    08$^h$\,46$^m$\,53$\fs$5    & +17$^{\circ}$\,52$'$\,46$''$   &	14.56& 44\tablefootmark{$b$}, 45\tablefootmark{$b$}, 46\tablefootmark{$b$}º	&  T	 \\
  PN\,G\,283.9+09.7  	&DS\,1 			& 120596335 	&    10$^h$\,54$^m$\,40$\fs$6    & -48$^{\circ}$\,47$'$\,03$''$   &	12.42& 36\tablefootmark{$b$}, 37\tablefootmark{$b$}	&  T	 \\
  PN\,G\,025.3+40.8  	&IC\,4593	&  119955971      &   16$^h$\,11$^m$\,44$\fs$5      &  +12$^{\circ}$\,04$'$\,17$''$  & 11.42	&  51\tablefootmark{$b$}		&  T \\
  PN\,G\,055.4+16.0		& Abell\,46  		&  423311936	& 18$^h$\,31$^m$\,18$\fs$3    & +26$^{\circ}$\,56$'$\,13$''$    &15.20& 26\tablefootmark{$b$}, 40\tablefootmark{$b$}	& T	\\ 
  PN\,G\,075.9+11.6		&  AMU\,1 	&  63373833	&  19$^h$\,31$^m$\,08$\fs$9 	& +43$^{\circ}$\,24$'$\,58$''$   &13.82& 15\tablefootmark{$b$}, 40\tablefootmark{$b$}, 41\tablefootmark{$c$}, 54\tablefootmark{$b$}, 55\tablefootmark{$c$}	& T	\\  
   PN\,G\,091.0+20.0		&WPS\,28\tablefootmark{$2$}  	&  229966905	& 19$^h$\,18$^m$\,23$\fs$9	    & +59$^{\circ}$\,59$'$\,55$''$    & 15.17 & (14--26, 40, 41, 47--56)\tablefootmark{$b,c$} 	& P	\\
    PN\,G\,053.9-33.2  		&Fr\,2-21	&  352568802      & 21$^h$\,26$^m$\,21$\fs$2        &   +00$^{\circ}$\,58$'$\,34$''$ & 14.50	&  55\tablefootmark{$b$}		&  P \\
PN\,G\,072.0-24.6  	&Pa\,165		&  367508287      &  21$^h$\,38$^m$\,52$\fs$8       &  +18$^{\circ}$\,40$'$\,15$''$   & 13.07	&  55\tablefootmark{$b$}	&  L \\
  PN\,G\,002.7-52.4		&IC\,5148/50	&  2027562015	&  21$^h$\,59$^m$\,35$\fs$1    	& -39$^{\circ}$\,23$'$\,08$''$		& 16.58 & 28\tablefootmark{$b$}& 	T	\\ 
      PN\,G\,306.5-31.1		&Fr\,1-4 		&  401795856	&   22$^h$\,21$^m$\,50$\fs$8		&  -84$^{\circ}$\,54$'$\,38$''$  		& 16.07 & 27\tablefootmark{$c$}, 39\tablefootmark{$b$}& 	P	\\
 \hline
 \multicolumn{8}{l}{\textit{Targets with severe contamination in all sectors}}\\
 \hline
 PN\,G\,146.1-08.1		& Fr\,2-23  		&  117502568	& 03$^h$\,14$^m$\,46$\fs$0    & +48$^{\circ}$\,12$'$\,06$''$    &14.60 & 18\tablefootmark{$c$}	&P \\  
  PN\,G\,150.9-10.1  		&Bode\,1 		&  431762266	& 03$^h$\,31$^m$\,12$\fs$0    & +43$^{\circ}$\,54$'$\,15$''$    &11.87 & 18\tablefootmark{$c$}	& P	 \\
     PN\,G\,158.9+17.8 	&PuWe\,1 		& 322716838	& 06$^h$\,19$^m$\,34$\fs$0    & +55$^{\circ}$\,36$'$\,44$''$   &15.89 & 20\tablefootmark{$c$}& 	T	 \\
        PN\,G\,205.1+14.2  	&Abell\,21 			& 247056967 	&    07$^h$\,29$^m$\,02$\fs$7    & +13$^{\circ}$\,14$'$\,49$''$   &	16.41& 44\tablefootmark{$c$}, 45\tablefootmark{$c$}, 46\tablefootmark{$c$}	&  T	 \\
  PN\,G\,227.3+12.9  	&Fr\,2-25 		& 803286817	&   08$^h$\,04$^m$\,04$\fs$4		&  -06$^{\circ}$\,30$'$\,57$''$	    	&16.65& 34\tablefootmark{$c$}	& L	 \\
   PN\,G\,191.4+33.1  	&TK\,1				&  400728564      & 08$^h$\,27$^m$\,05$\fs$5         & +31$^{\circ}$\,30$'$\,08$''$  &  16.16	&  44\tablefootmark{$c$}, 45\tablefootmark{$c$}, 46\tablefootmark{$c$}, 47\tablefootmark{$c$}
  	&  L \\
 PN\,G\,238.0+34.8  	&Abell\,33		& 841940392	&   09$^h$\,39$^m$\,09$\fs$1		&  -02$^{\circ}$\,48$'$\,32$''$	    	&16.26& 35\tablefootmark{$c$}	& L	 \\
   PN\,G\,274.3+09.1  	&Lo\,4				&  870093894   & 10$^h$\,05$^m$\,45$\fs$8         &  -44$^{\circ}$\,21$'$\,34$''$  & 17.01	&  36\tablefootmark{$c$}	&  T   \\
    PN\,G\,063.1+13.9		&NGC\,6720 	&  1714951233	& 18$^h$\,53$^m$\,35$\fs$1    & +33$^{\circ}$\,01$'$\,45$''$    & 15.85 & 40\tablefootmark{$c$}, 53\tablefootmark{$c$}, 54\tablefootmark{$c$}	 & T	\\
PN\,G\,046.8+03.8  	&Sh\,2-78		&  1802361495      &   19$^h$\,03$^m$\,10$\fs$1     &  +14$^{\circ}$\,06$'$\,59$''$   & 17.77	&  54\tablefootmark{$c$}	&  T \\
  PN\,G\,032.1-06.2  		&FP\,J1912-0331&  49472361       &  19$^h$\,12$^m$\,31$\fs$5       &  -03$^{\circ}$\,31$'$\,32$''$  & 14.28	&  54\tablefootmark{$c$}		&  P \\
  PN\,G\,041.8-02.9  		&NGC\,6781	&  131389775      &  19$^h$\,18$^m$\,28$\fs$1       &  +06$^{\circ}$\,32$'$\,19$''$  & 16.57	&  54\tablefootmark{$c$}	&  T \\
   PN\,G\,077.6+14.7  	&Abell\,61 			& 1716997043	&    19$^h$\,19$^m$\,10$\fs$2    & +46$^{\circ}$\,14$'$\,52$''$   &	17.72& 40\tablefootmark{$c$}, 41\tablefootmark{$c$}, 54\tablefootmark{$c$}, 55\tablefootmark{$c$} 	&  T	 \\
   PN\,G\,053.8-03.0		& Abell\,63  		&  1842385646	&  19$^h$\,42$^m$\,10$\fs$3 	& +17$^{\circ}$\,05$'$\,14$''$   &15.03& 40\tablefootmark{$c$}, 41\tablefootmark{$c$}, 54\tablefootmark{$c$}	& T	\\
  PN\,G\,059.1-00.7		&Kn\,9 		&  1850055136	& 19$^h$\,44$^m$\,58$\fs$0    & +22$^{\circ}$\,45$'$\,48$''$    & 16.56 & 14\tablefootmark{$c$}, 41\tablefootmark{$c$}, 54\tablefootmark{$c$}	& L	\\
   PN\,G\,060.8-03.6  	&NGC\,6853		&  1849363609	 & 19$^h$\,59$^m$\,36$\fs$4         &  +22$^{\circ}$\,43$'$\,16$''$  & 	14.49 &  41\tablefootmark{$c$}, 54\tablefootmark{$c$}&  T   \\
  PN\,G\,042.5-14.5  		&NGC\,6852	&  1984811130     &  20$^h$\,00$^m$\,39$\fs$2      &   +01$^{\circ}$\,43$'$\,41$''$ & 18.16	&  54\tablefootmark{$c$}	&  T \\
 PN\,G\,107.0+21.3  	&K\,1-6			&  237292969   & 20$^h$\,04$^m$\,14$\fs$3         &  +74$^{\circ}$\,25$'$\,36$''$  & 11.65	&  47\tablefootmark{$c$}, 49\tablefootmark{$c$}, 50\tablefootmark{$c$}, 52\tablefootmark{$c$}	&
    T   \\
            	& 	&  	&   	&    &  &  53\tablefootmark{$c$}, 54\tablefootmark{$c$}, 56\tablefootmark{$c$}	 \\   
 PN\,G\,061.4-09.5		&NGC\,6905  	&  402913811	& 20$^h$\,22$^m$\,23$\fs$0    & +20$^{\circ}$\,06$'$\,16$''$    &14.77 & 14\tablefootmark{$c$}, 54\tablefootmark{$c$}, 55\tablefootmark{$c$}	& T	\\
    PN\,G\,014.2-03.4		& Pa\,39  		&  277013493	& 20$^h$\,25$^m$\,34$\fs$6    & +53$^{\circ}$\,12$'$\,45$''$    &13.99 & 16\tablefootmark{$c$}, 55\tablefootmark{$c$}, 56\tablefootmark{$c$}	& L\\  
PN\,G\,058.8-16.9  		&Kn\,121		&  341066367      &  20$^h$\,42$^m$\,01$\fs$9        &  +13$^{\circ}$\,51$'$\,15$''$  & 16.00	&  55\tablefootmark{$c$}	&  T \\
     PN\,G\,059.7-18.7  		&Abell\,72		&  291153543      & 20$^h$\,50$^m$\,02$\fs$1		&    +13$^{\circ}$\,33$'$\,30$''$        & 16.46	&  55\tablefootmark{$c$}	&  T \\
  PN\,G\,072.7-17.1  		&Abell\,74	&  1951127516     &  21$^h$\,16$^m$\,52$\fs$3        &    +24$^{\circ}$\,08$'$\,51$''$  & 17.46	&  55\tablefootmark{$c$}		&  T \\
    PN\,G\,080.3-10.4  	&V*\,V2027\,Cyg 	& 117070953	& 21$^h$\,17$^m$\,08$\fs$3    & +34$^{\circ}$\,12$'$\,27$''$   &13.50 & 15\tablefootmark{$c$}, 55\tablefootmark{$c$}	& T	 \\
  PN\,G\,105.9-20.0		& Fr\,1-6  		&  352444061	& 23$^h$\,27$^m$\,15$\fs$9 	& +40$^{\circ}$\,01$'$\,23$''$   &15.08 & 16\tablefootmark{$c$},17\tablefootmark{$c$}	& P	\\
   \hline
 \multicolumn{8}{l}{\textit{Targets with variability  from Cycle 1}}\\
 \hline
  PN\,G\,118.8-74.7  	&NGC\,246		& 3905338      &  00$^h$\,47$^m$\,03$\fs$3         & -11$^{\circ}$\,52$'$\,19$''$   & 12.31	&  3\tablefootmark{$c$} , 30\tablefootmark{$c$} 	& T  \\
   PN\,G\,215.5-30.8 	 	&Abell\,7 			& 169305167 	&    05$^h$\,03$^m$\,07$\fs$5    & -15$^{\circ}$\,36$'$\,23$''$   &	15.85& 5\tablefootmark{$a$}, 32\tablefootmark{$a$}	&  T	 \\
    PN\,G\,219.2+07.5 	&RWT\,152		& 65145453 	&    07$^h$\,29$^m$\,58$\fs$5     & -02$^{\circ}$\,06$'$\,38$''$    &	13.11& 7\tablefootmark{$a$}, 34\tablefootmark{$b$}	& T  \\
   PN\,G\,278.1-05.9  	&NGC\,2867				& 387196603\tablefootmark{$*$}      &  09$^h$\,21$^m$\,25$\fs$4         & -58$^{\circ}$\,18$'$\,41$''$   & 14.66	&  9\tablefootmark{$c$}, 10\tablefootmark{$c$}, 35\tablefootmark{$c$}, 37\tablefootmark{$c$}	& T  \\
 &PG\,1034+001				& 124598476      &  10$^h$\,37$^m$\,03$\fs$8         & -00$^{\circ}$\,08$'$\,19$''$   & 13.66	&  9\tablefootmark{$a$}, 35\tablefootmark{$a$}, 45\tablefootmark{$a$}, 46\tablefootmark{$a$}& Ionised ISM  \\
PN\,G\,307.2-03.4  	&NGC\,5189	&  341689253      & 13$^h$\,33$^m$\,32$\fs$9         &  -65$^{\circ}$\,58$'$\,27$''$  & 14.49	&  11\tablefootmark{$c$}, 38\tablefootmark{$c$}		&  T \\
  PN\,G\,036.1-57.1		&NGC\,7293 		& 69813909  	&    22$^h$\,29$^m$\,38$\fs$5    & -20$^{\circ}$\,50$'$\,14$''$    & 13.95& 2\tablefootmark{$a$}, 28\tablefootmark{$a$}, 42\tablefootmark{$a$}	&  T	\\
  
  \end{longtable} 
  \tablefoot{
\tablefoottext{a}{Negligible contamination}
\tablefoottext{b}{Minimal contamination ($\le$ 10\%) in the aperture mask}
\tablefoottext{c}{Severe contamination ($\ge$ 10\%) in the aperture mask}\\
\tablefoottext{*}{Object with two identificators in the TIC Catalogue}
\tablefoottext{1}{The TPFs of K\,1-16 show minimal contamination from other stars in all the sectors except three (sectors 23, 26 and 55), which have negligible contamination.}
\tablefoottext{2}{The TPFs of WPS\,28 show severe contamination from other stars in all the sectors except one (sector 21), which has minimal contamination.}
}
}
\FloatBarrier

\begin{figure*}
          \includegraphics[width=1\textwidth]{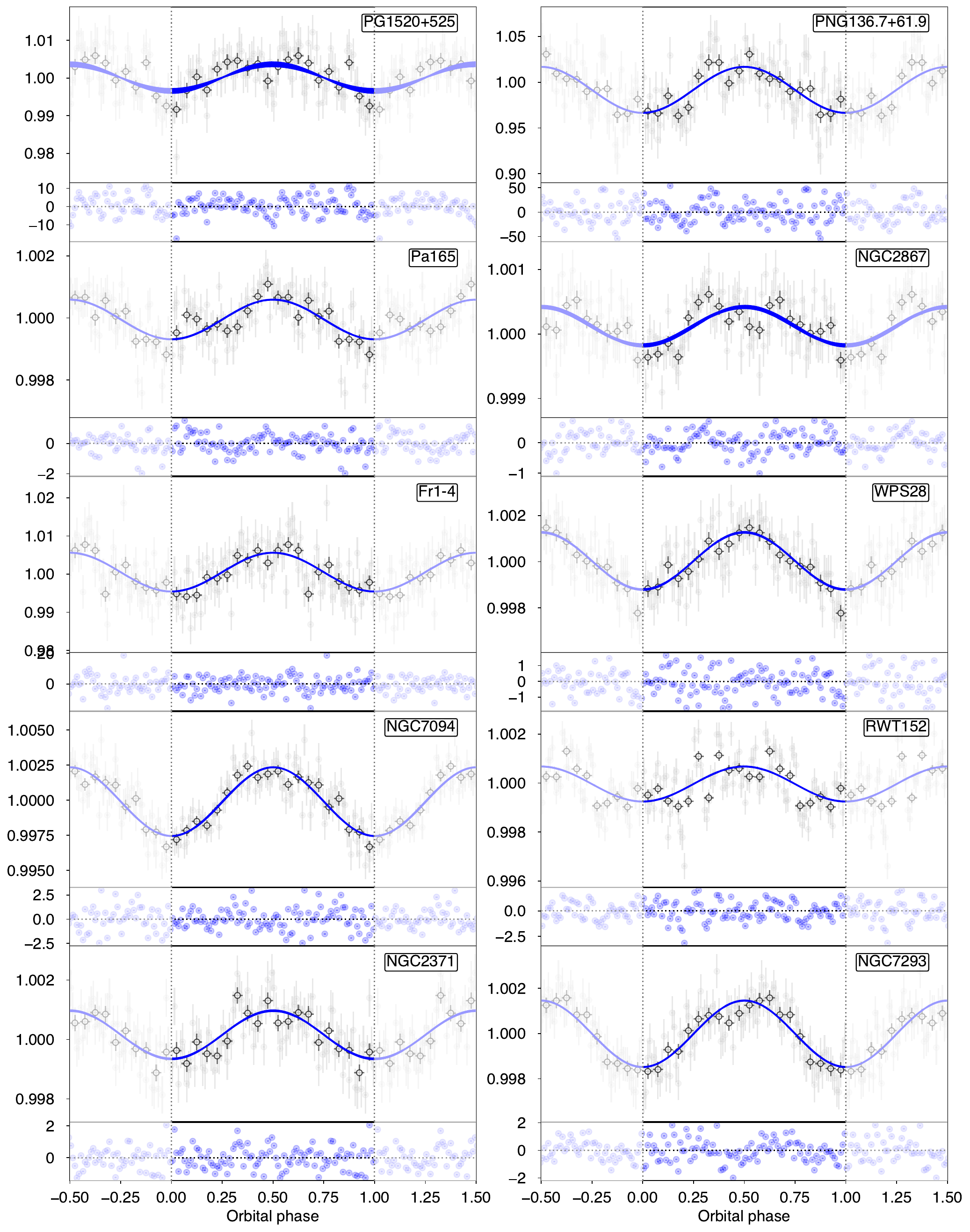}
       \caption{Results from the MCMC analysis for the proposed binary candidates. 1E solution is plotted in blue and 3E solution in red. Two different bin sizes are shown with black (bin size of 0.01 in phase) and grey (bin size of 0.05 in phase) symbols. The period derived from the fitting in each case can be found in Tables\,\ref{table:results_fit_candidates_1E} and \ref{table:results_fit_candidates_3E}.}
\label{fig:phase_new_candidates}
\end{figure*}

  \addtocounter{figure}{-1}
  
\begin{figure*}      \includegraphics[width=1\textwidth]{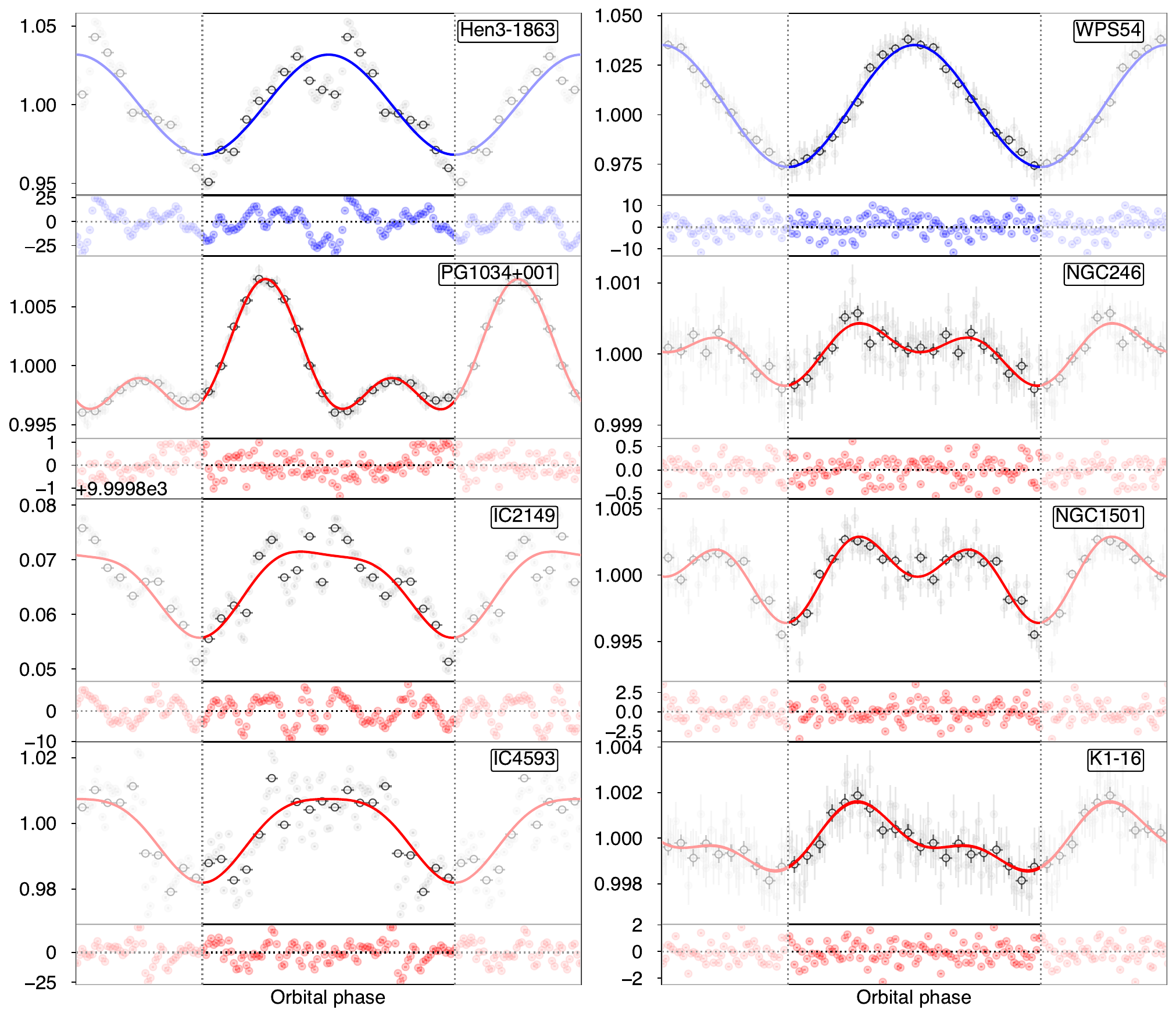}
       \caption{continued.}
\label{fig:phase_new_candidates}
\end{figure*}

\clearpage

\section{Figures}
\FloatBarrier
  \begin{figure}[h!]
     \includegraphics[width=0.335\textwidth]{Figures/TPF_Gaia_TIC471015303_S20_A28_maglim6.pdf}
      \includegraphics[width=0.335\textwidth]{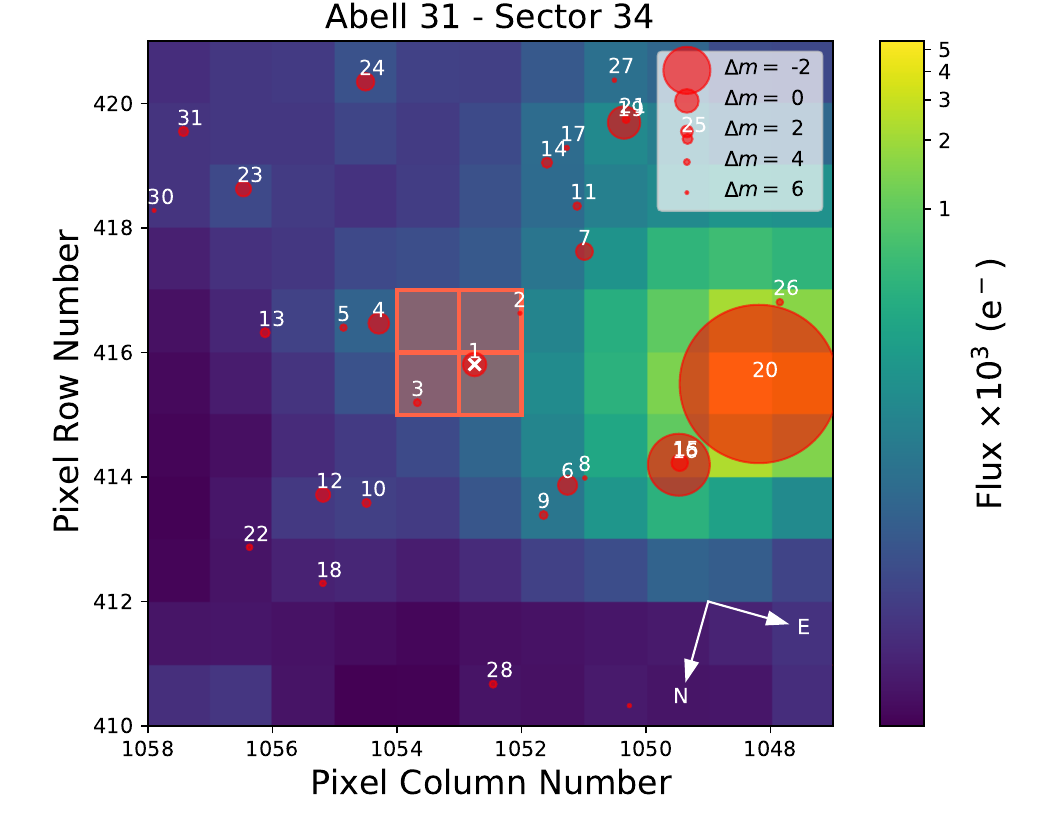}
      \includegraphics[width=0.335\textwidth]{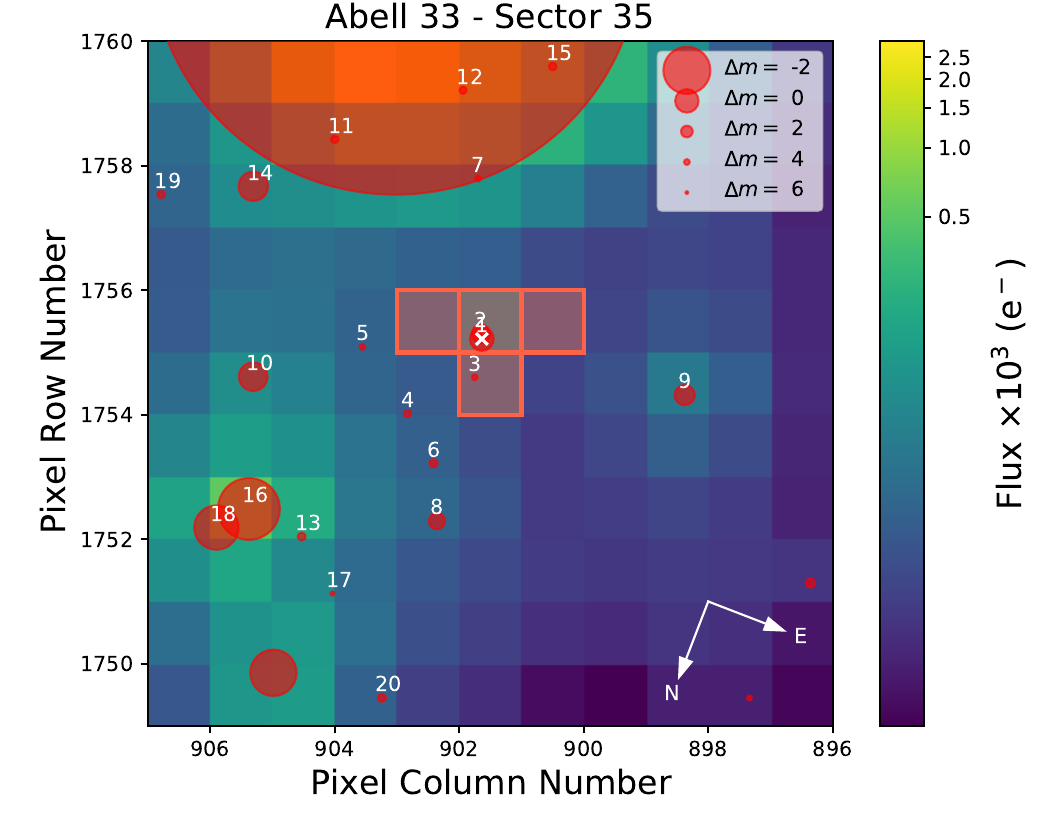}
     \includegraphics[width=0.335\textwidth]{Figures/TPF_Gaia_TIC423311936_S26_A46_maglim6.pdf}     
     \includegraphics[width=0.335\textwidth]{Figures/TPF_Gaia_TIC342025025_S14_A63_maglim3.pdf}
     \includegraphics[width=0.335\textwidth]{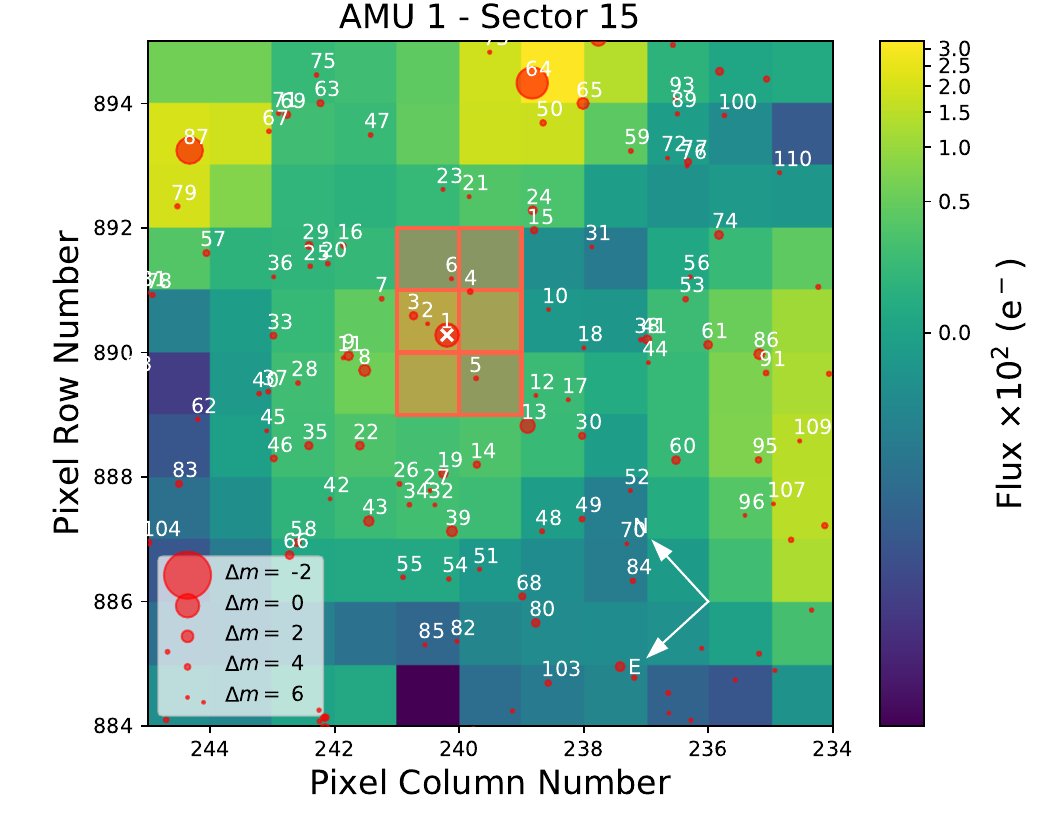}
      \includegraphics[width=0.335\textwidth]{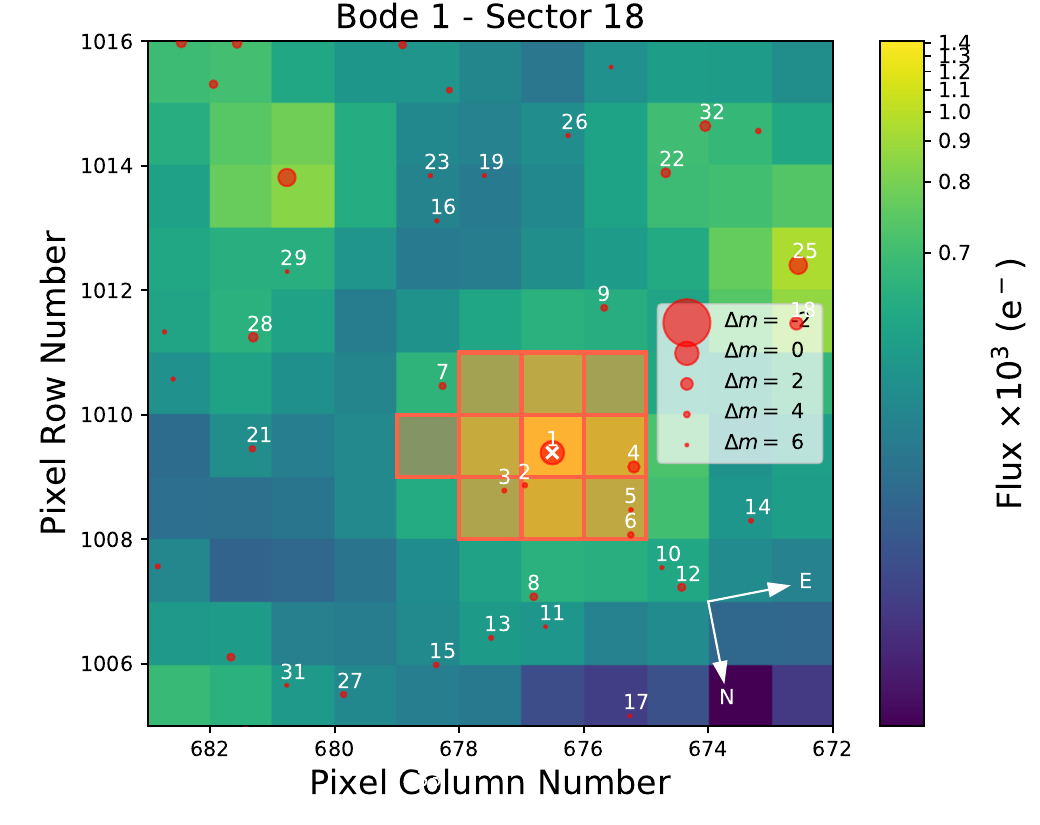}     
          \includegraphics[width=0.335\textwidth]{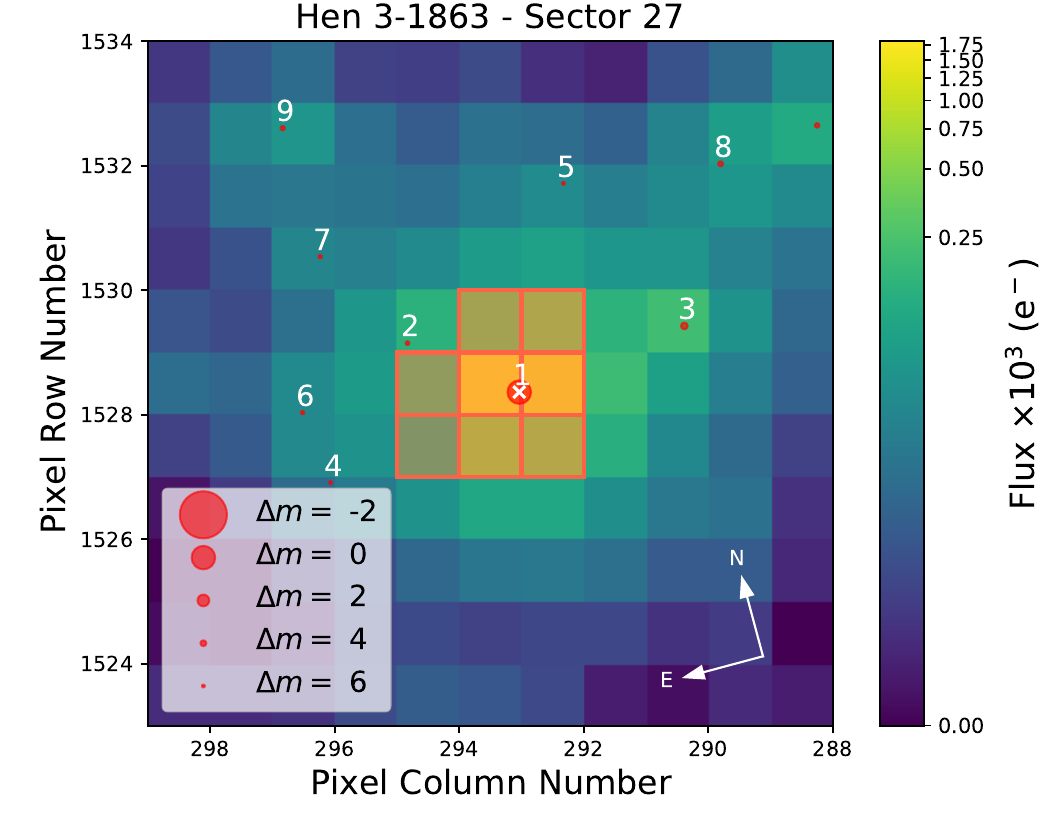}
           \includegraphics[width=0.335\textwidth]{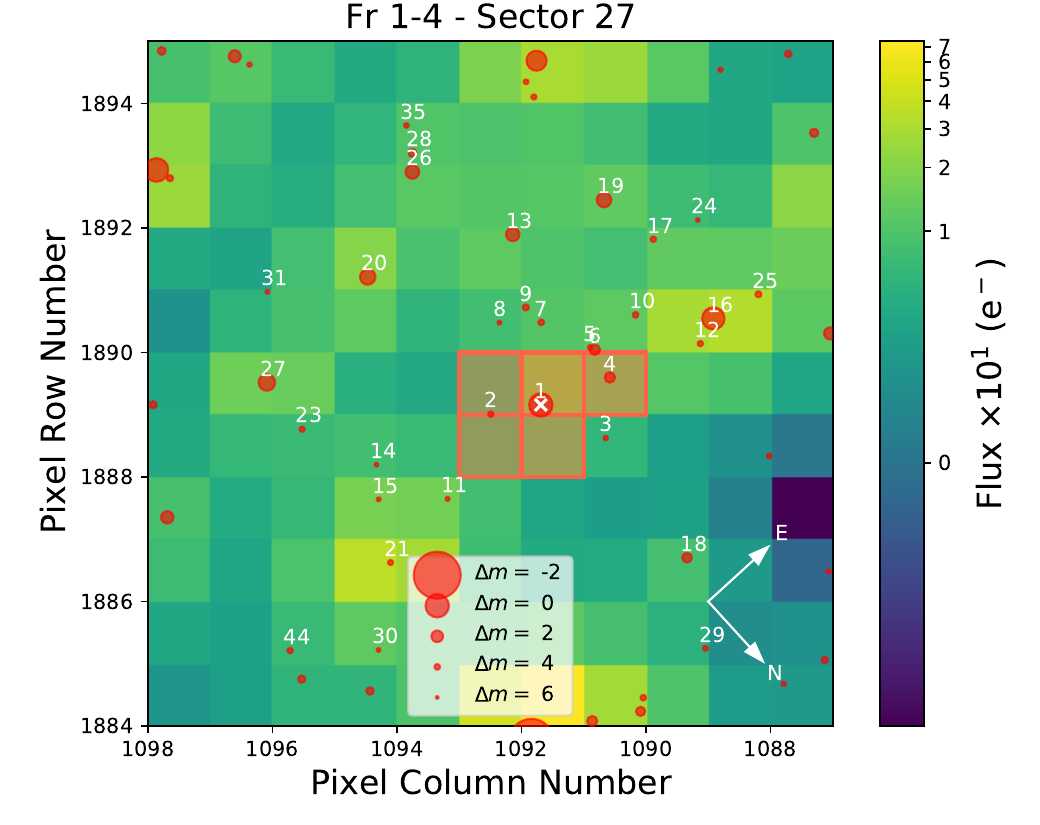}
      \includegraphics[width=0.335\textwidth]{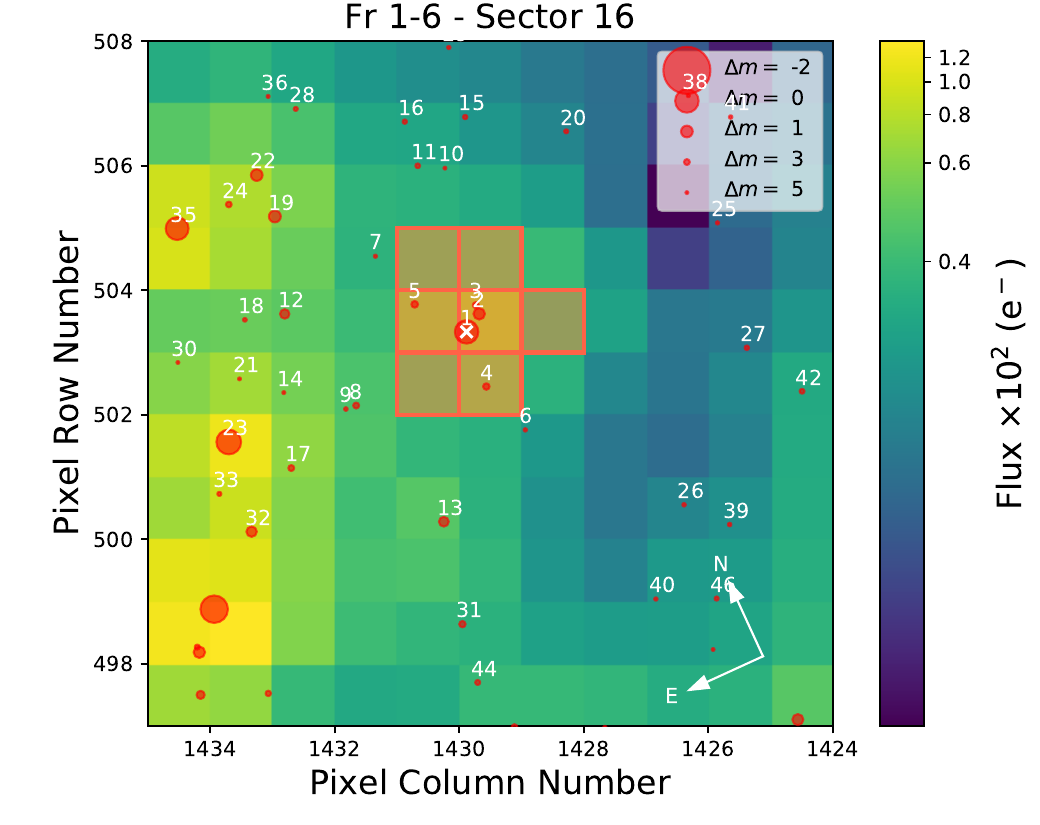}
       \includegraphics[width=0.335\textwidth]{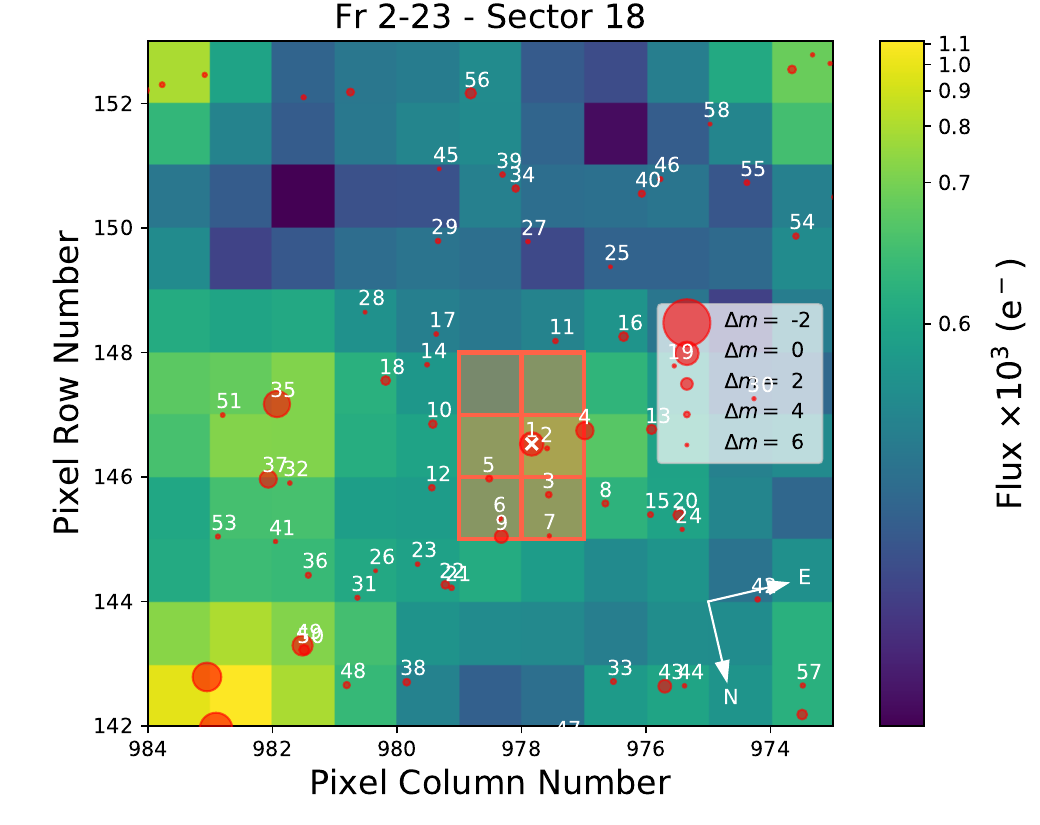}
            \includegraphics[width=0.335\textwidth]{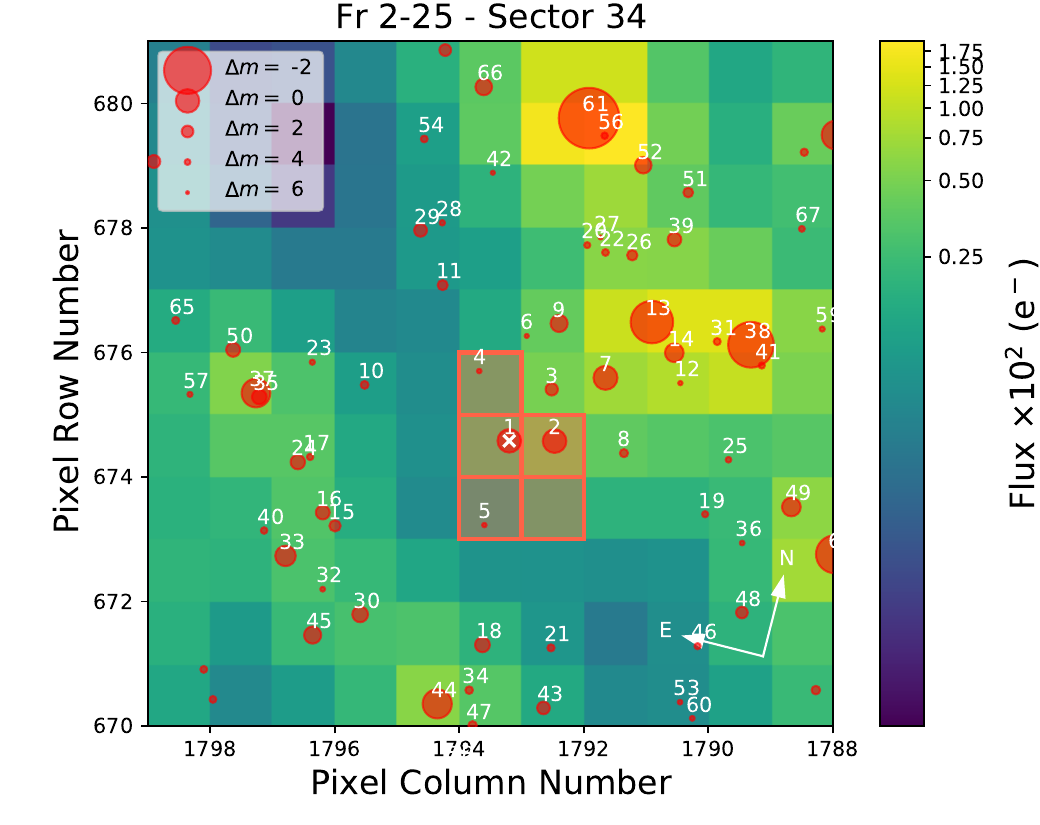}

\caption{Target pixel files (TPFs) of each central star in the sample (marked with white crosses) obtained with {\sc tpfplotter}. The red circles are the sources of the \textit{Gaia} DR3 catalogue in the field with scaled magnitudes (see legend). The aperture mask used by the pipeline to extract the photometry is also marked. Pixel scale is 21 arcsec\,pixel$^{-1}$.}
\label{fig:TPFs}
\end{figure}
\FloatBarrier

\addtocounter{figure}{-1}

  \begin{figure*}
               \includegraphics[width=0.335\textwidth]{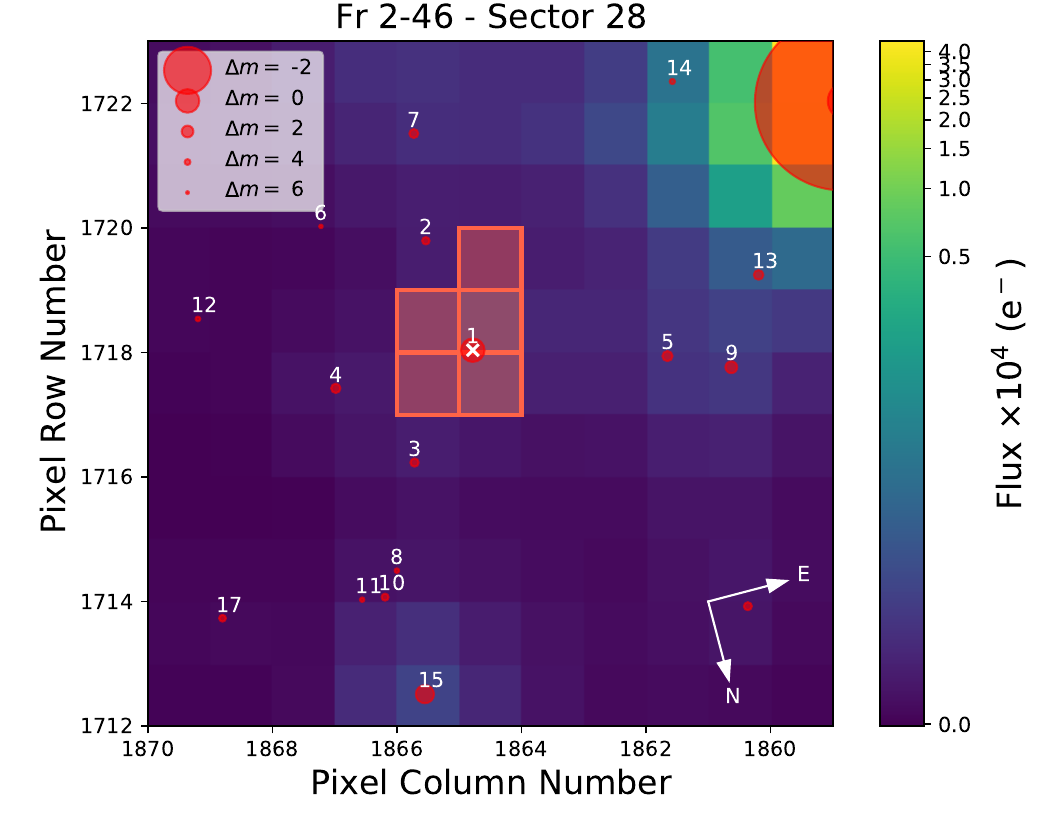}          
     \includegraphics[width=0.335\textwidth]{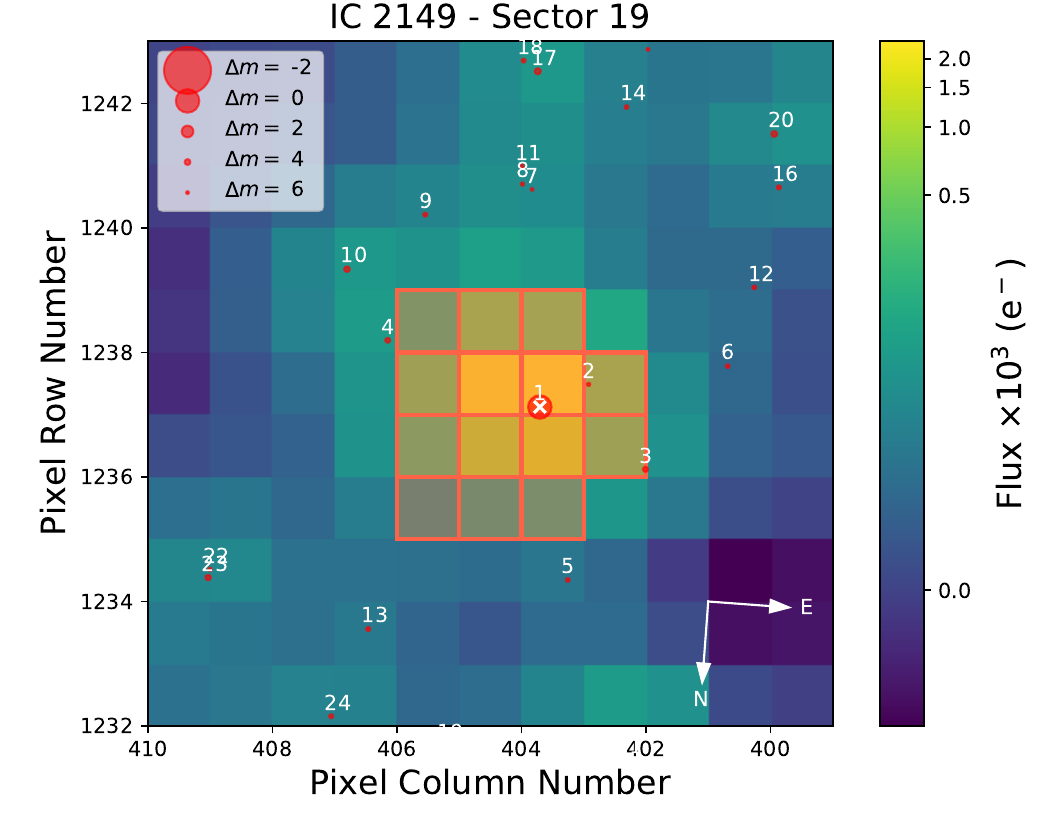}
             \includegraphics[width=0.335\textwidth]{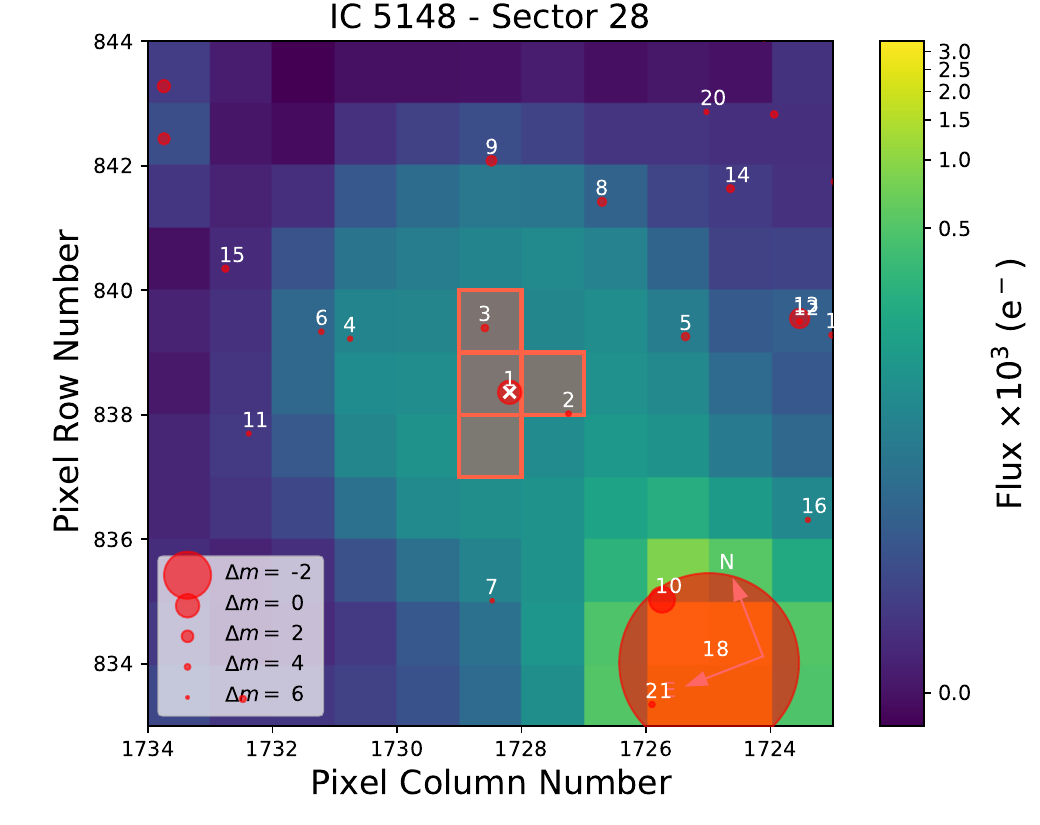}
 \includegraphics[width=0.335\textwidth]{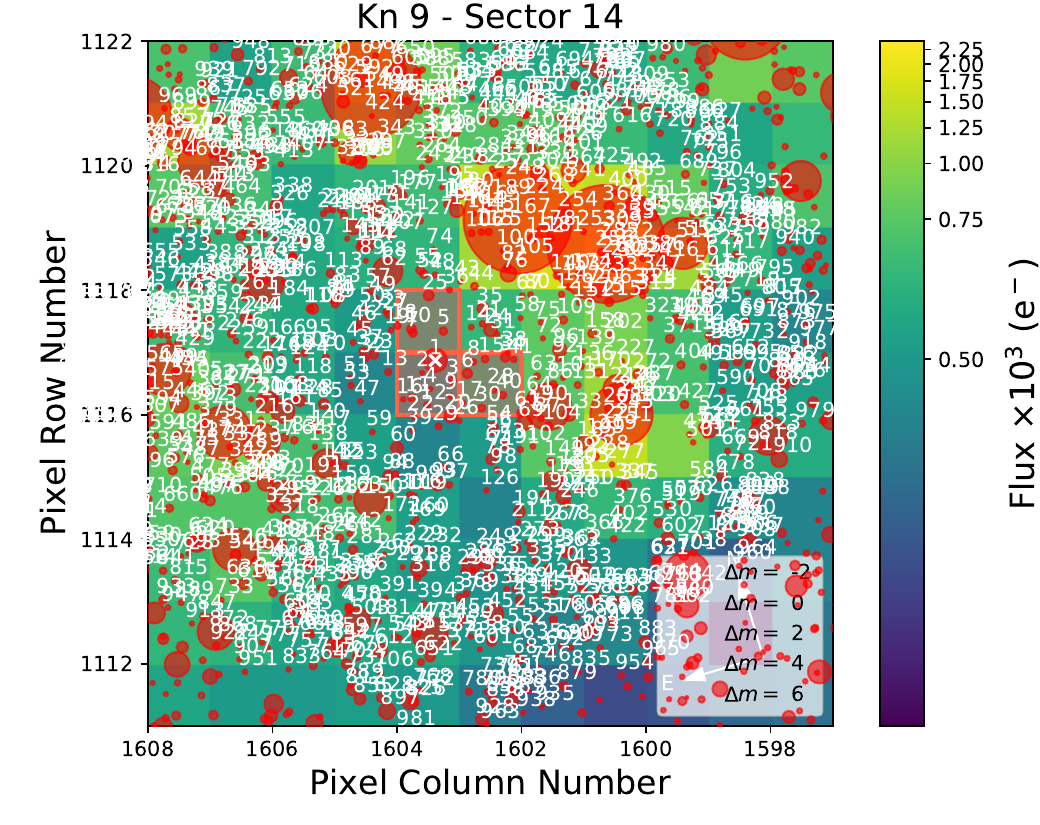}
\includegraphics[width=0.335\textwidth]{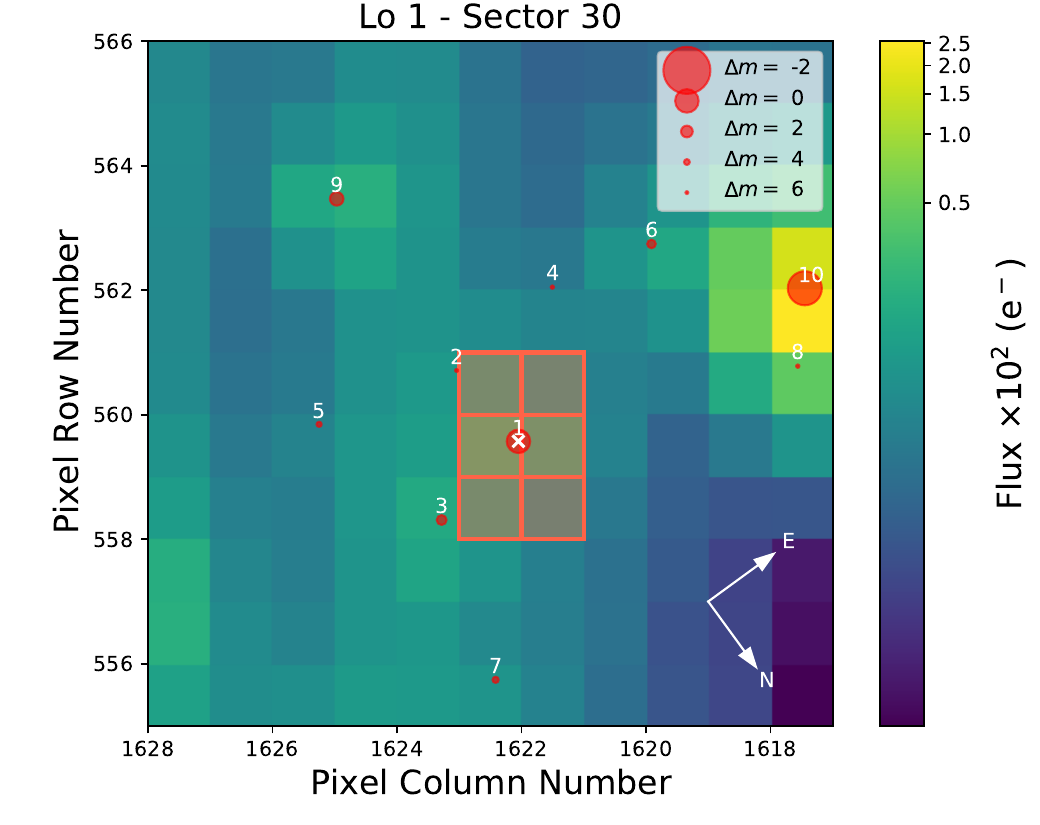}
       \includegraphics[width=0.335\textwidth]{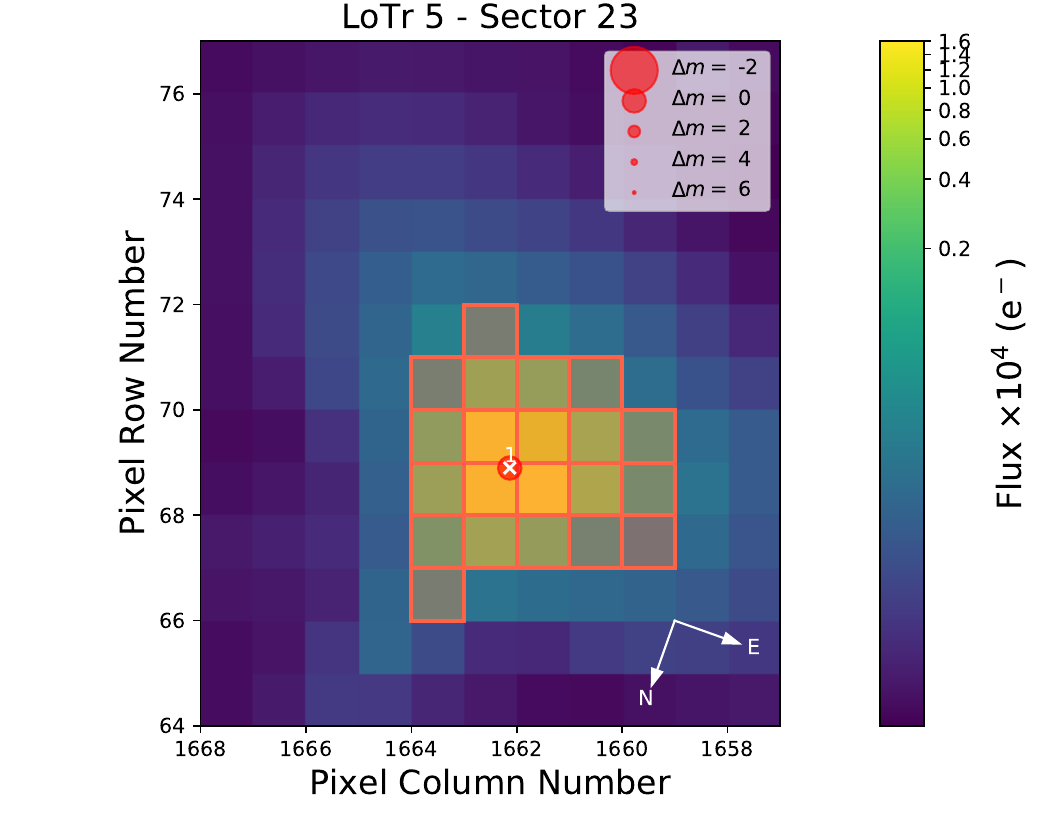}   
     \includegraphics[width=0.335\textwidth]{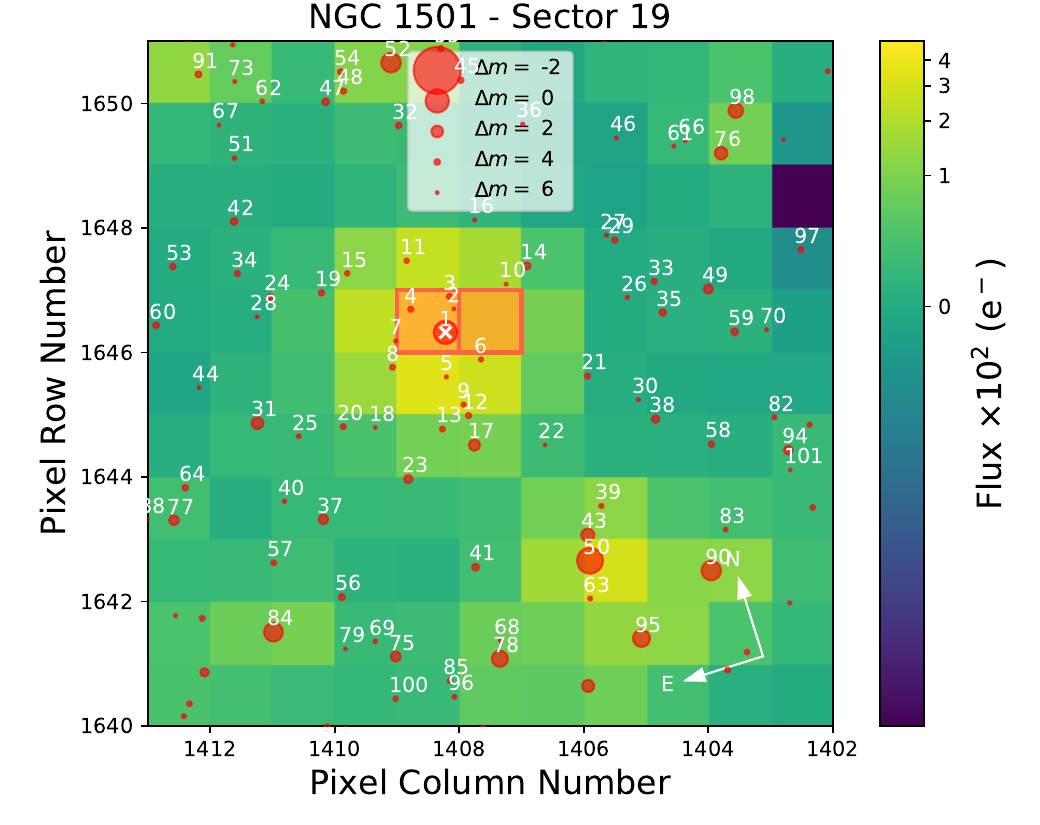}
     \includegraphics[width=0.335\textwidth]{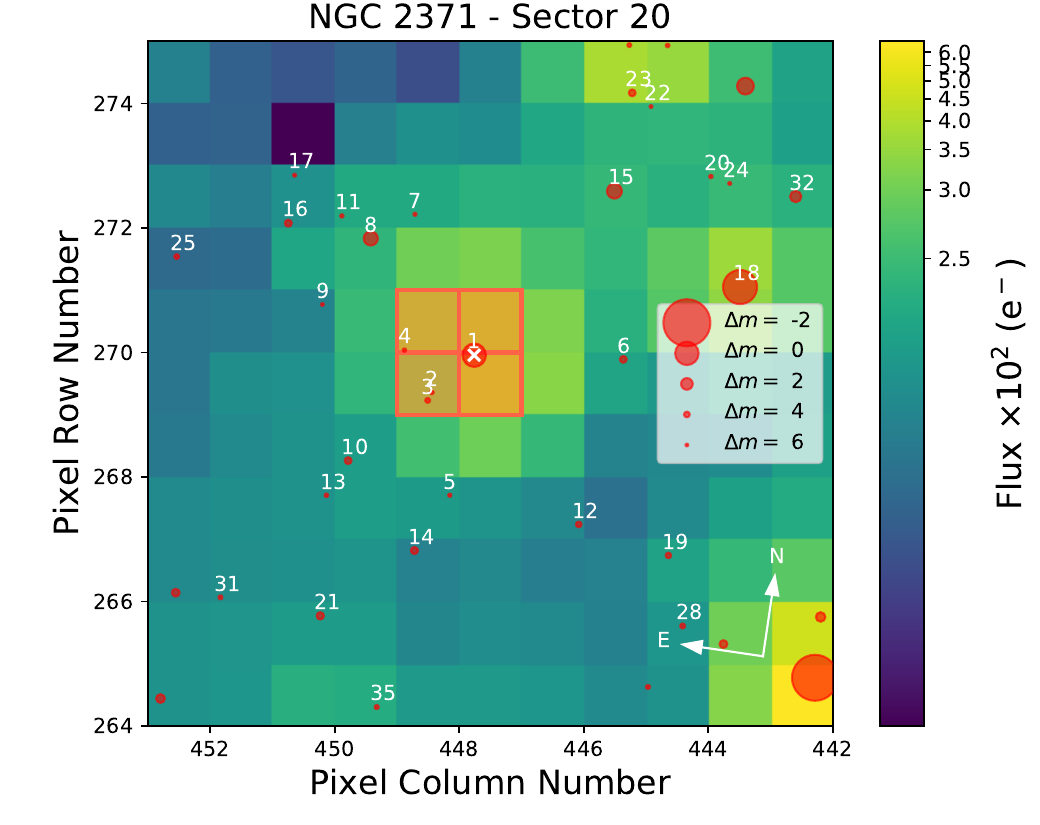}
      \includegraphics[width=0.335\textwidth]{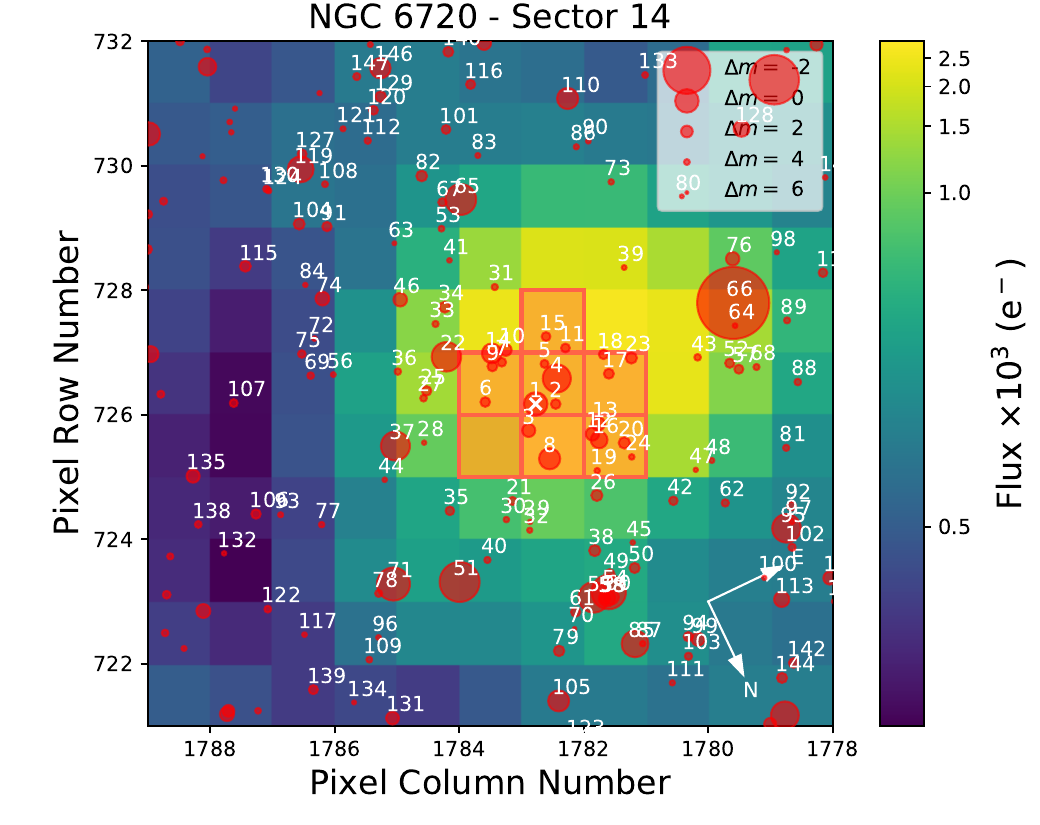}     
     \includegraphics[width=0.335\textwidth]{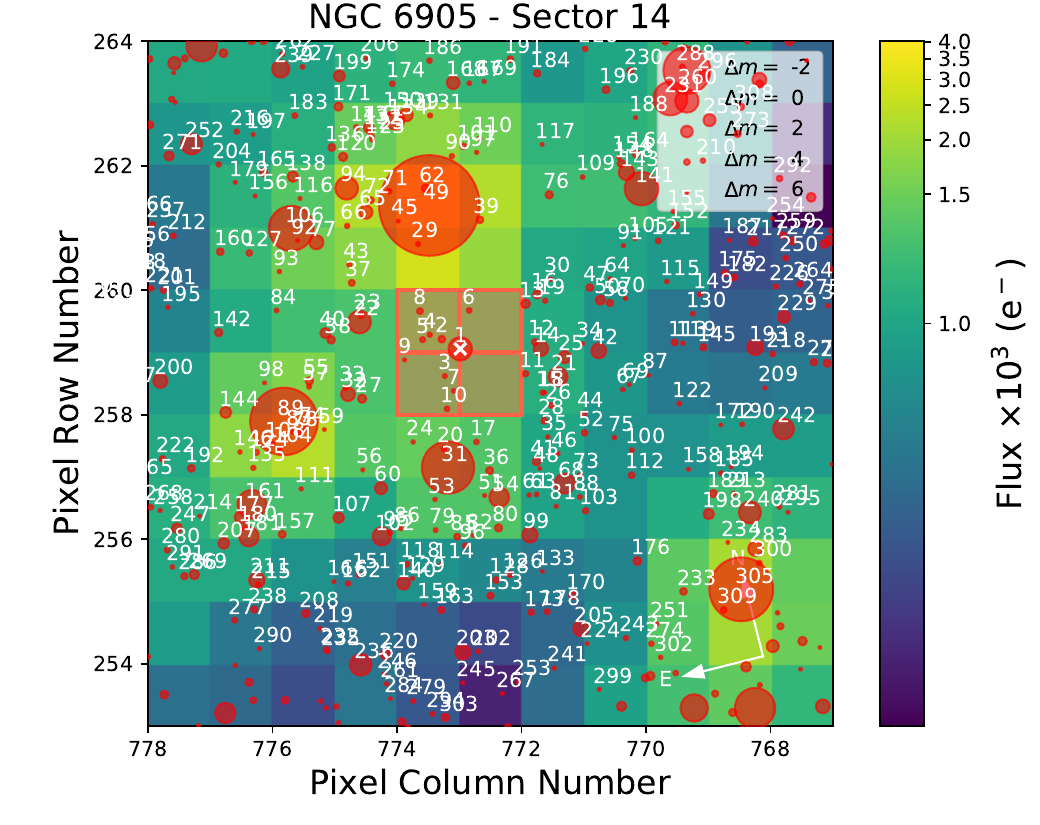}
        \includegraphics[width=0.335\textwidth]{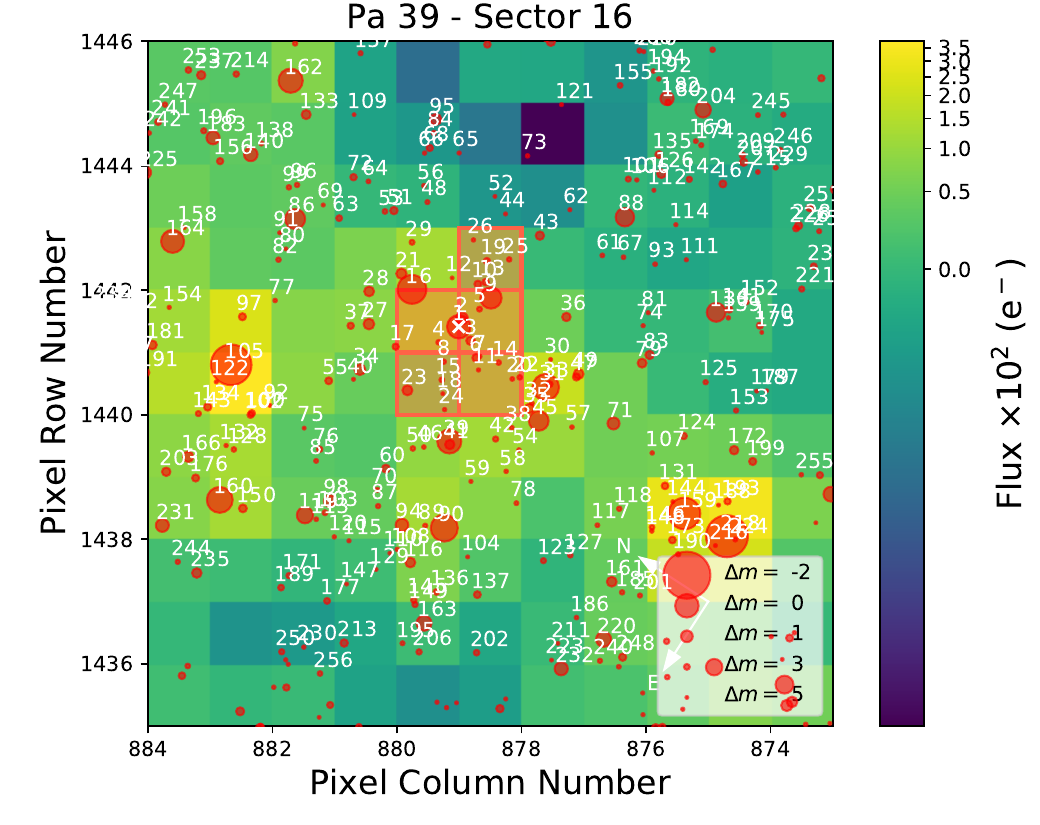}
         \includegraphics[width=0.335\textwidth]{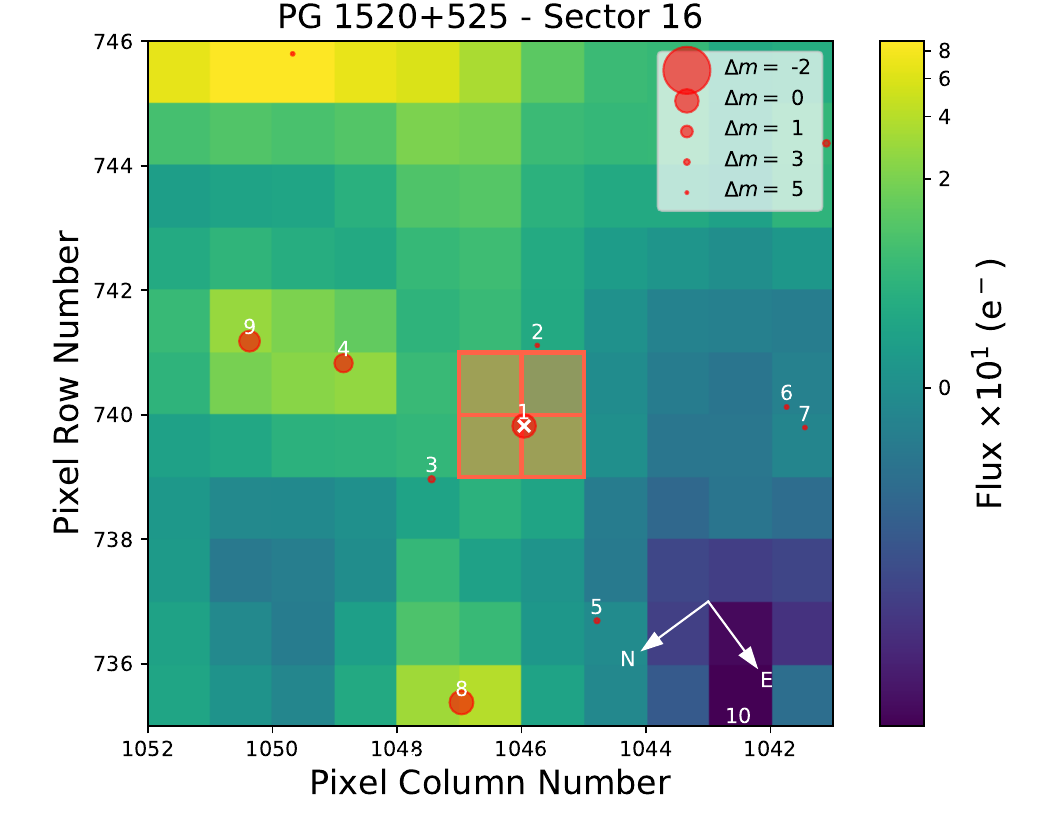}
             \includegraphics[width=0.33\textwidth]{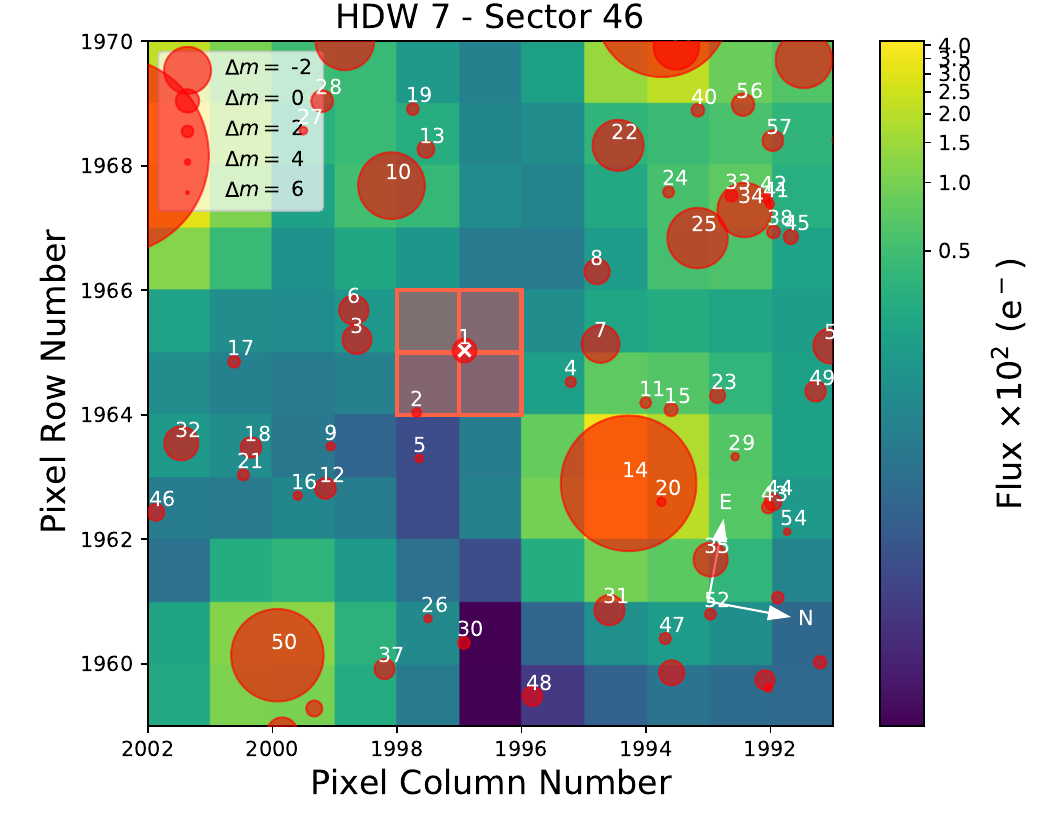}
                \includegraphics[width=0.33\textwidth]{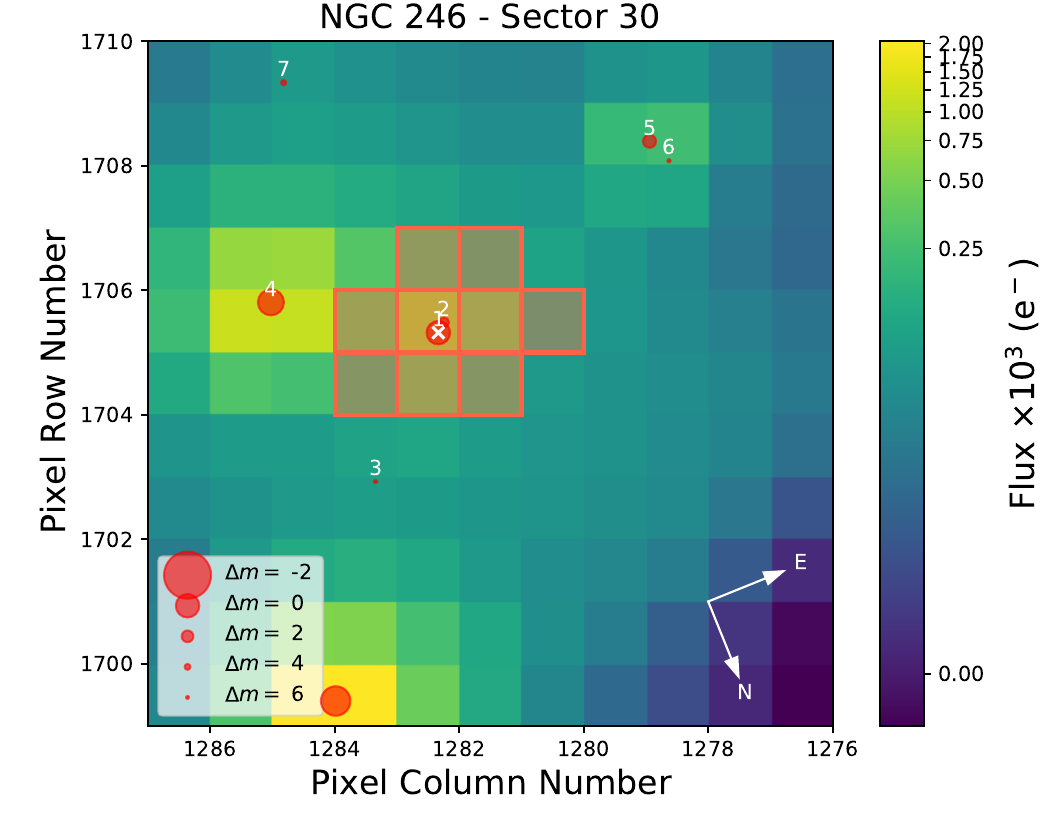}
    \includegraphics[width=0.33\textwidth]{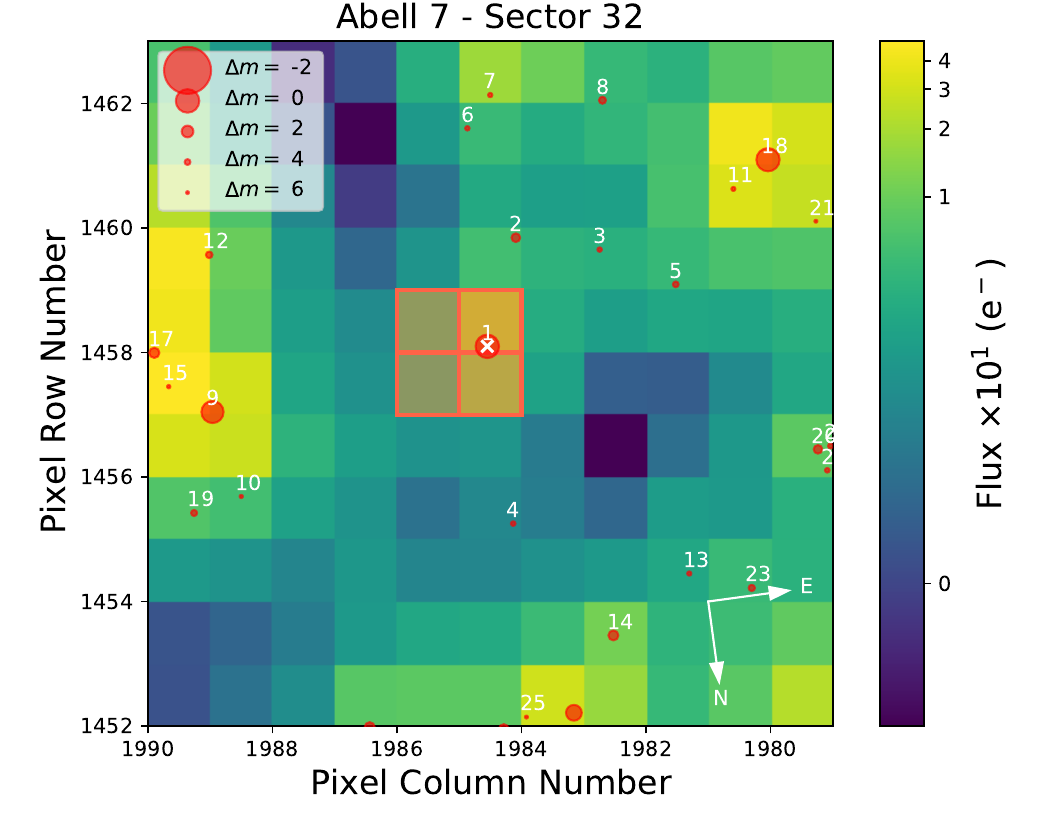}
\caption{continued.}
\label{fig:TPFs}
\end{figure*}

\addtocounter{figure}{-1}

  \begin{figure*}
             \includegraphics[width=0.33\textwidth]{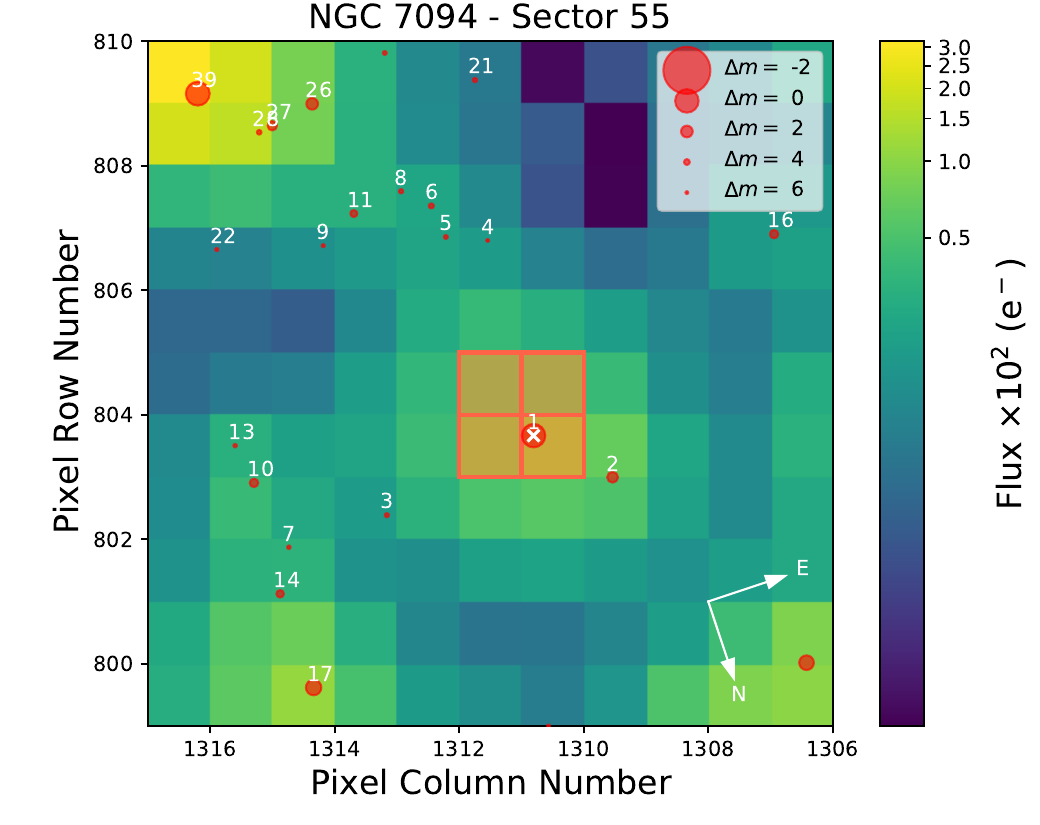}
      \includegraphics[width=0.33\textwidth]{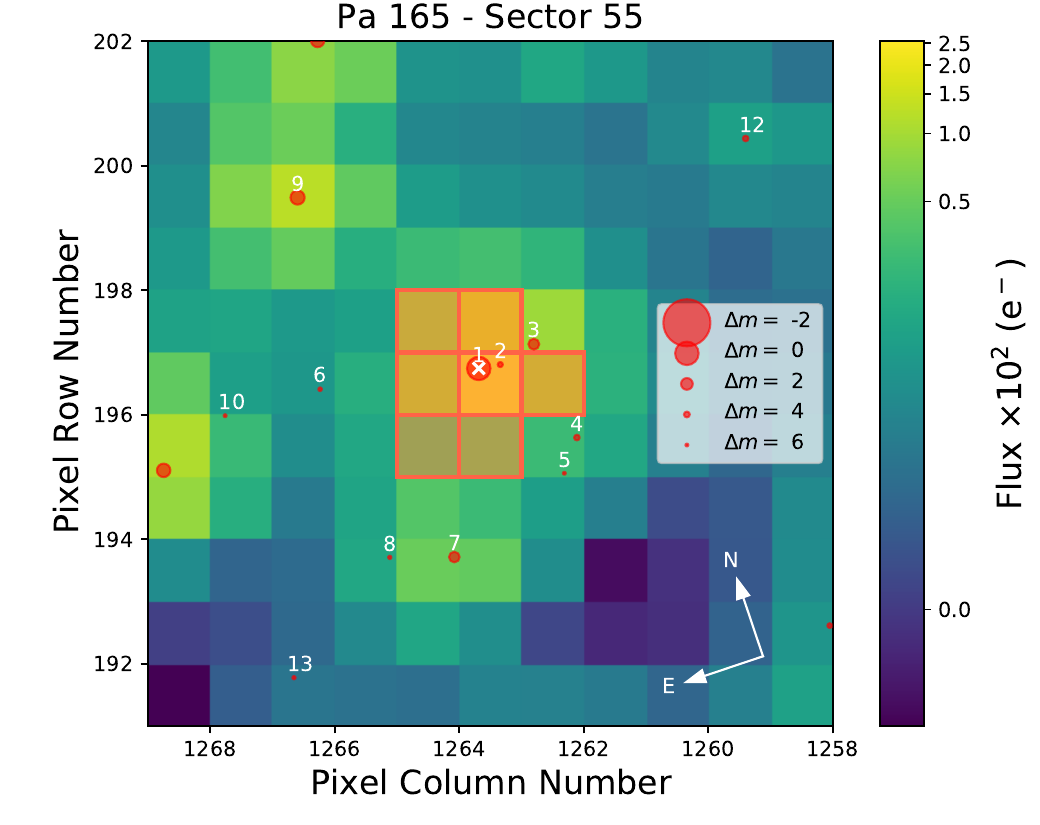}
      \includegraphics[width=0.33\textwidth]{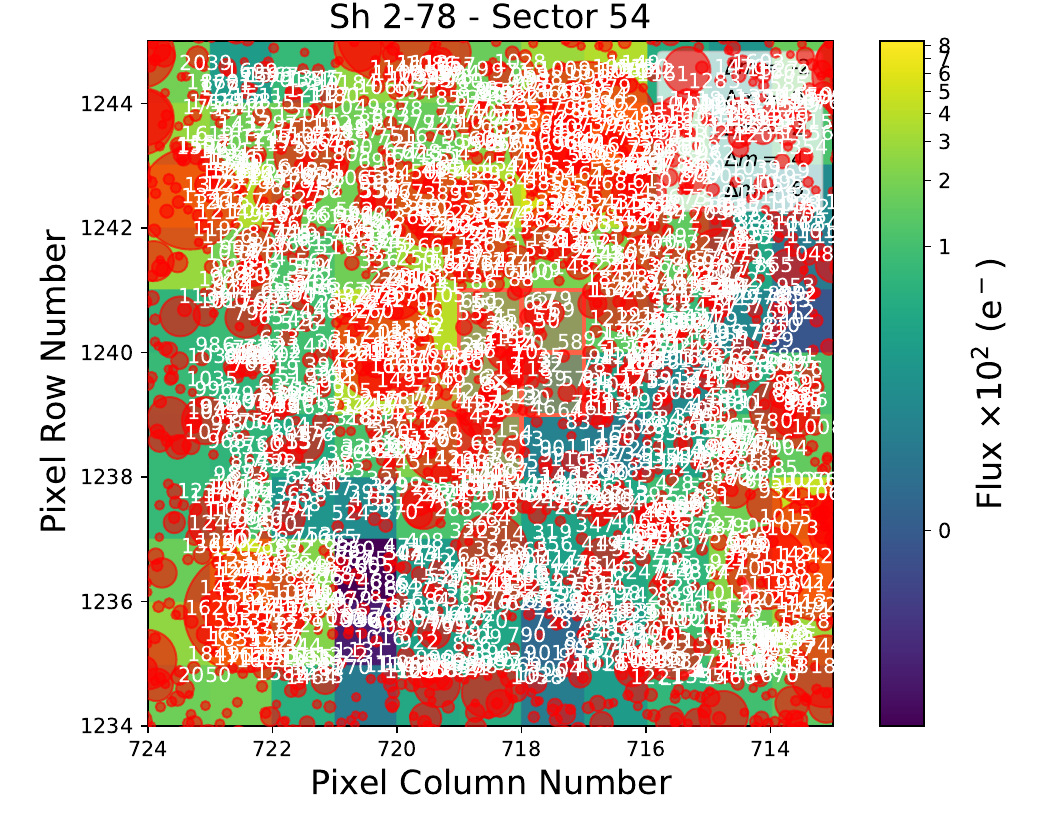}
            \includegraphics[width=0.335\textwidth]{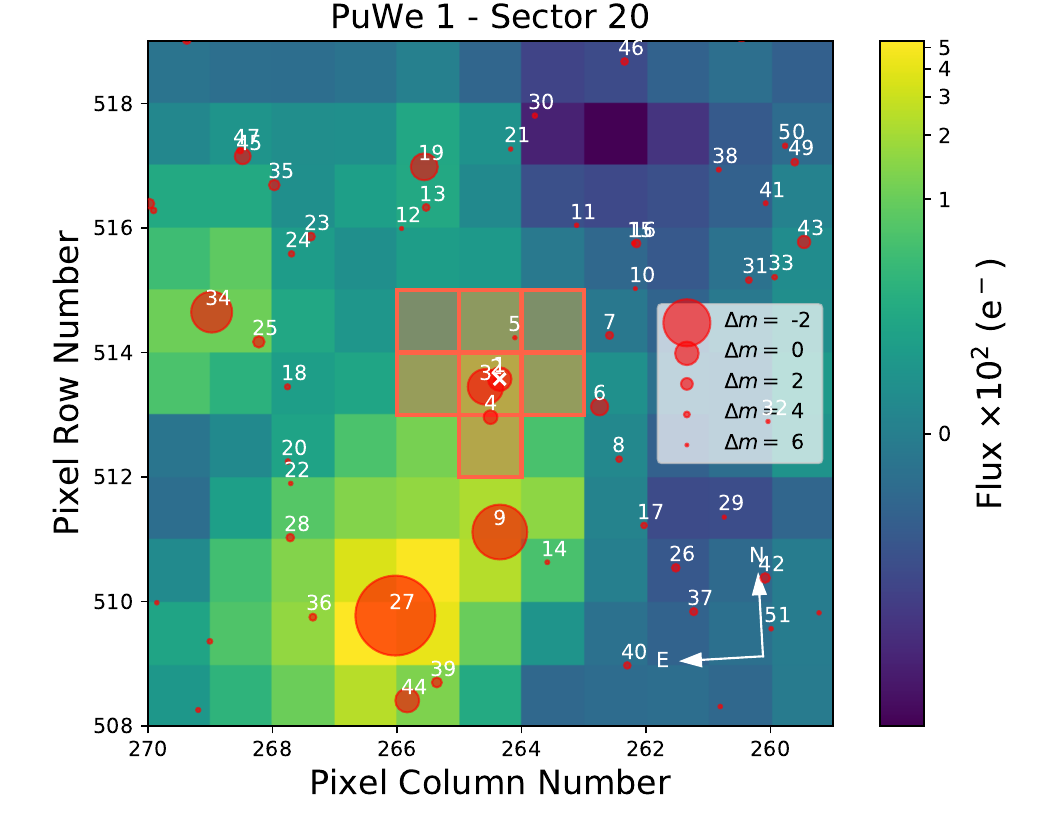}
         \includegraphics[width=0.335\textwidth]{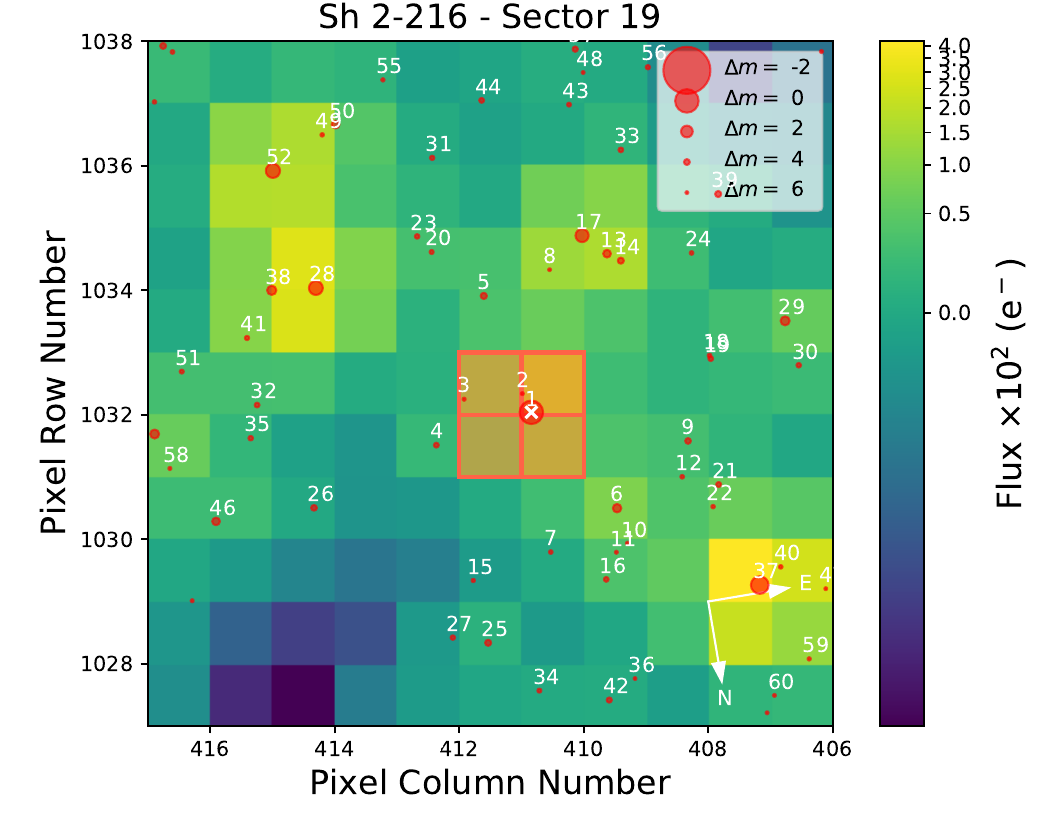}
          \includegraphics[width=0.335\textwidth]{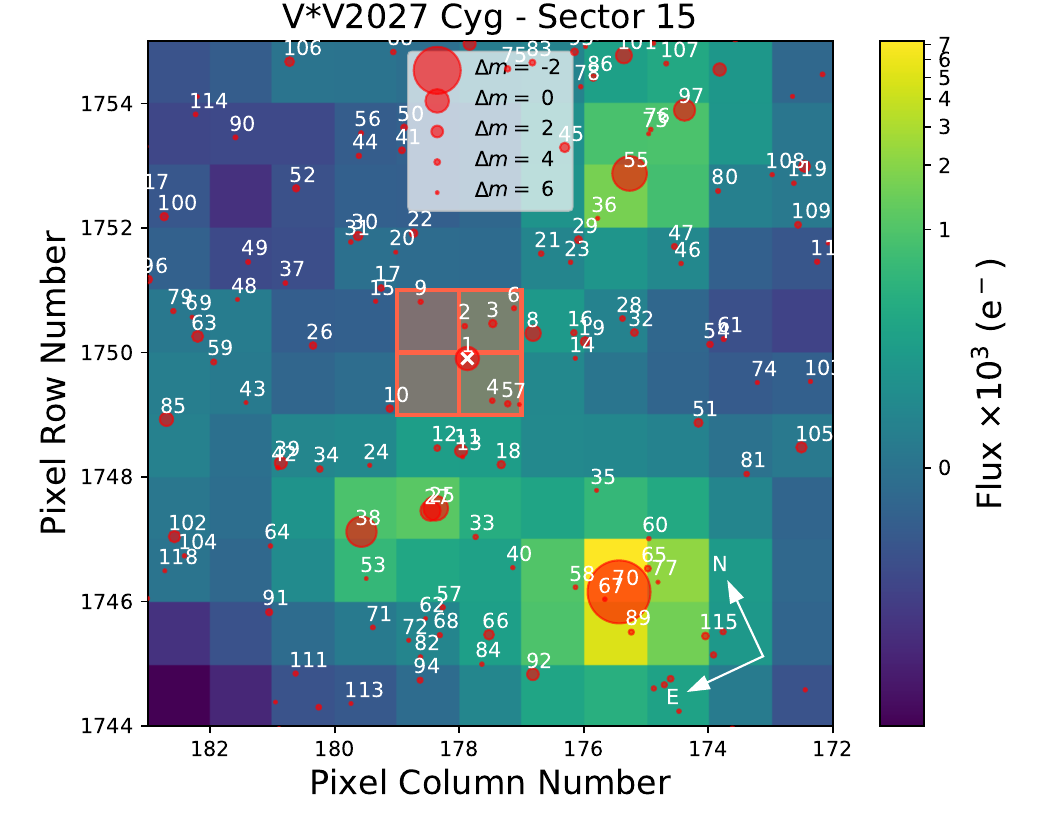}     
     \includegraphics[width=0.335\textwidth]{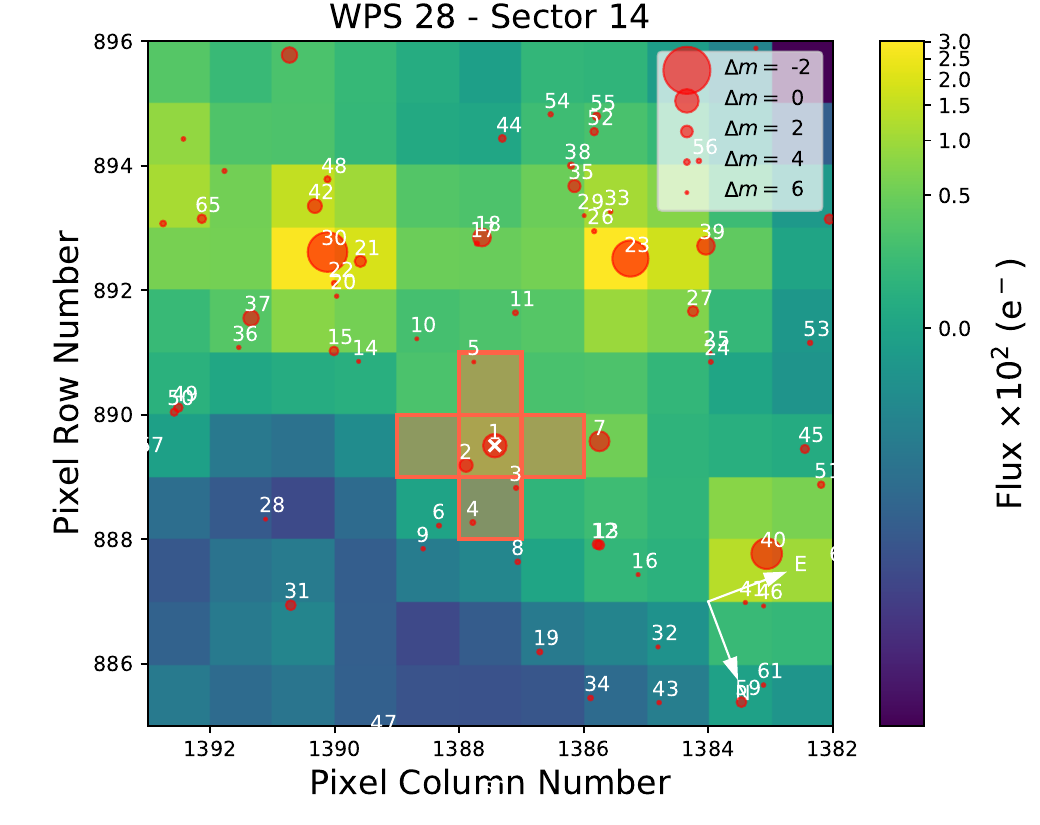}
     \includegraphics[width=0.335\textwidth]{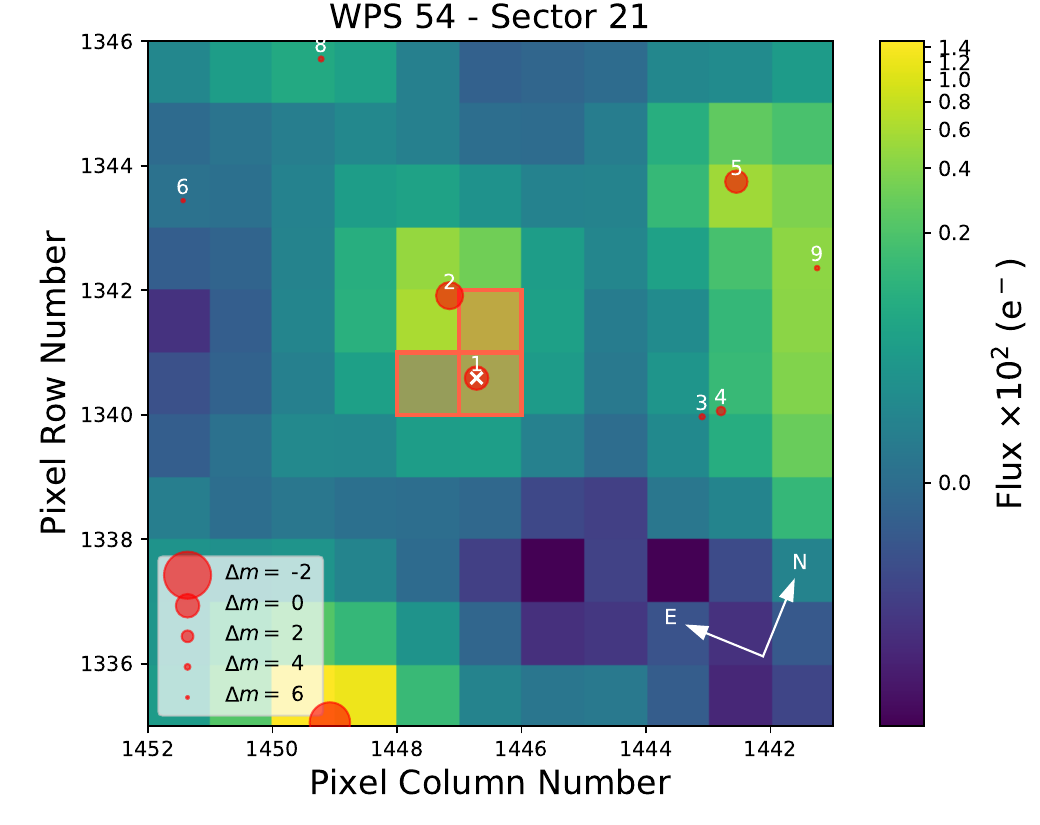}
     \includegraphics[width=0.335\textwidth]{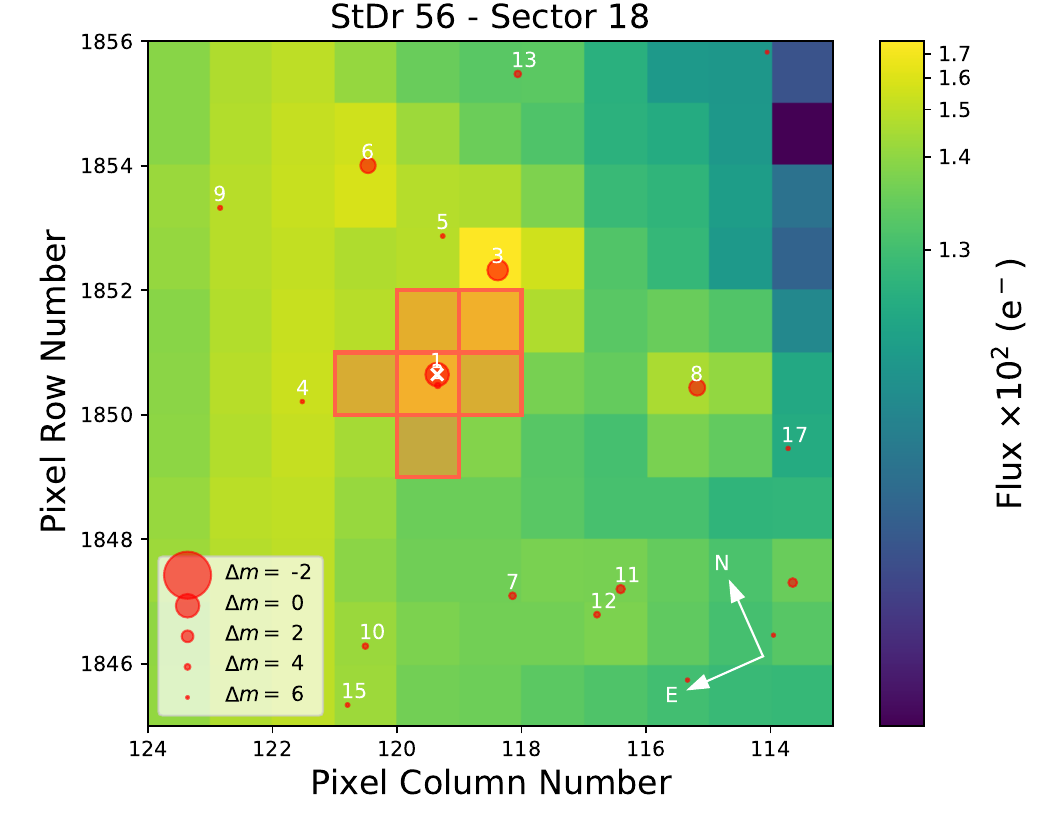}
        \includegraphics[width=0.33\textwidth]{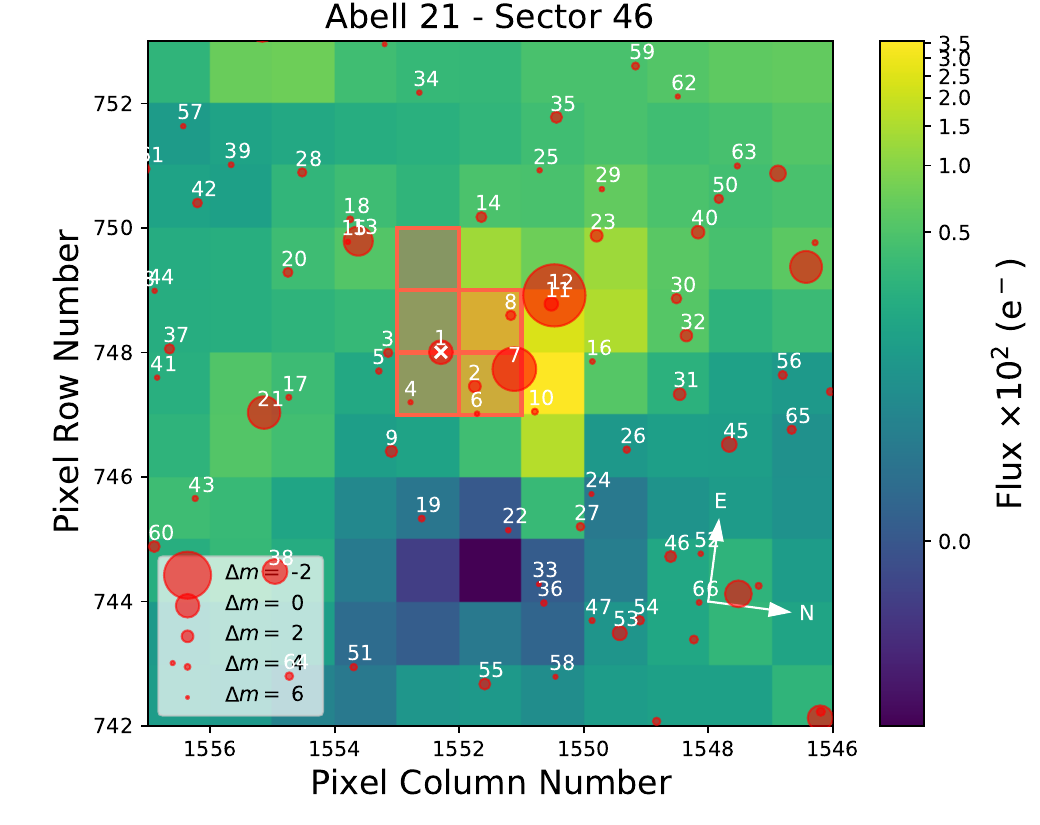}
              \includegraphics[width=0.33\textwidth]{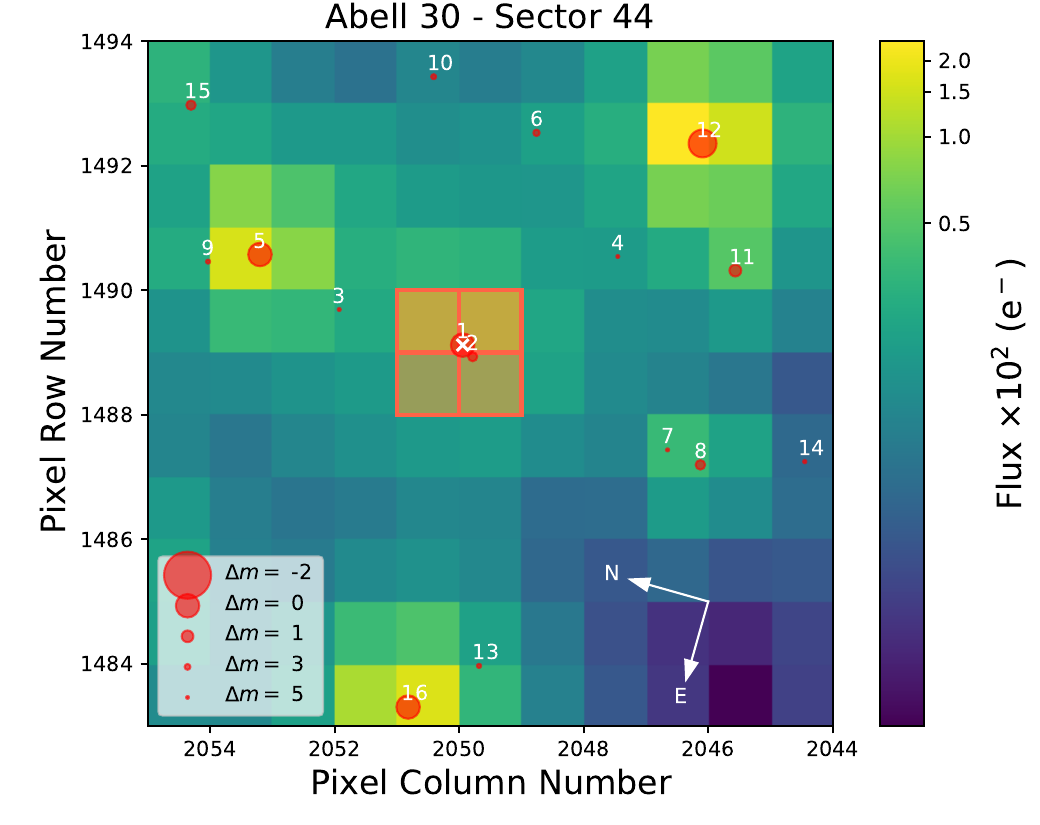}
     \includegraphics[width=0.33\textwidth]{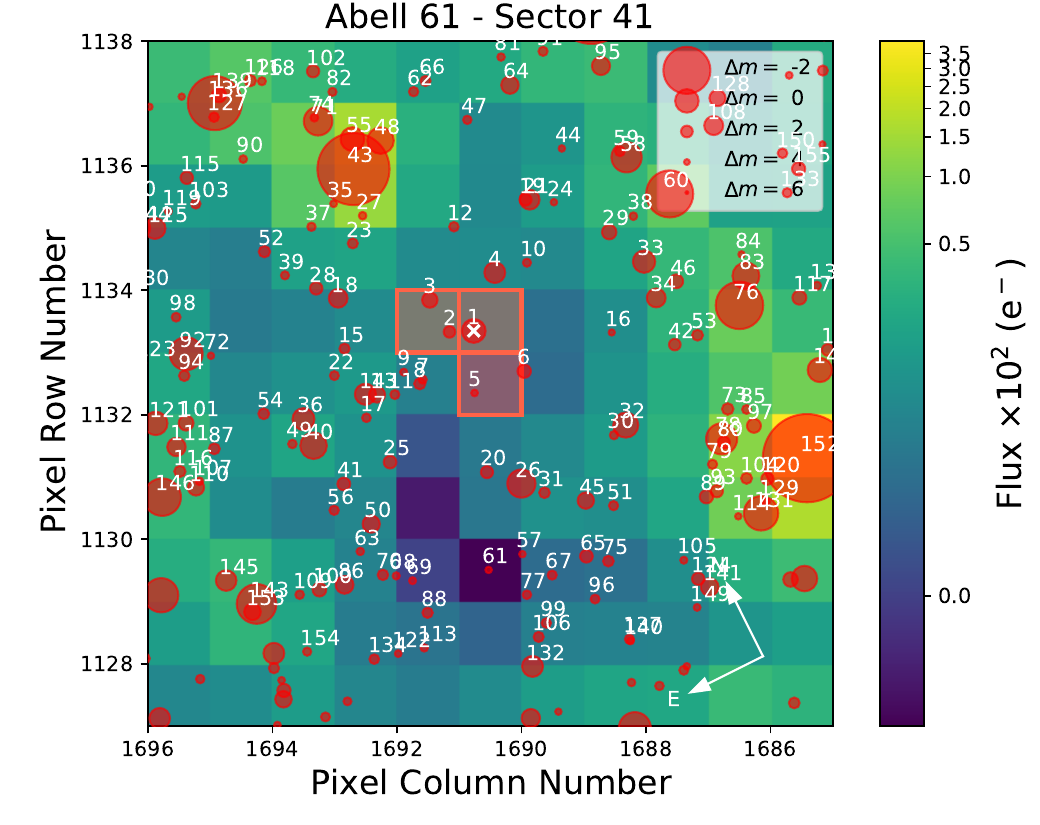}
    \includegraphics[width=0.33\textwidth]{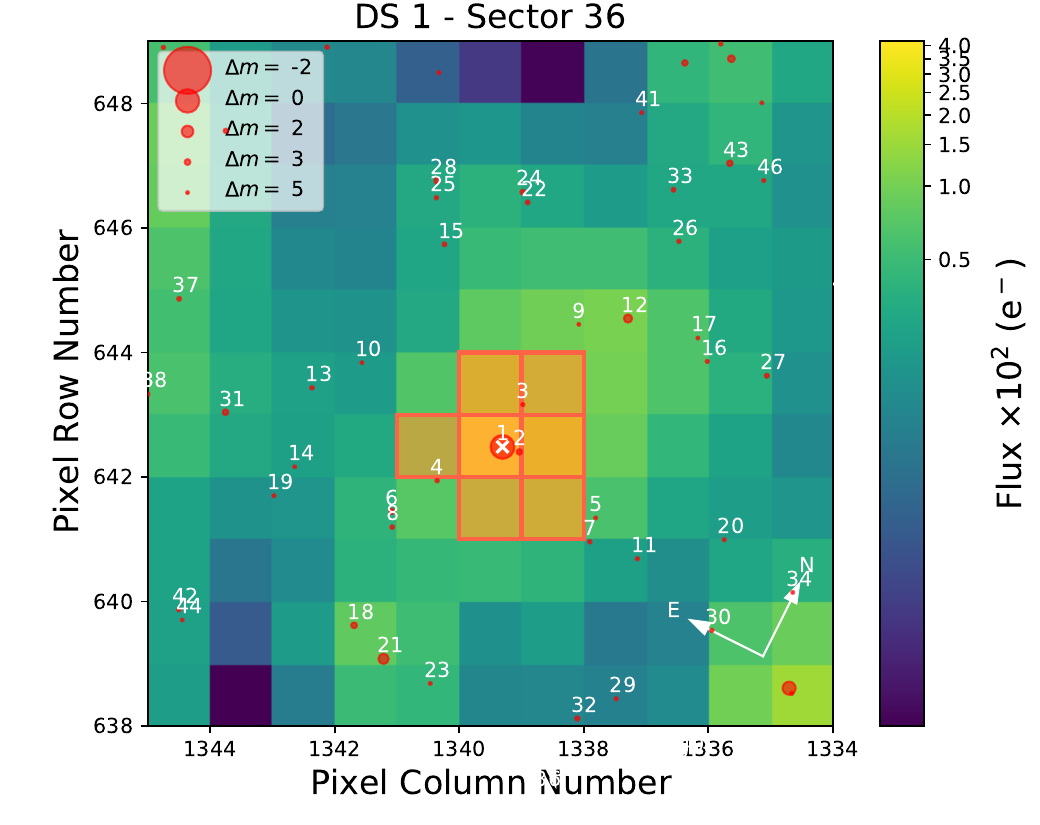}
     \includegraphics[width=0.33\textwidth]{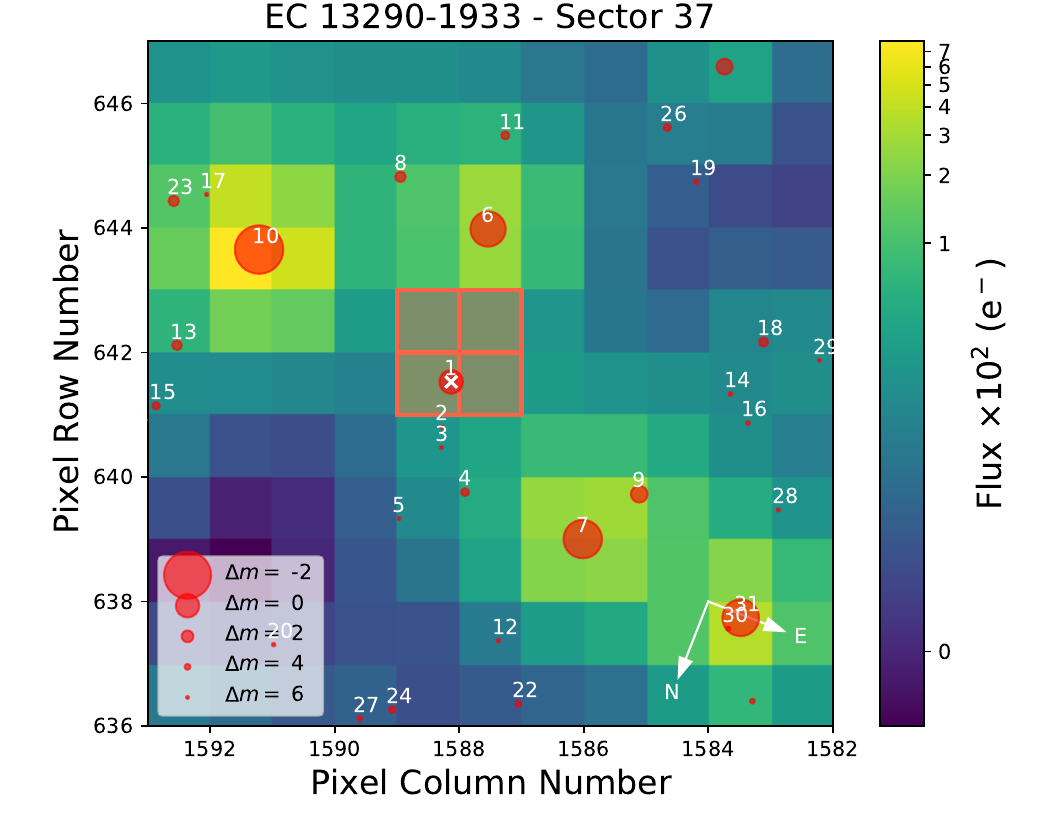}
        \includegraphics[width=0.33\textwidth]{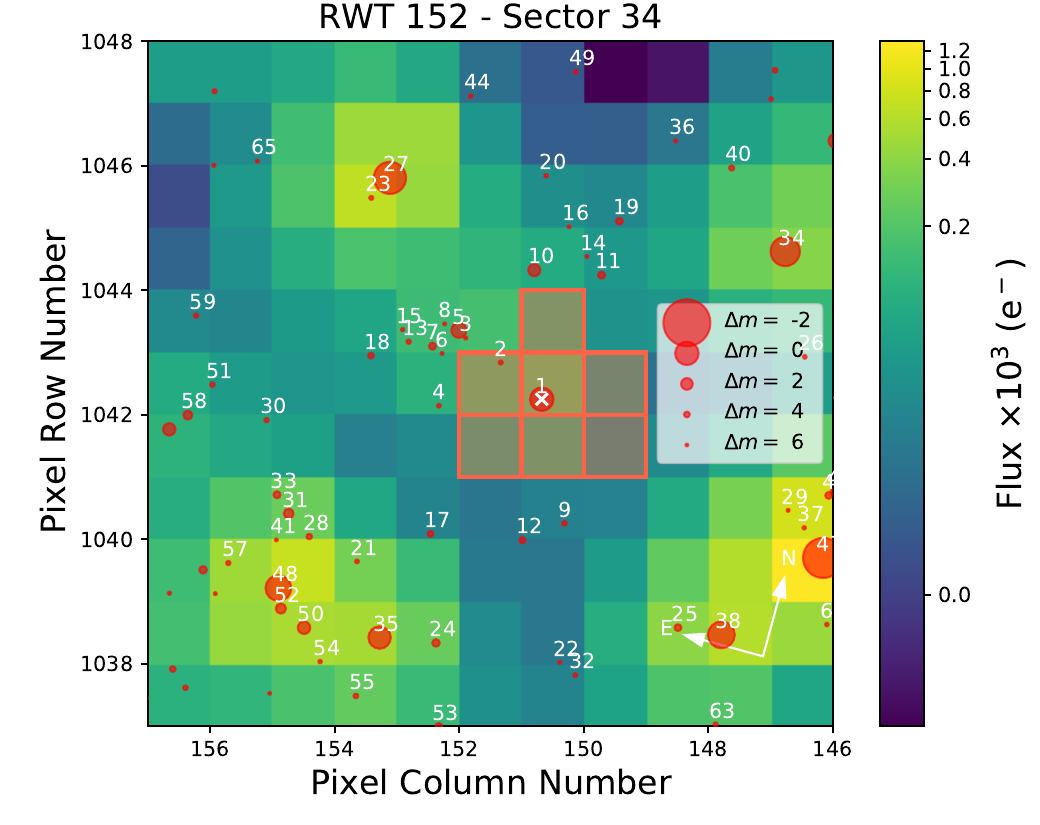}
\caption{continued.}
\label{fig:TPFs}
\end{figure*}

\addtocounter{figure}{-1}

  \begin{figure*}
  \centering
        \includegraphics[width=0.33\textwidth]{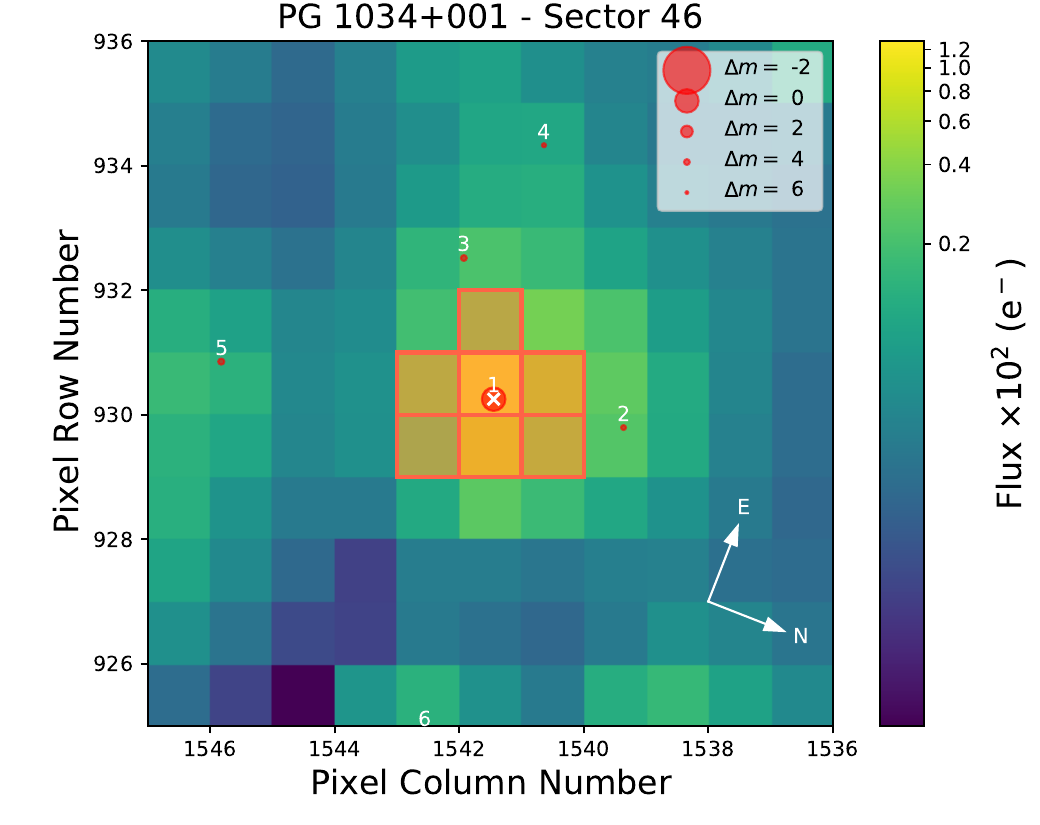}
    \includegraphics[width=0.33\textwidth]{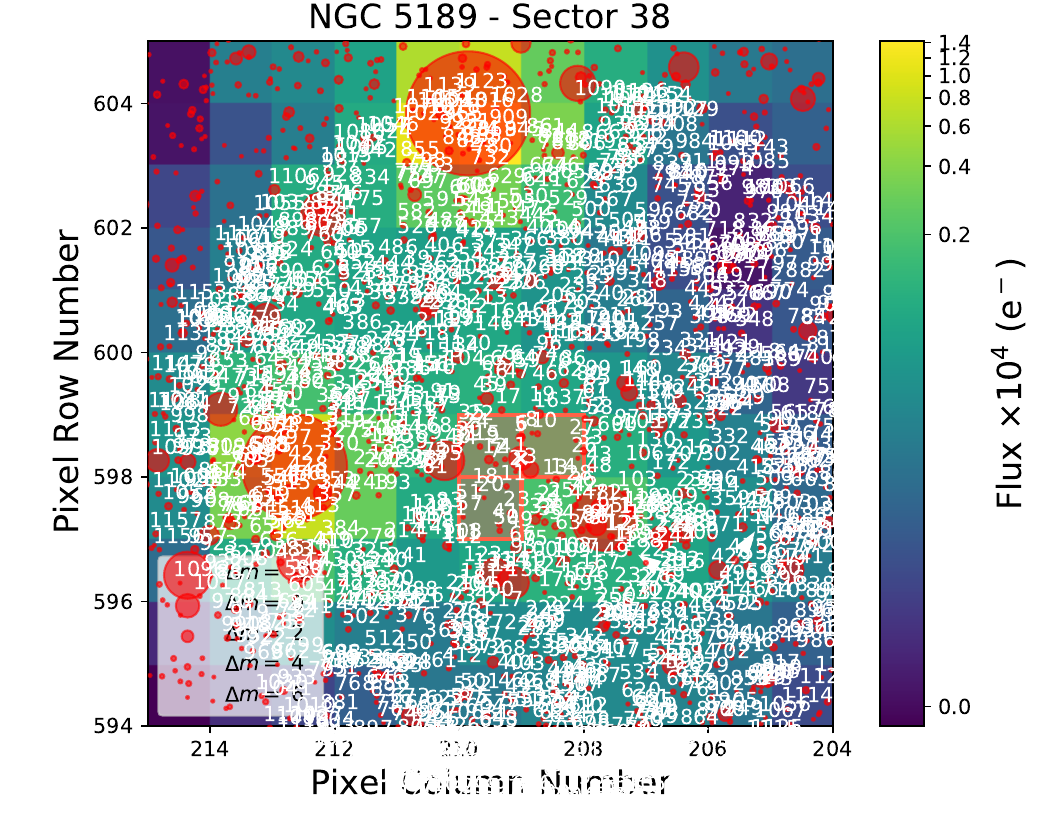}
         \includegraphics[width=0.33\textwidth]{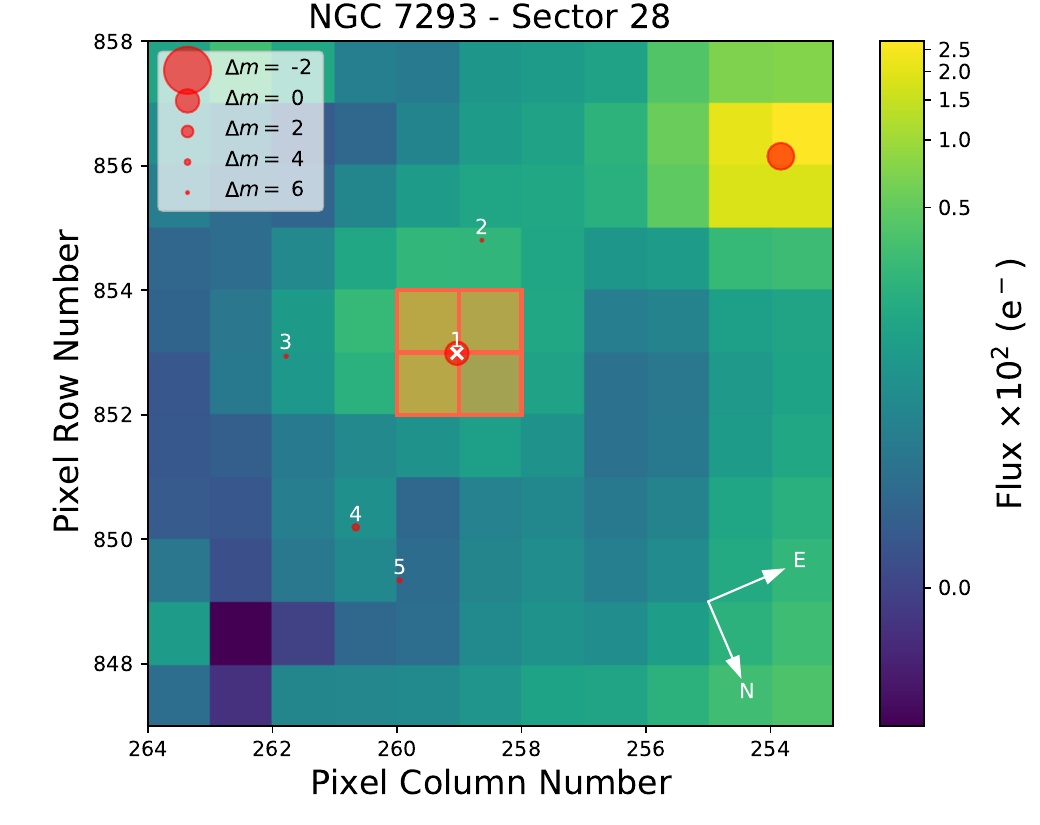}
      \includegraphics[width=0.33\textwidth]{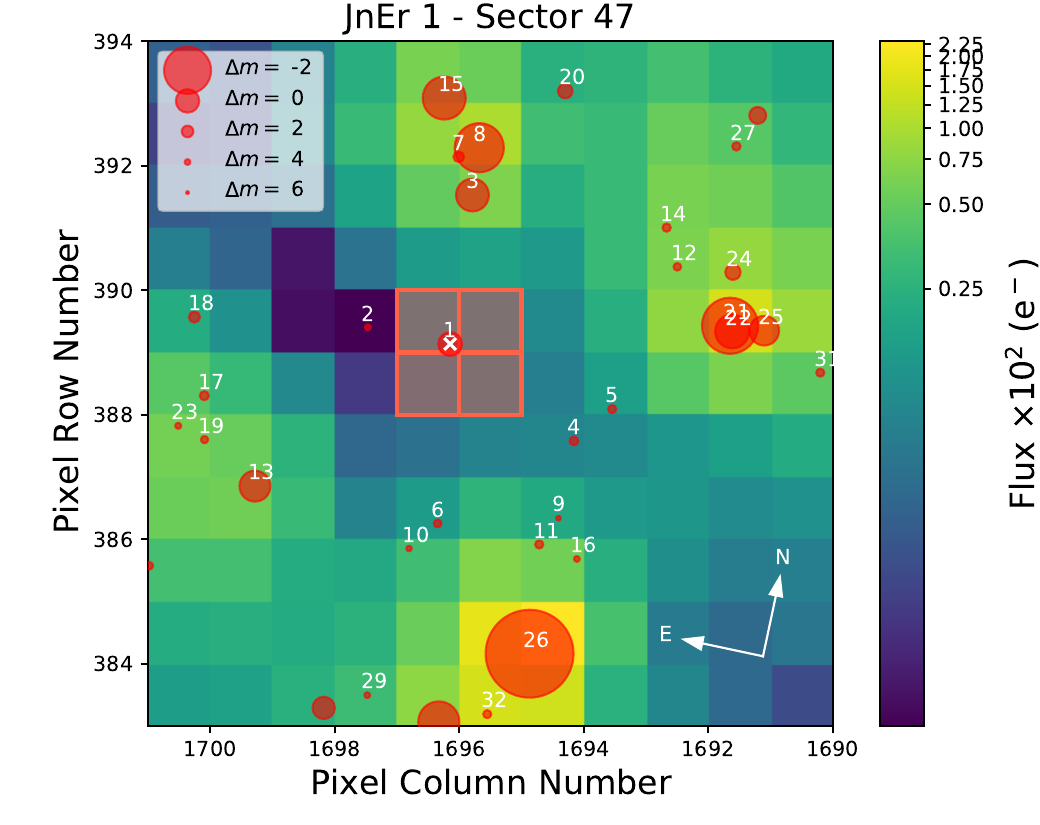}
     \includegraphics[width=0.33\textwidth]{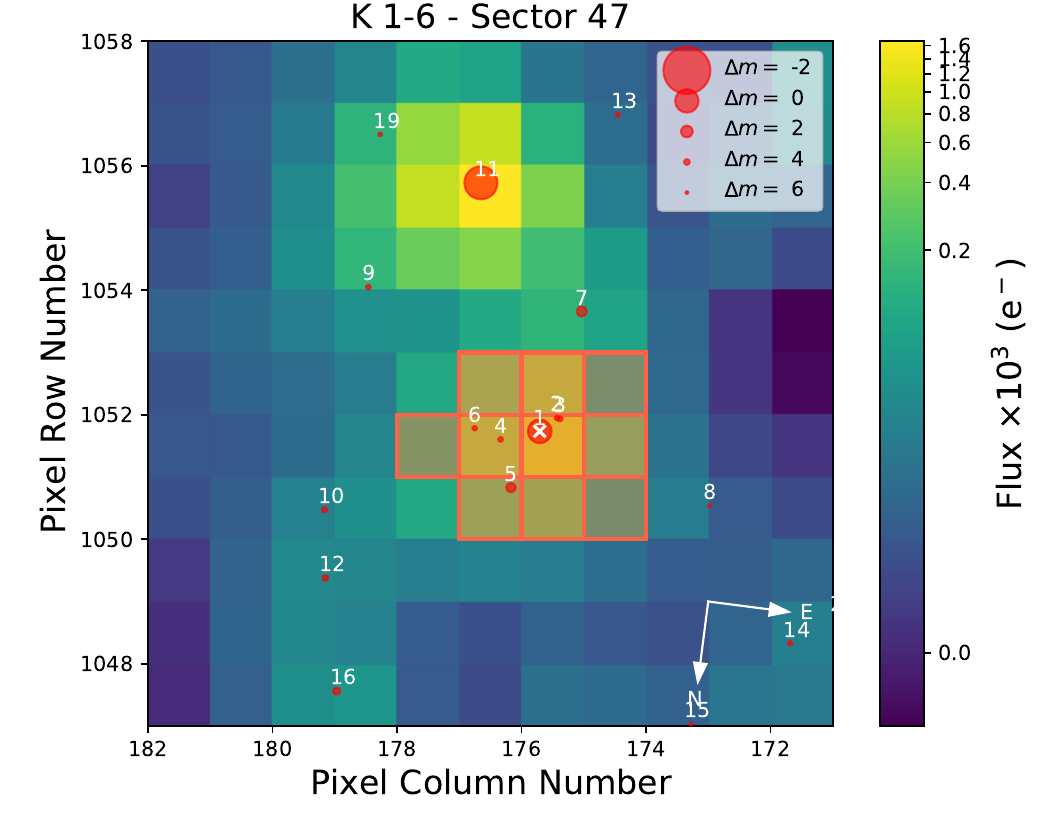}
      \includegraphics[width=0.33\textwidth]{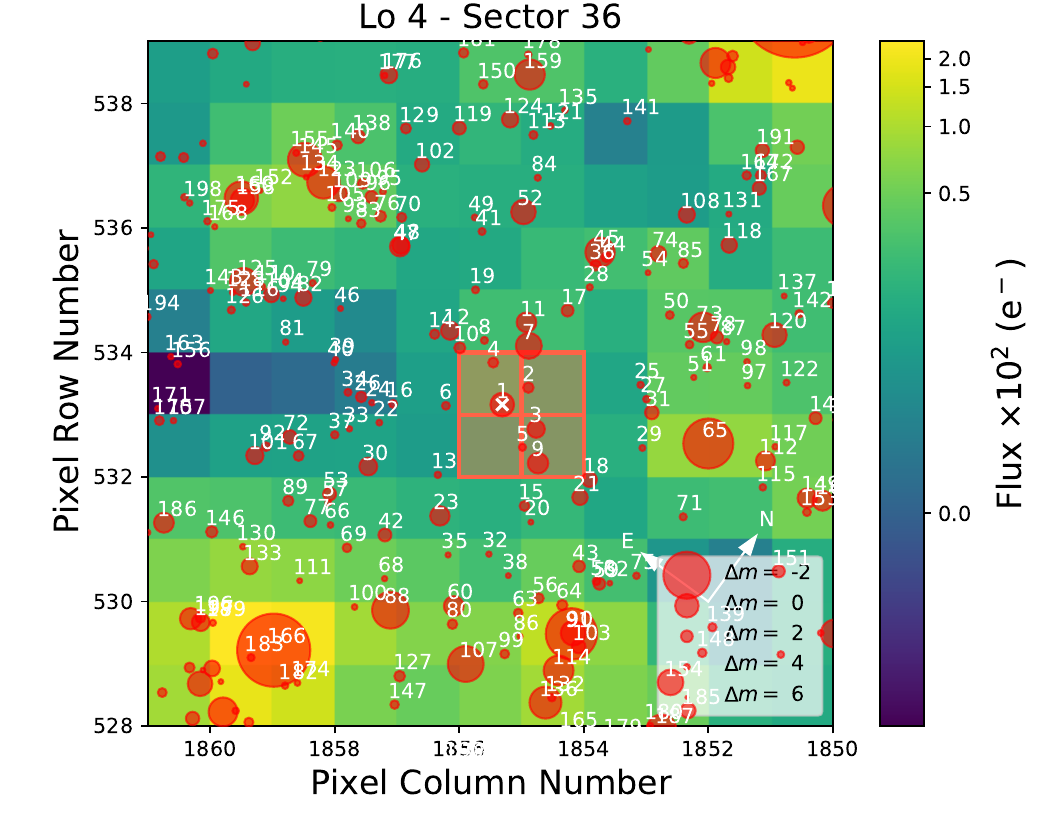}
     \includegraphics[width=0.33\textwidth]{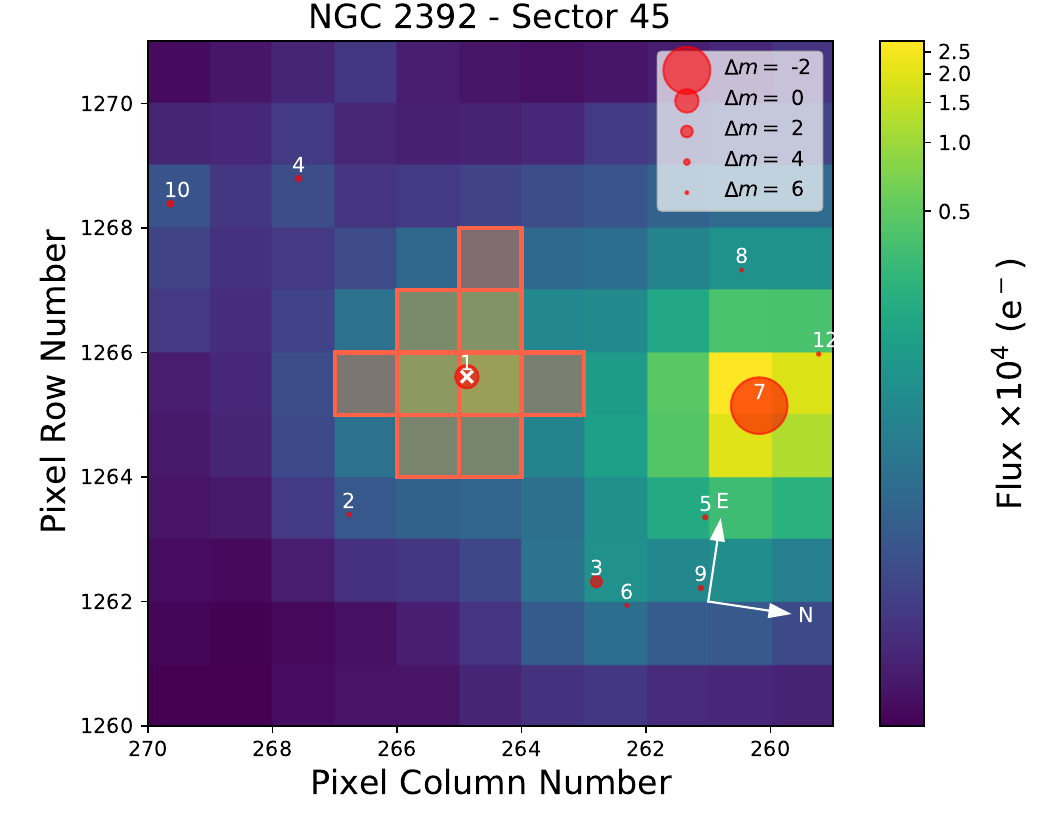}
     \includegraphics[width=0.33\textwidth]{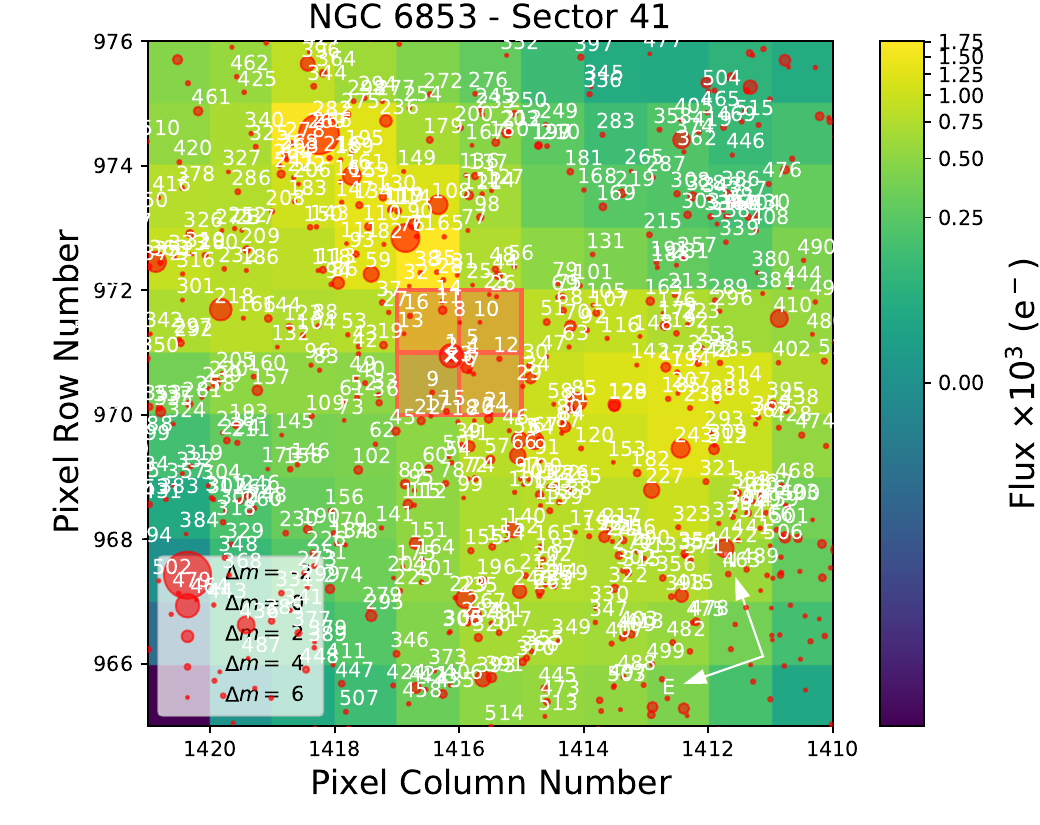}
    \includegraphics[width=0.33\textwidth]{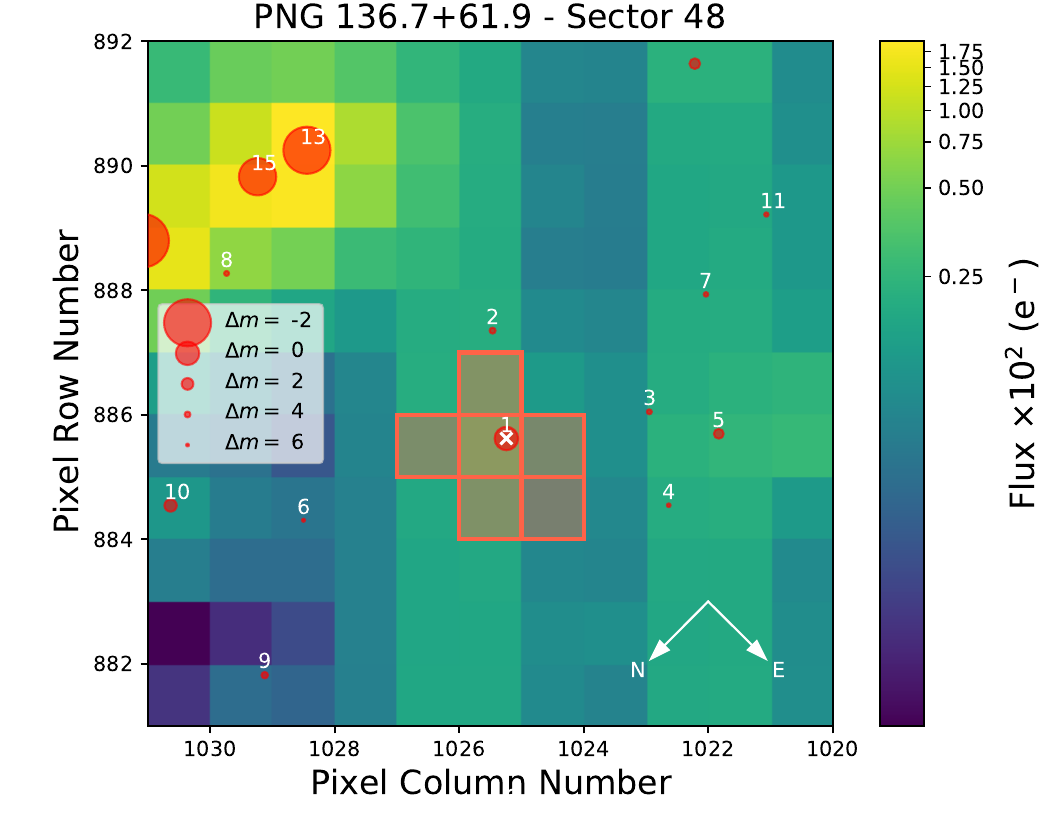}
       \includegraphics[width=0.33\textwidth]{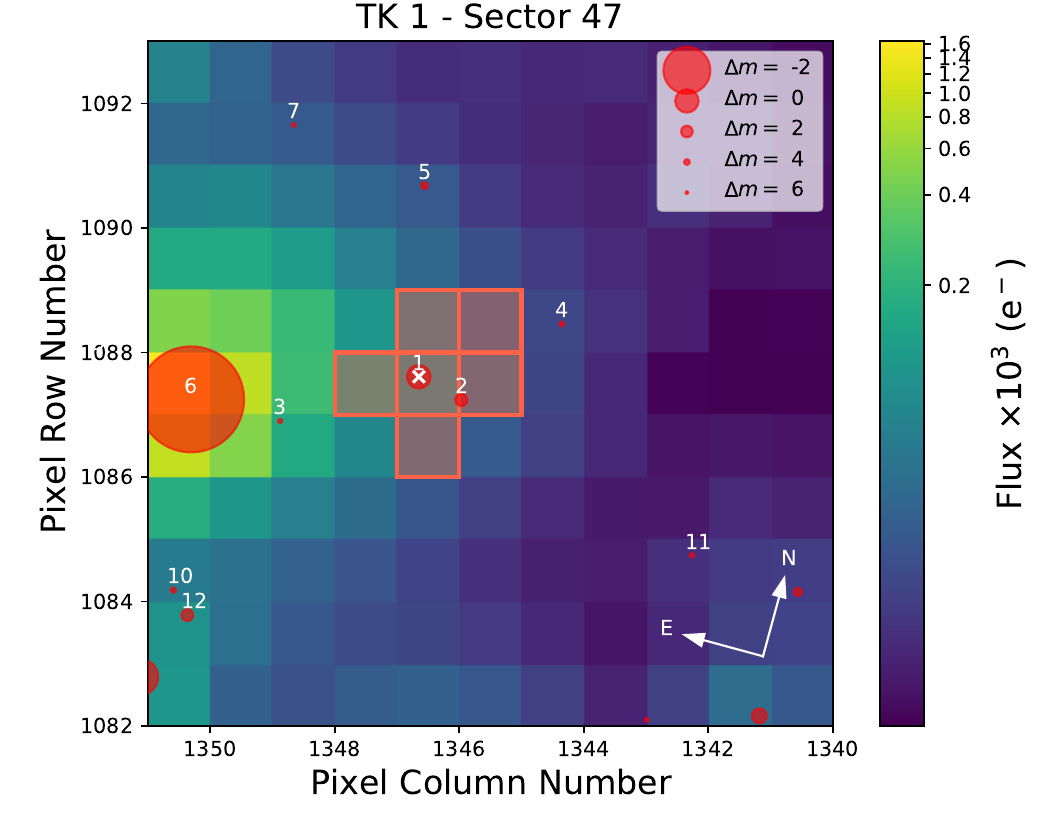}
     \includegraphics[width=0.33\textwidth]{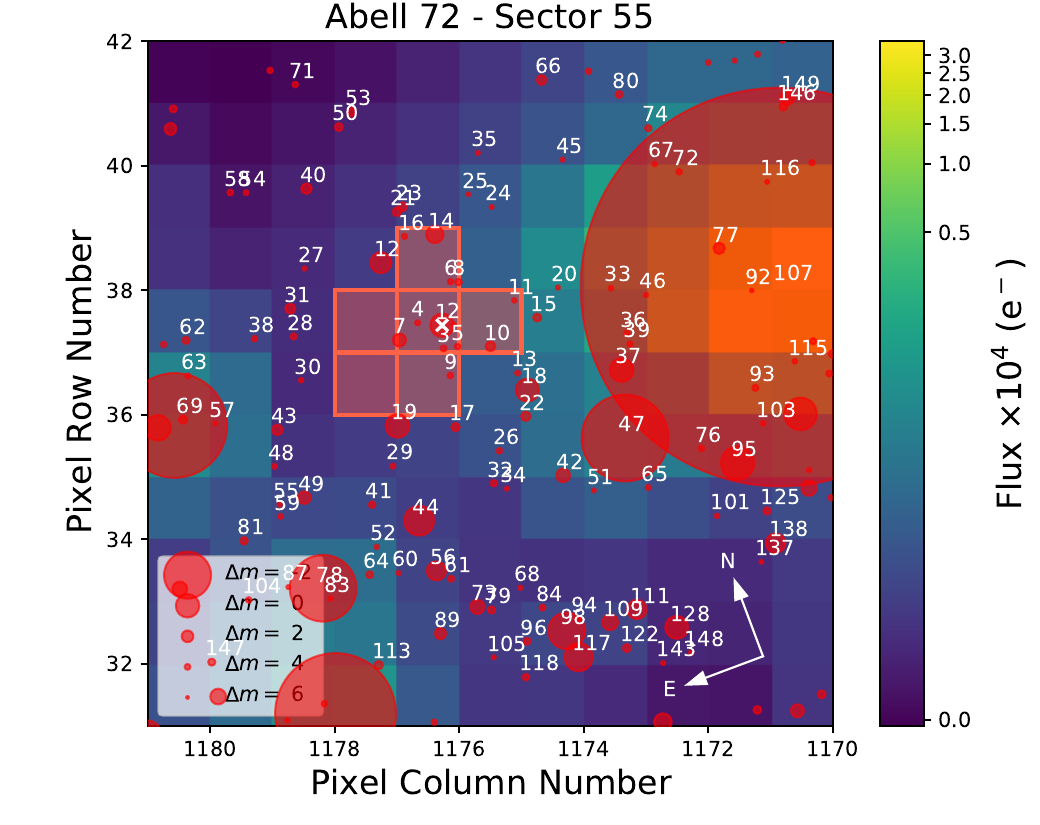}
     \includegraphics[width=0.33\textwidth]{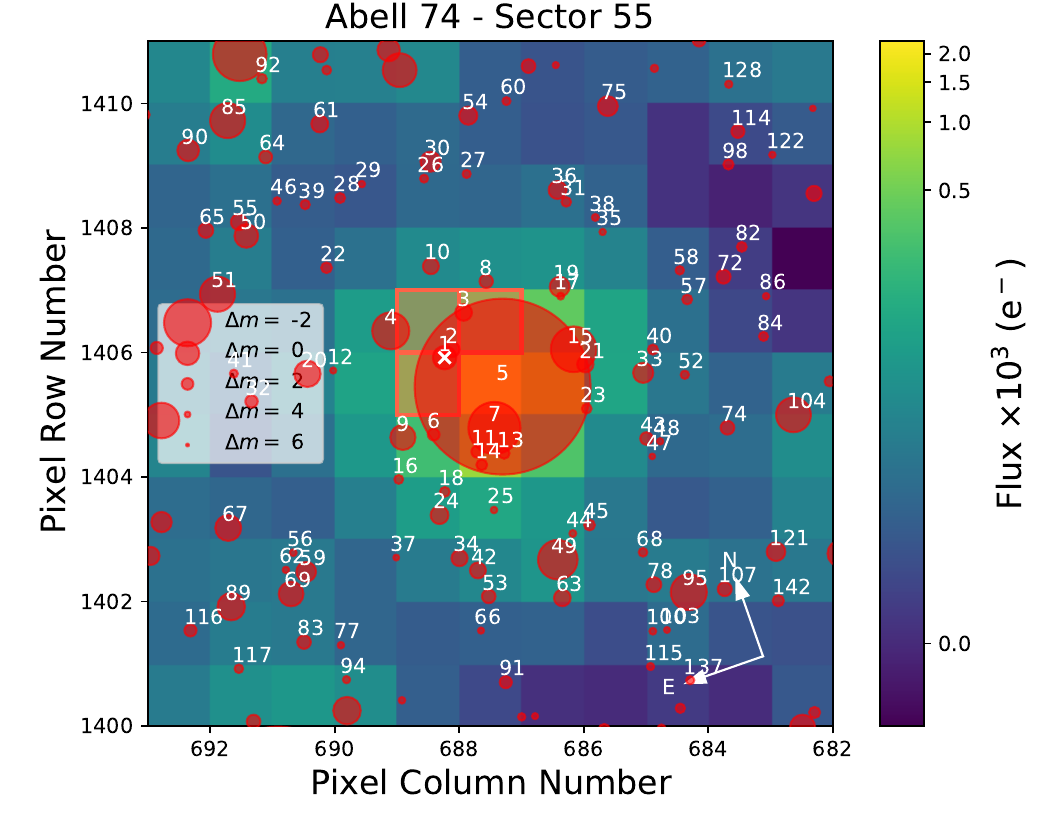}
     \includegraphics[width=0.33\textwidth]{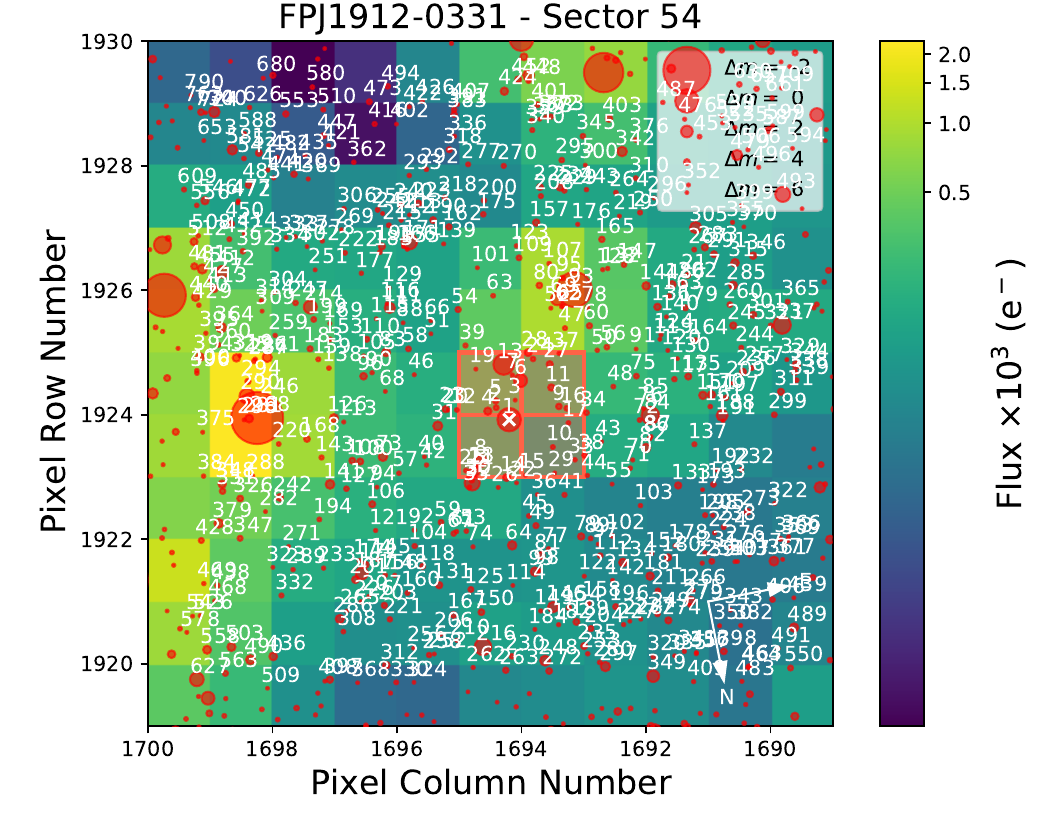}
    \includegraphics[width=0.33\textwidth]{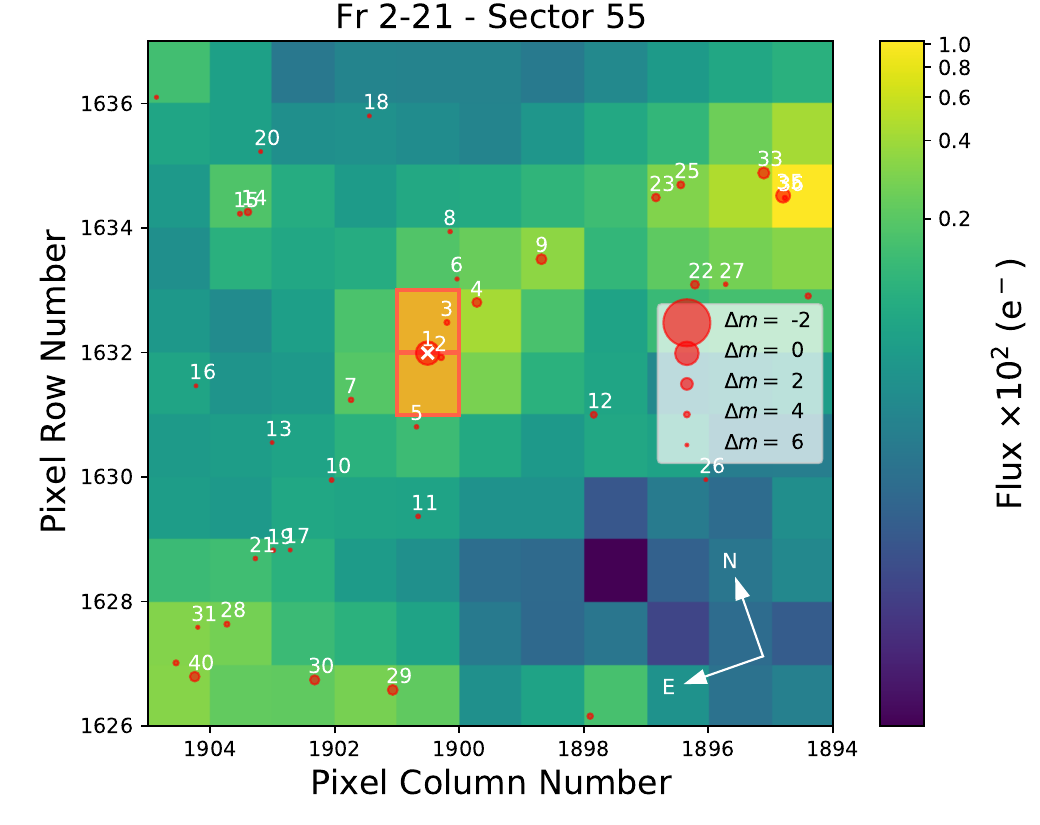}
     \includegraphics[width=0.33\textwidth]{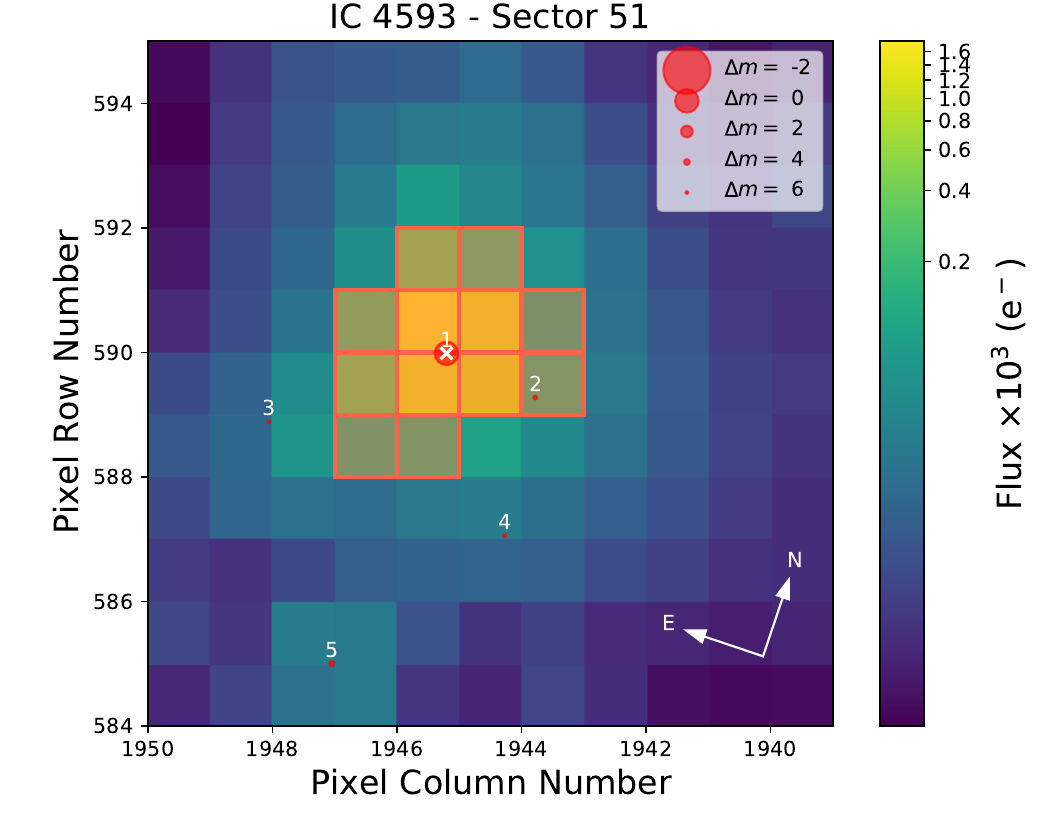}
\caption{continued.}
\label{fig:TPFs}
\end{figure*}

\addtocounter{figure}{-1}

  \begin{figure*}
      \includegraphics[width=0.33\textwidth]{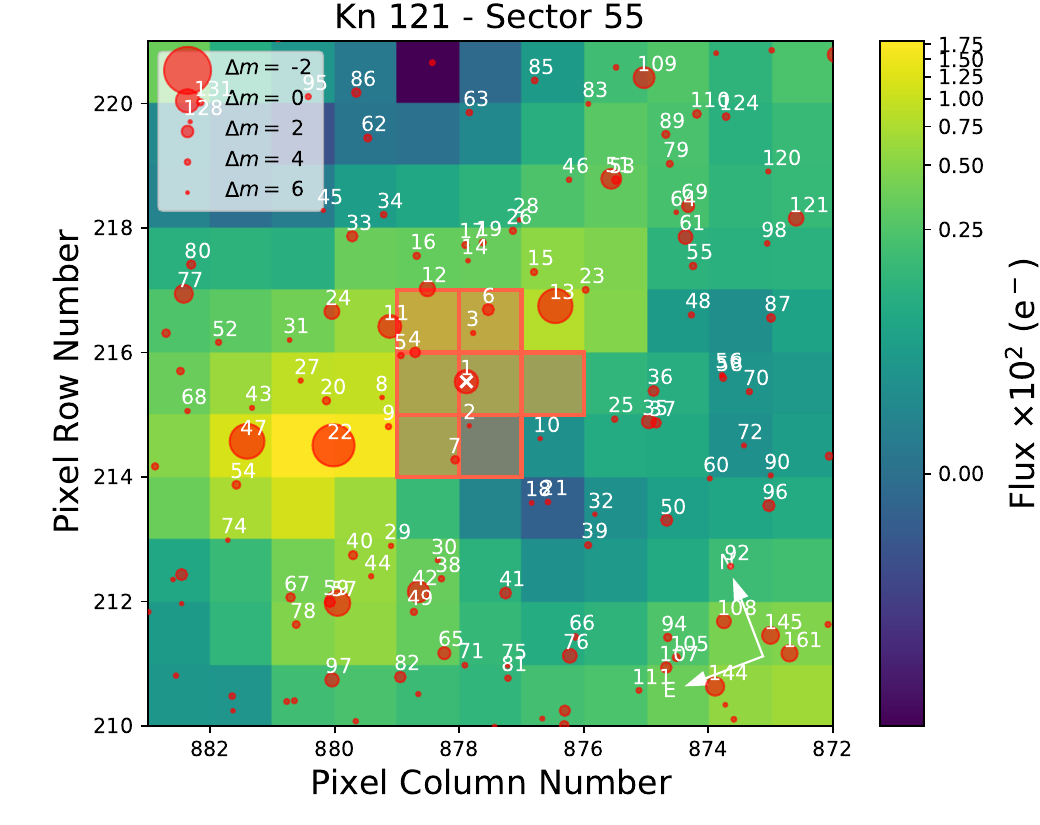}
      \includegraphics[width=0.33\textwidth]{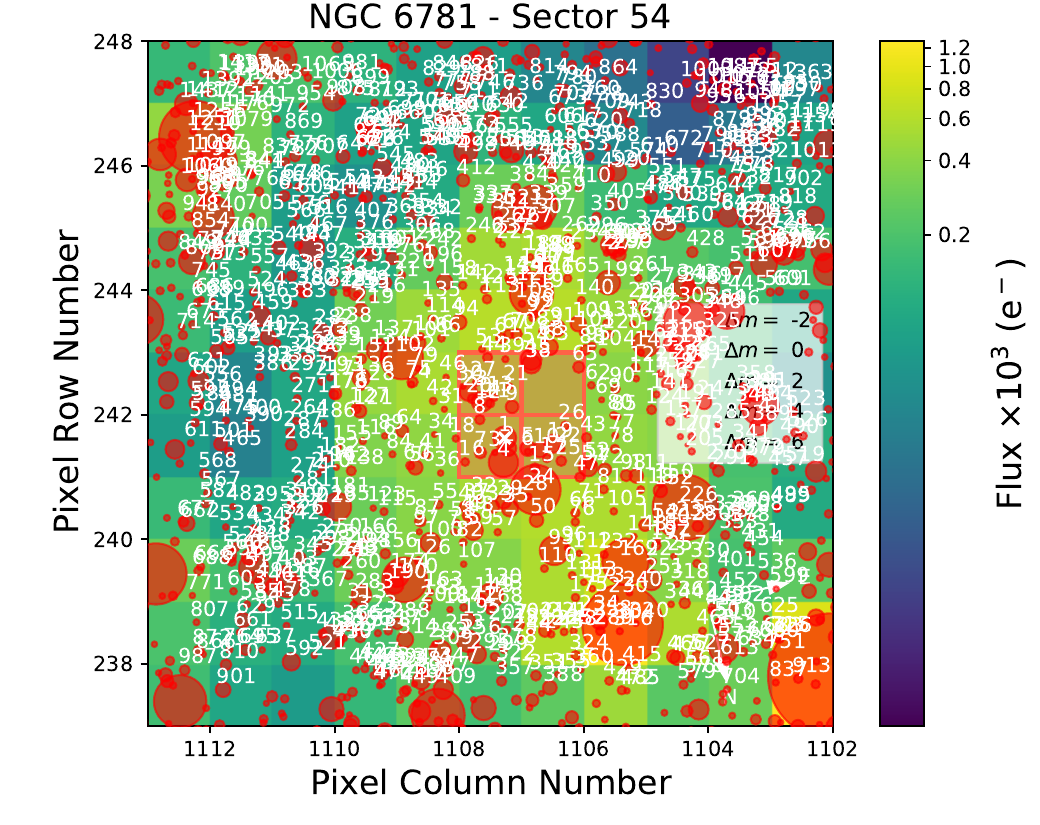}
       \includegraphics[width=0.33\textwidth]{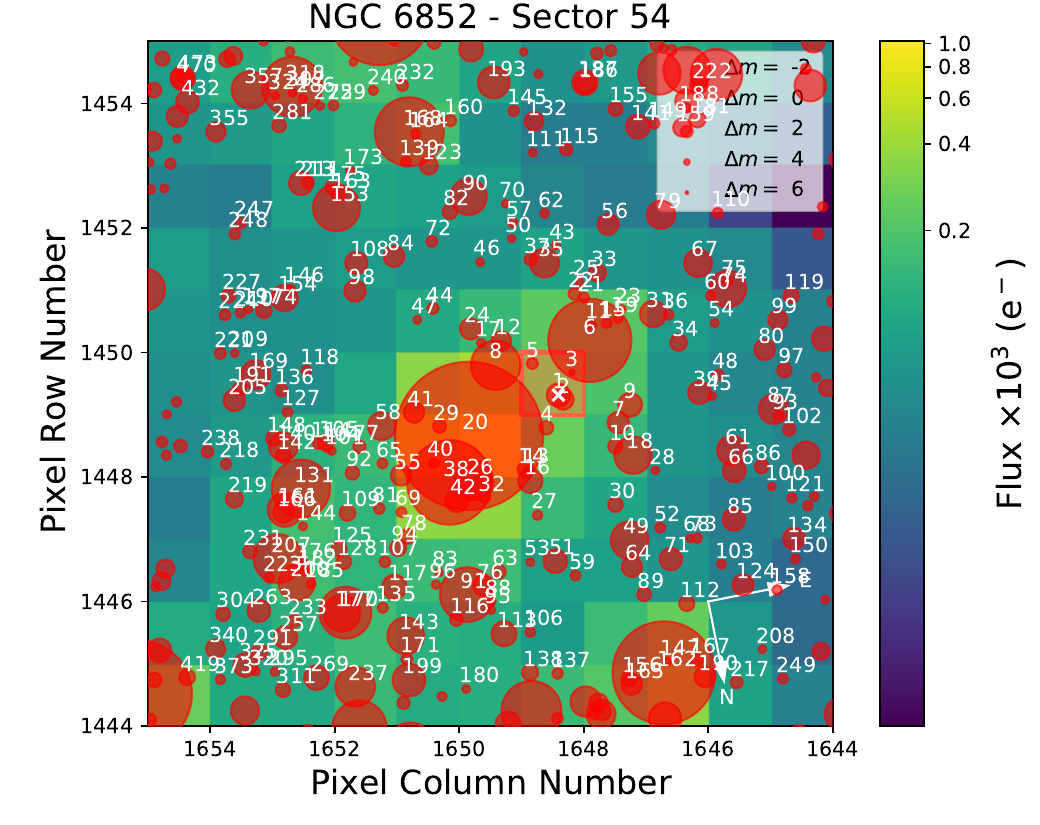}
     \includegraphics[width=0.335\textwidth]{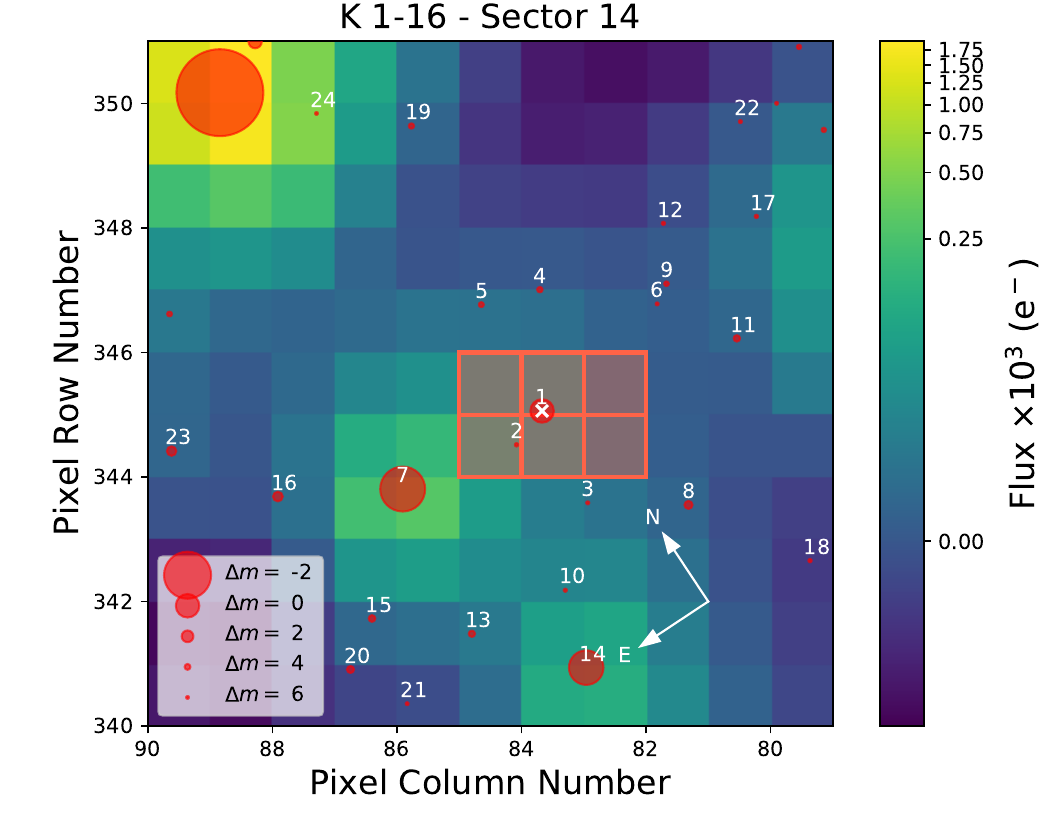}
     \includegraphics[width=0.335\textwidth]{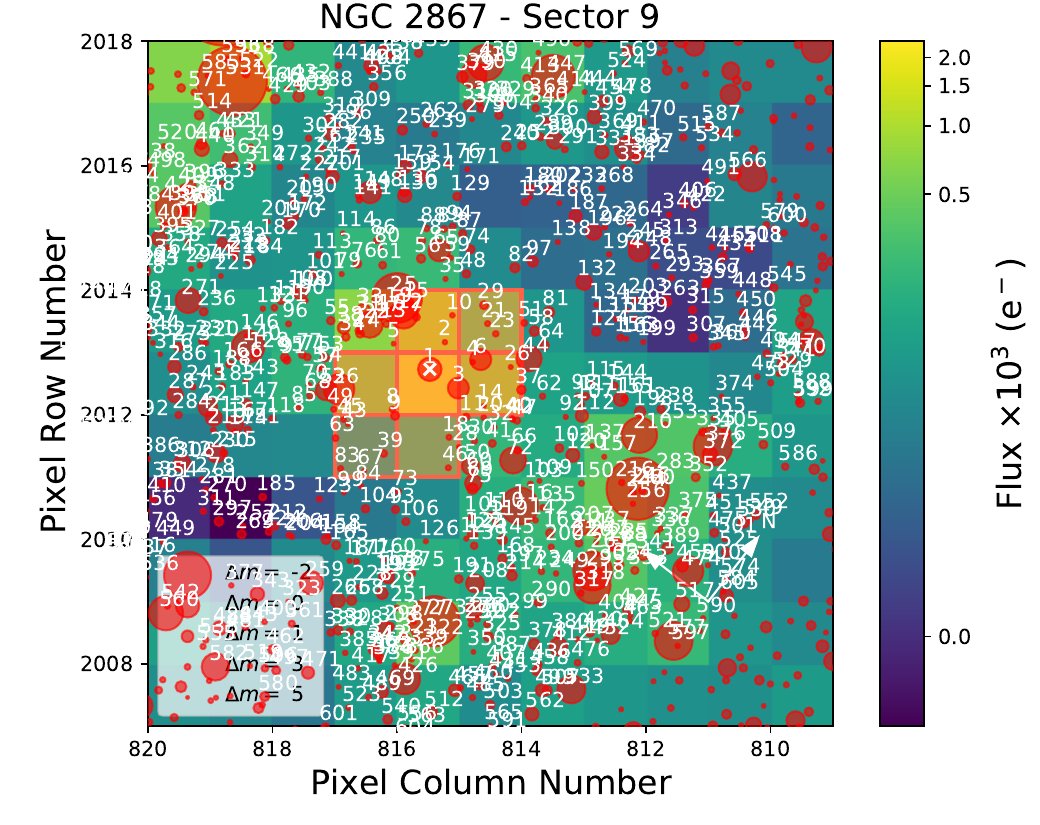}
\caption{continued.}
\label{fig:TPFs}
\end{figure*}

\begin{figure*}
    \includegraphics[width=0.95\textwidth]{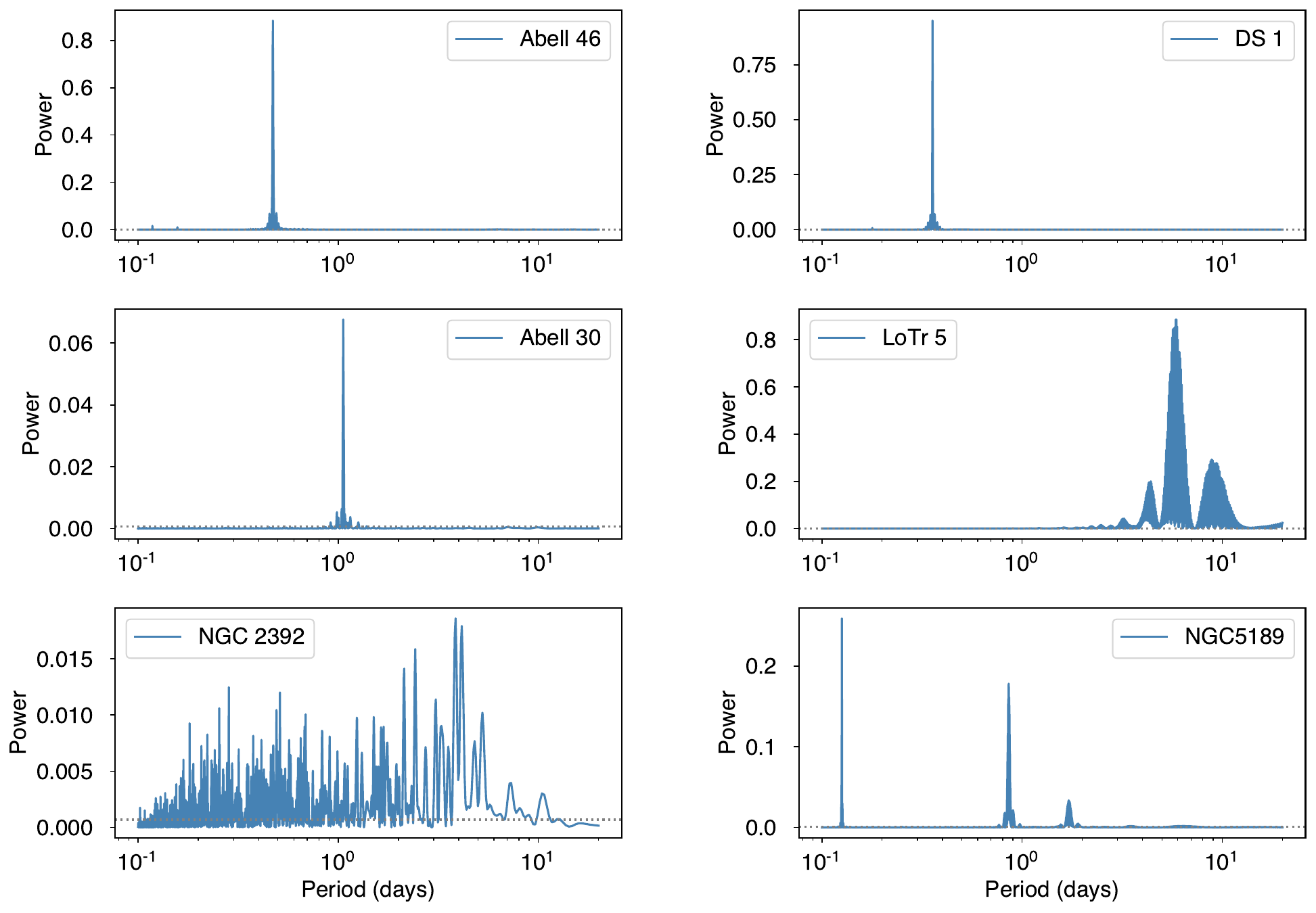}
   \caption{Periodograms of the known binary central stars Abell\,46, DS\,1, Abell\,30, LoTr\,5, NGC\,5189, and NGC\,2392. Grey horizontal dotted lines represent the False Alarm Probabilities (FAPs) at 10\%, 1\% and 0.1\%.}
\label{fig:periodograms_known}
\end{figure*}

\begin{figure*}
    \includegraphics[width=0.95\textwidth]{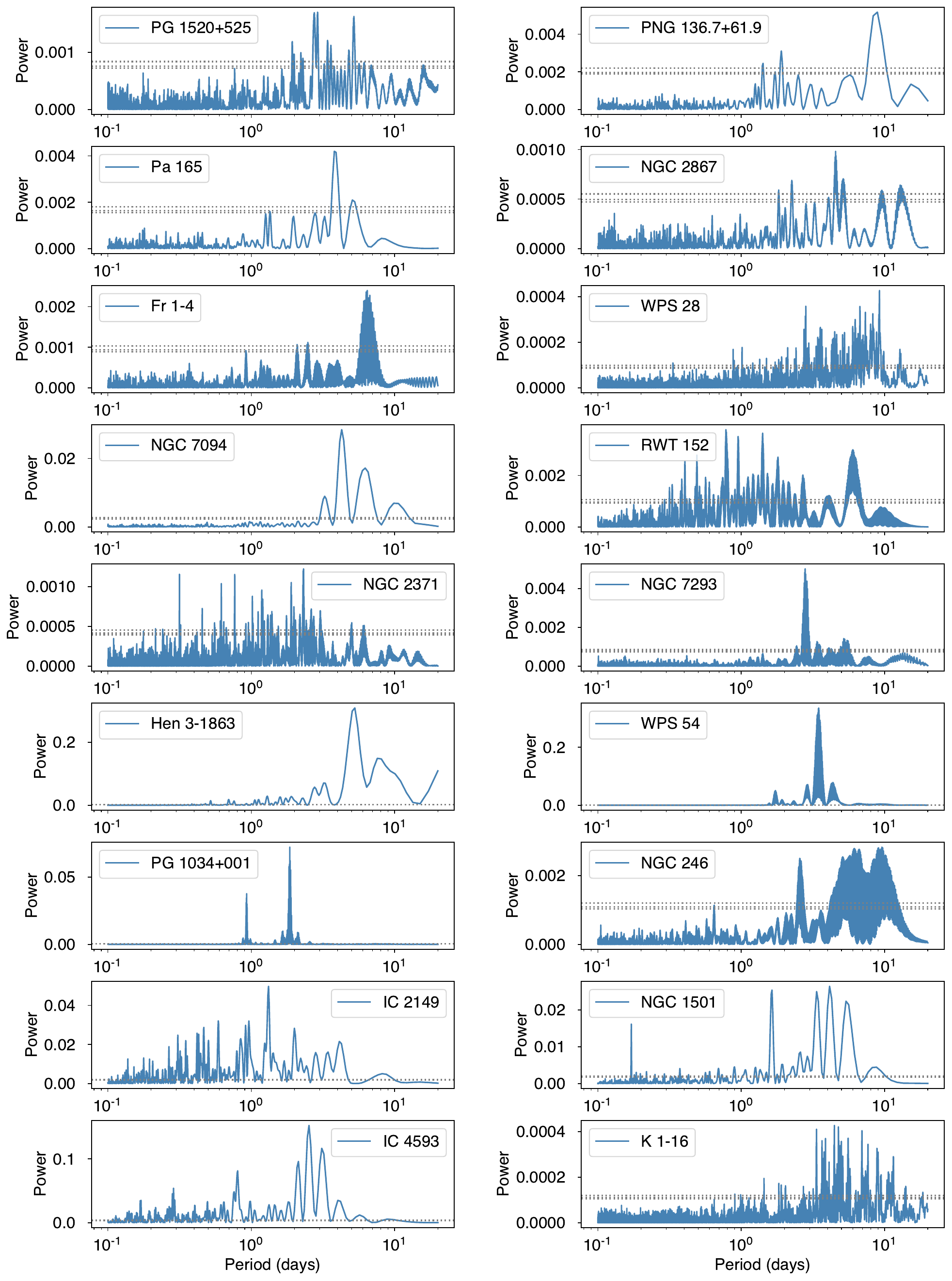}
   \caption{Periodograms of those central stars which show significant variability in their light curves. Grey horizontal dotted lines represent the False Alarm Probabilities (FAPs) at 10\%, 1\% and 0.1\%.}
\label{fig:periodograms_candidates}
\end{figure*}

 \begin{figure*}
    \includegraphics[width=0.95\textwidth]{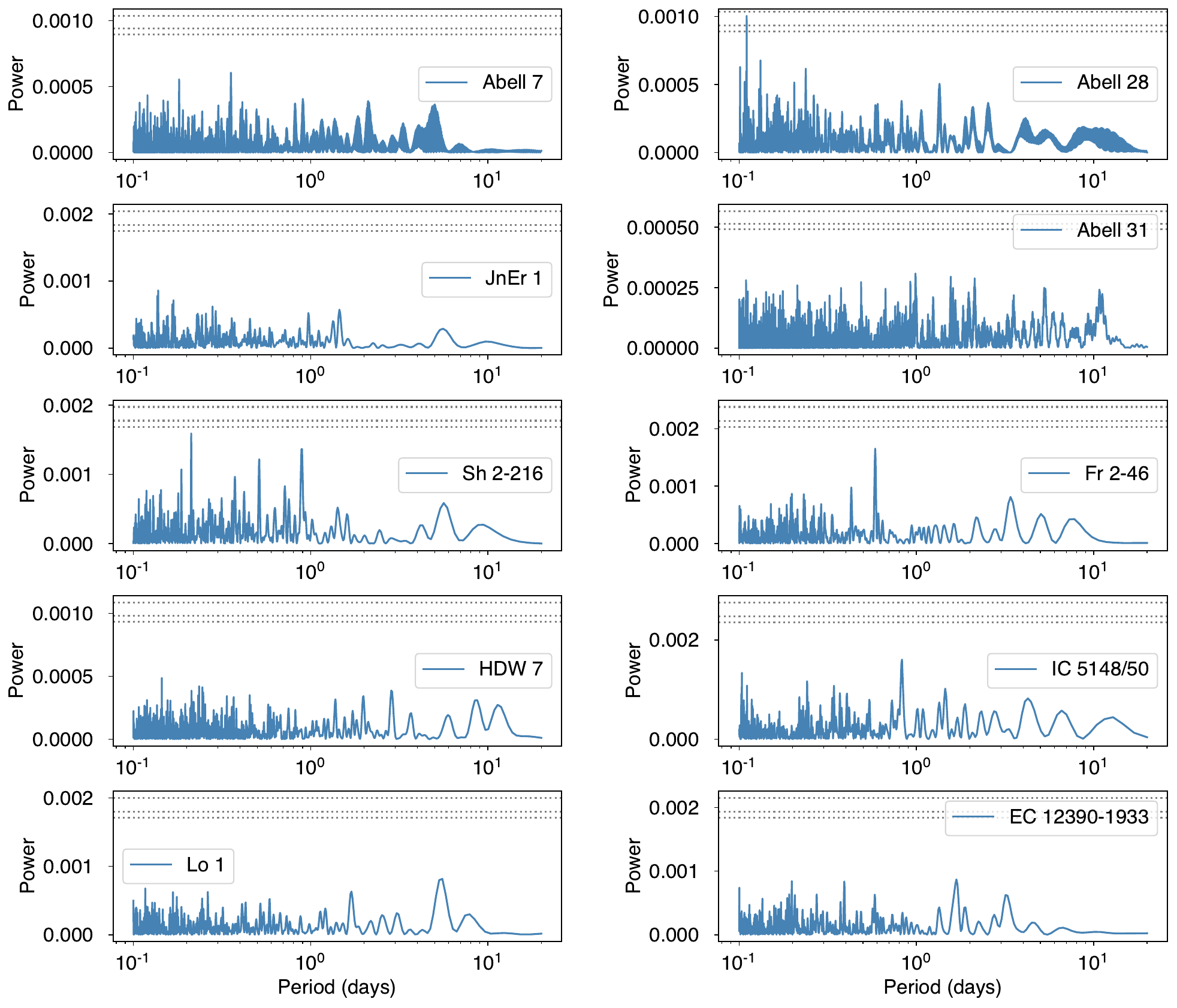}
   \caption{Periodograms of those central stars with no evidence in the MCMC analysis: Abell\,7, Abell\,28, JnEr\,1, Abell\,31, Sh\,2-16, Fr\,2-46, HDW\,7, IC\,5148/50, Lo\,1 and EC\,12390-1933. Grey horizontal dotted lines represent the False Alarm Probabilities (FAPs) at 10\%, 1\% and 0.1\% .}
\label{fig:periodograms}
\end{figure*}

\begin{figure*}
     \includegraphics[width=0.42\textwidth]{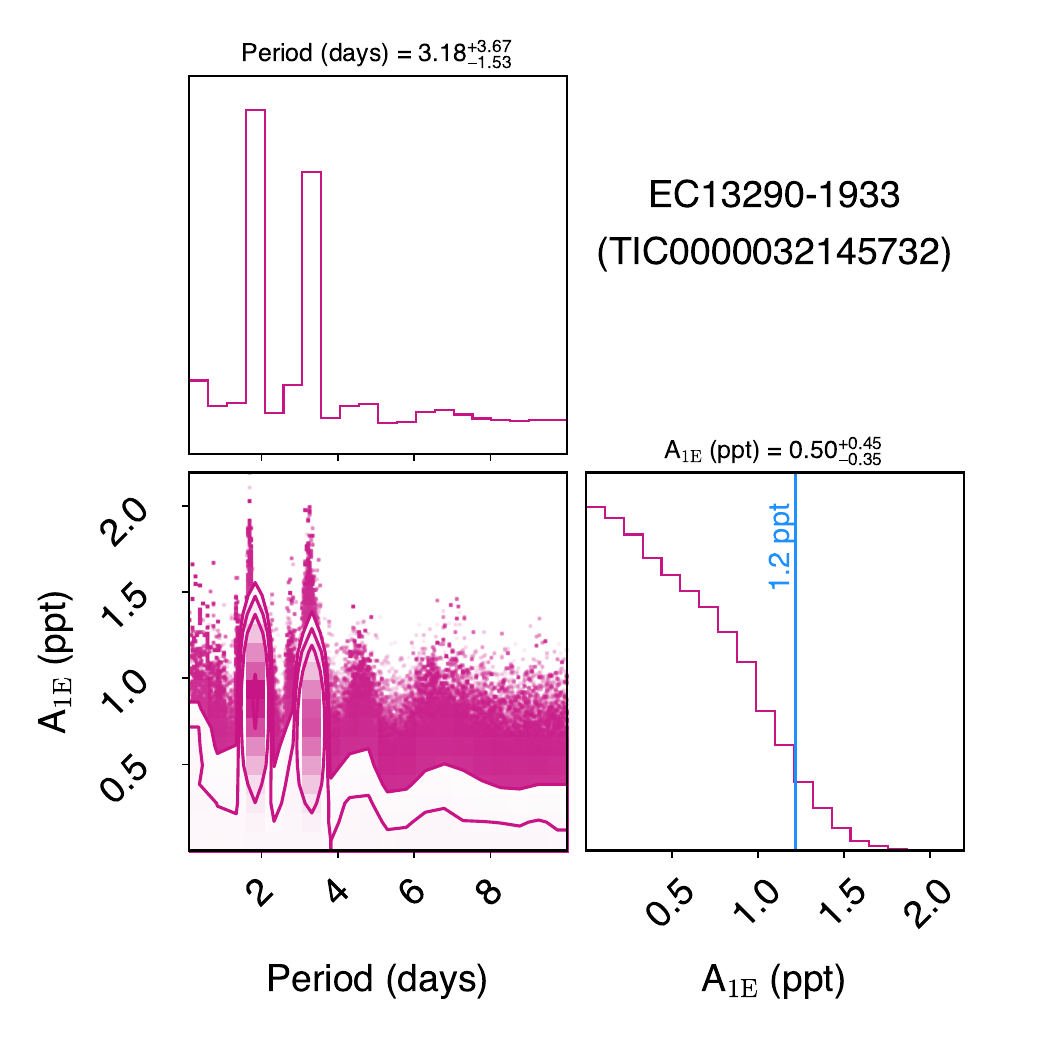}
      \includegraphics[width=0.42\textwidth]{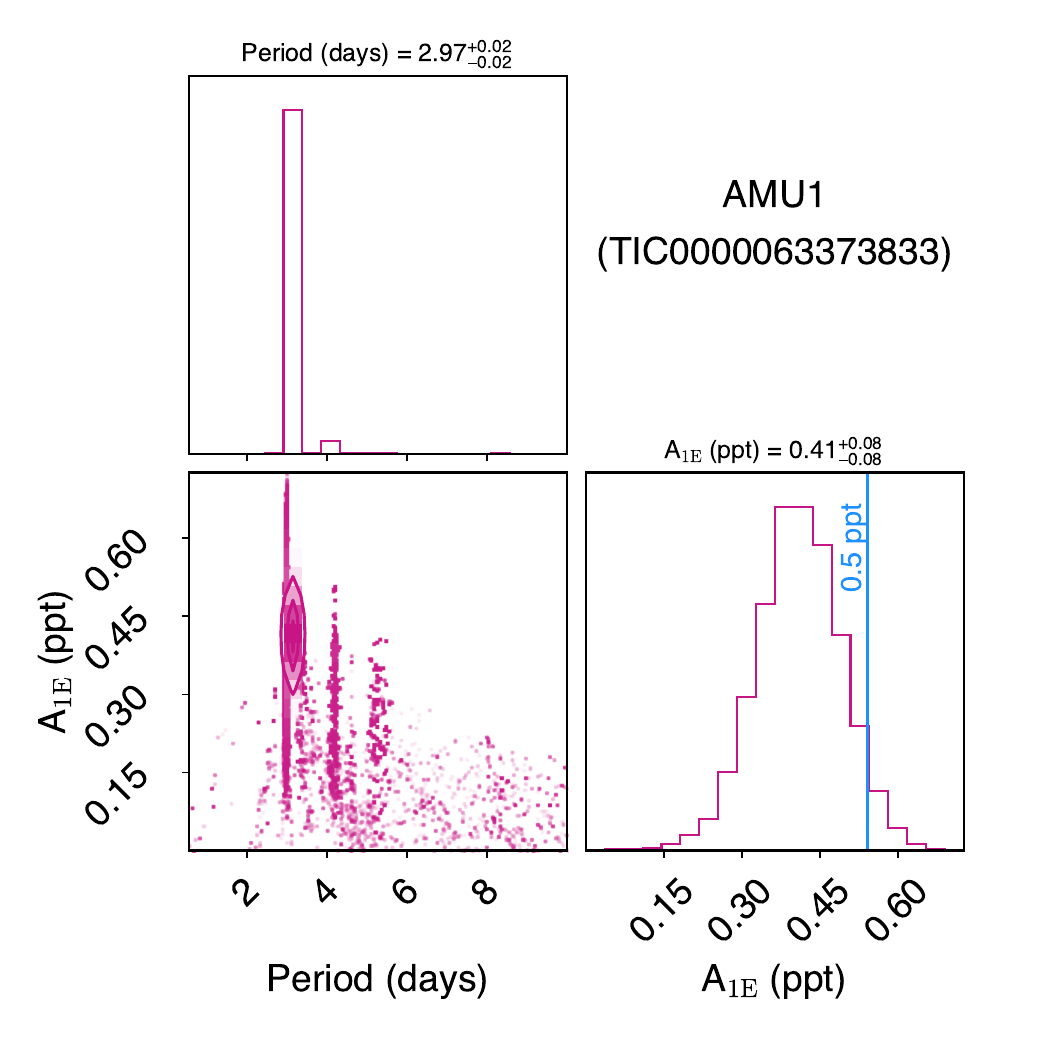}
      \includegraphics[width=0.42\textwidth]{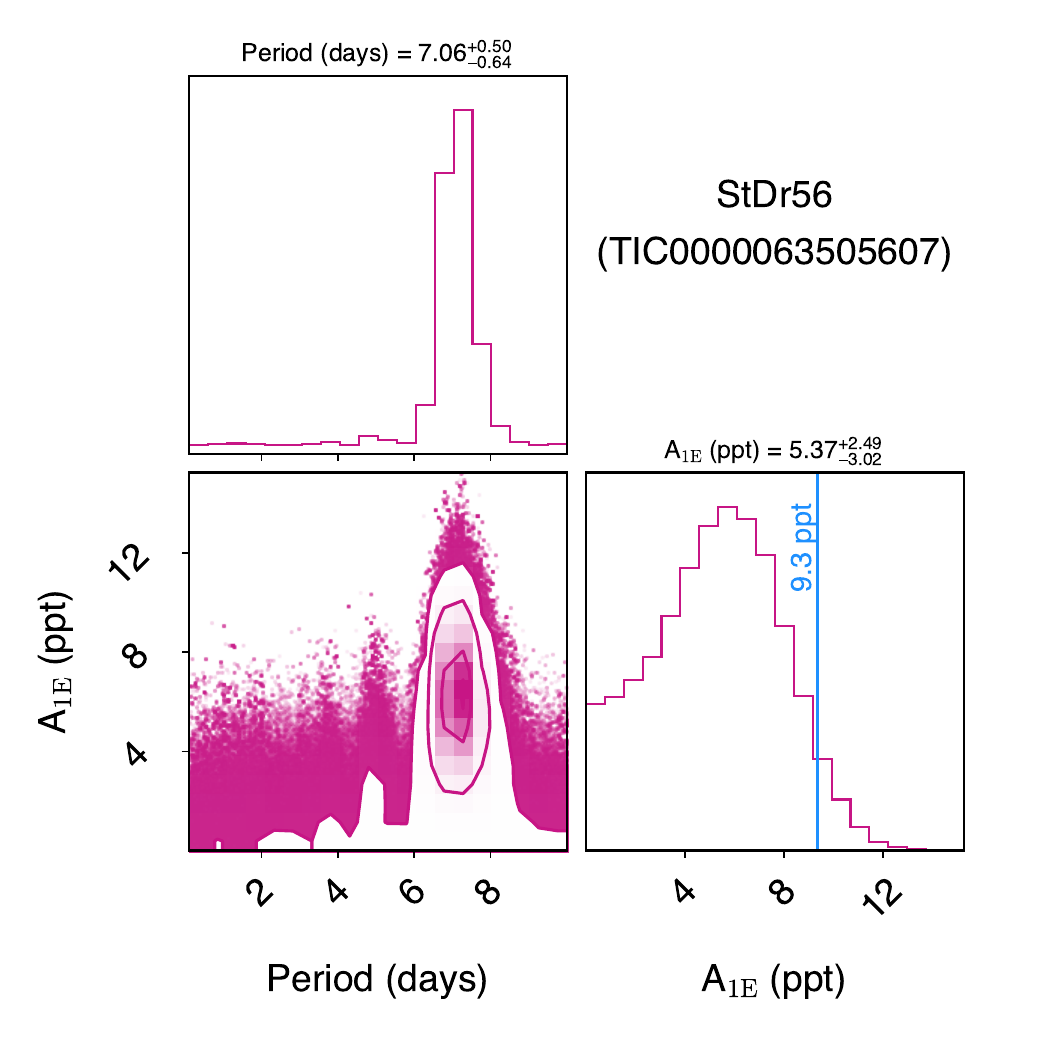}
     \includegraphics[width=0.42\textwidth]{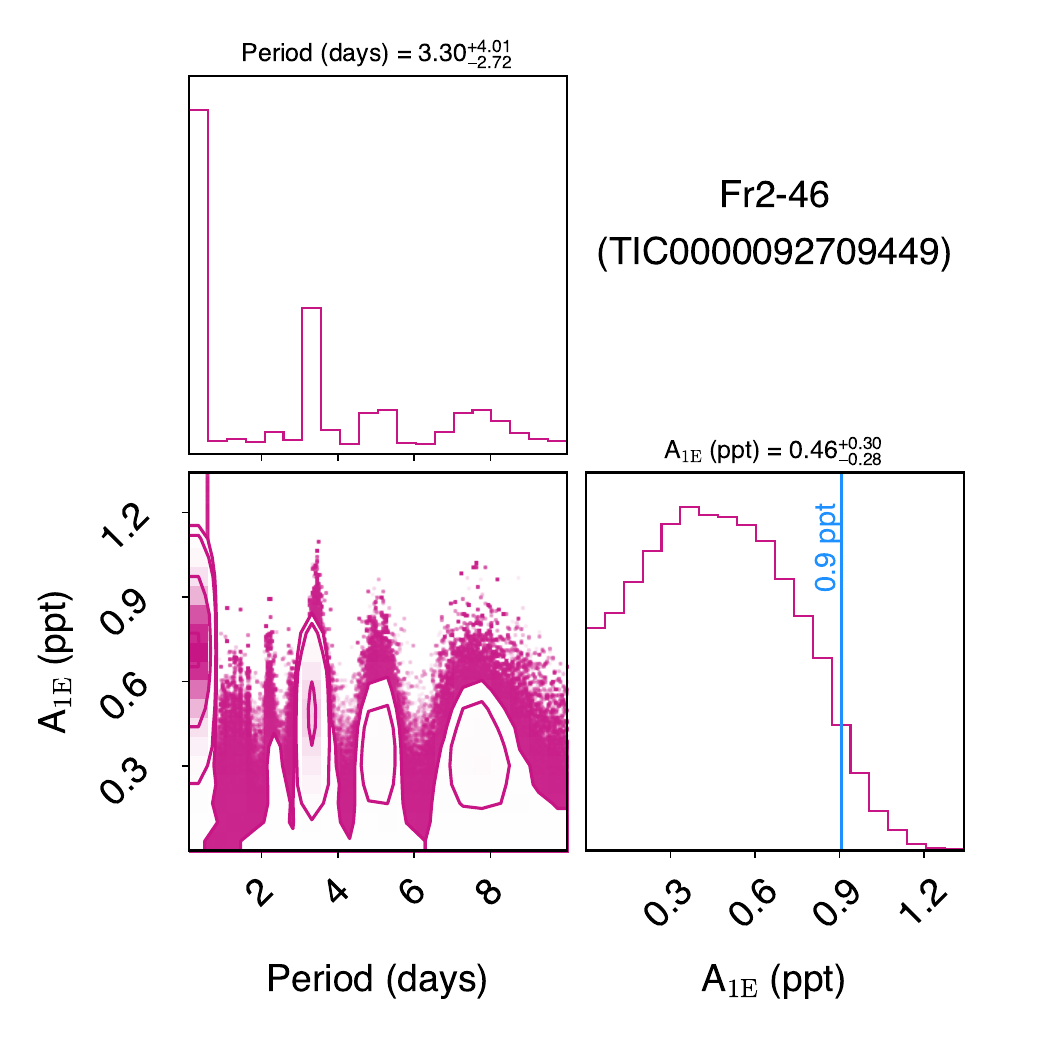}     
     \includegraphics[width=0.42\textwidth]{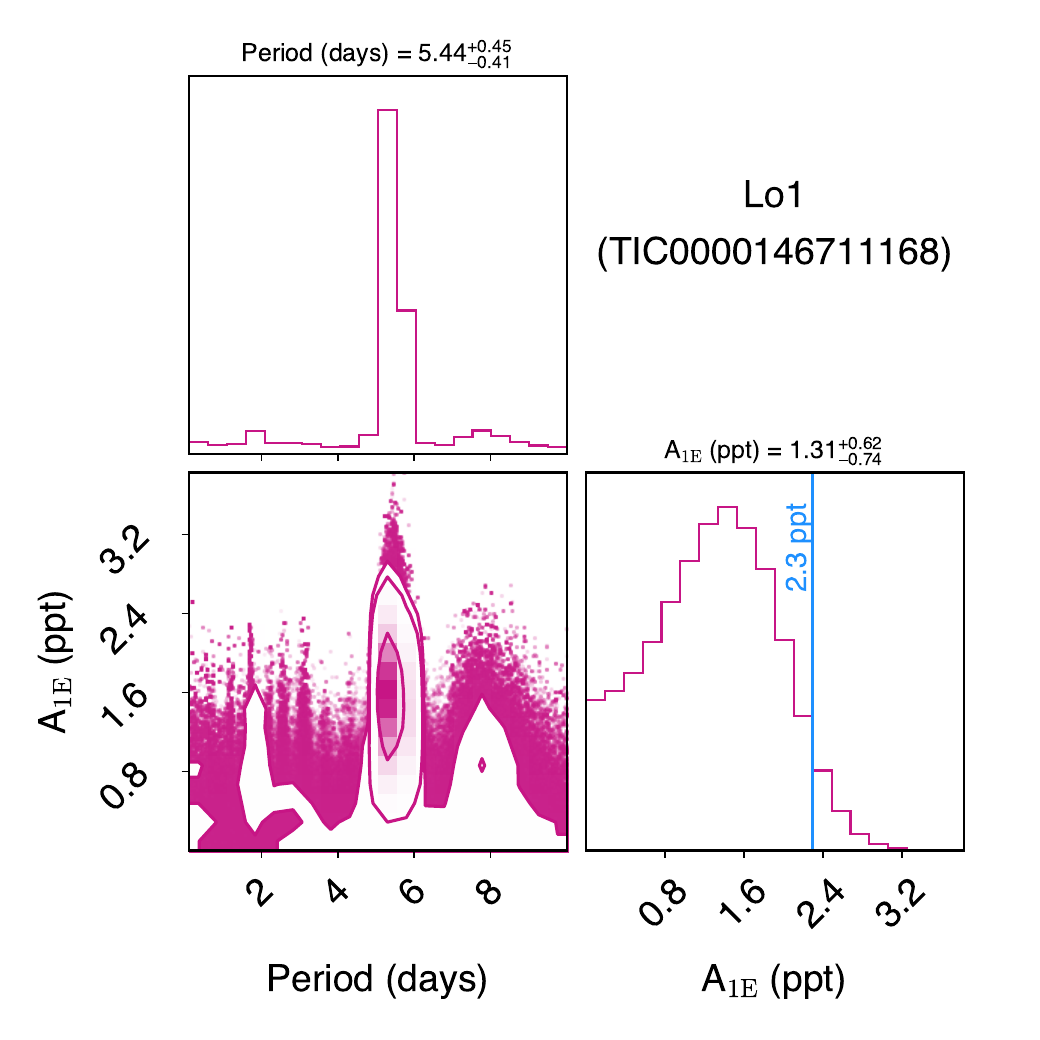}
     \includegraphics[width=0.42\textwidth]{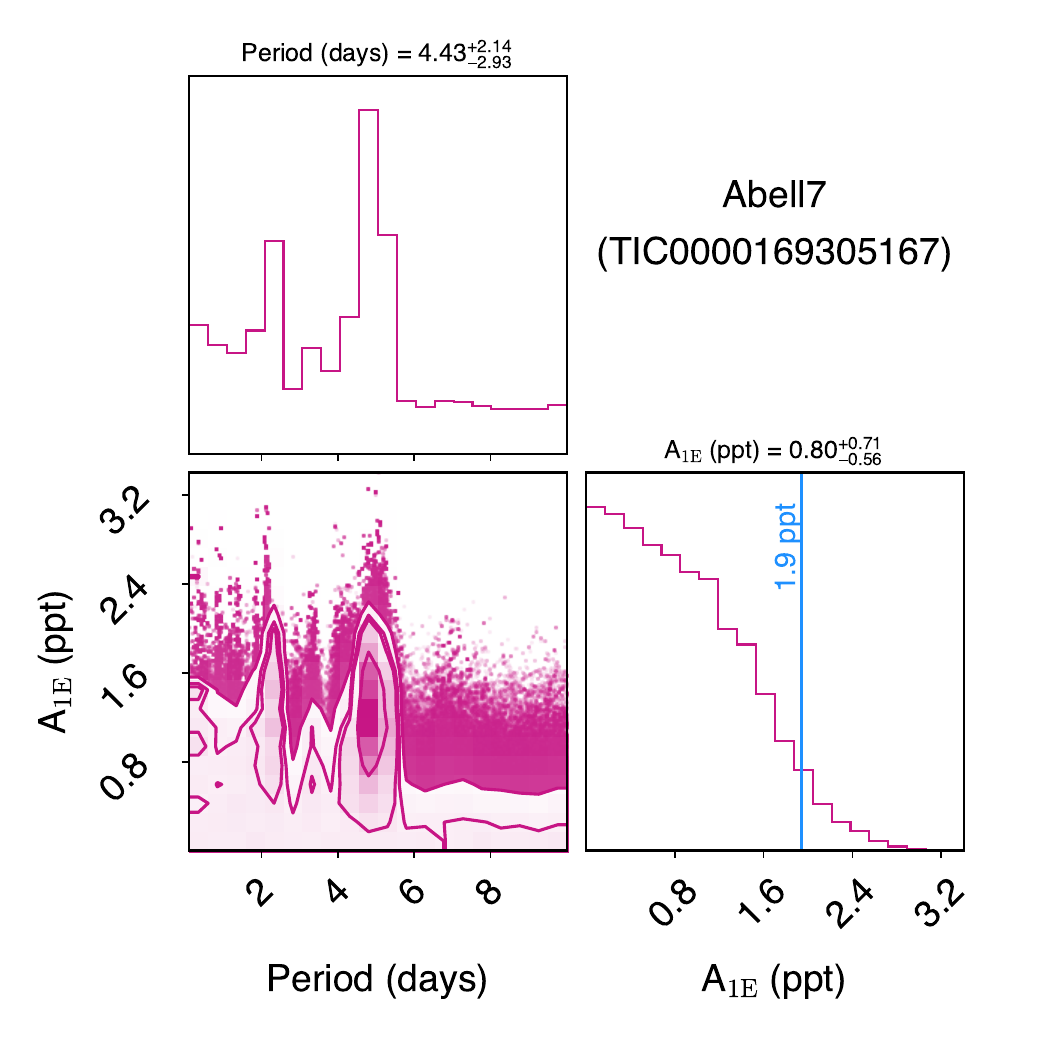}
       
\caption{Corner plots corresponding to the posterior distributions of the light curve modeling for targets with no or weak evidence in the 1-effect model. In each panel shows the marginalised posterior distributions of the orbital period and the amplitude of the model ($A_{\rm 1E}$) in the diagonal, while the 2D contour plot shows the dependency between both parameters. In the case of the amplitude ($A_{\rm 1E}$), we also include a vertical line showing the 95\% percentile, that we assume as the upper limit for any periodicity testable with the TESS data. }
\label{fig:corner_plots_flats}
\end{figure*}

\addtocounter{figure}{-1}

  \begin{figure*}
  \centering   
      \includegraphics[width=0.42\textwidth]{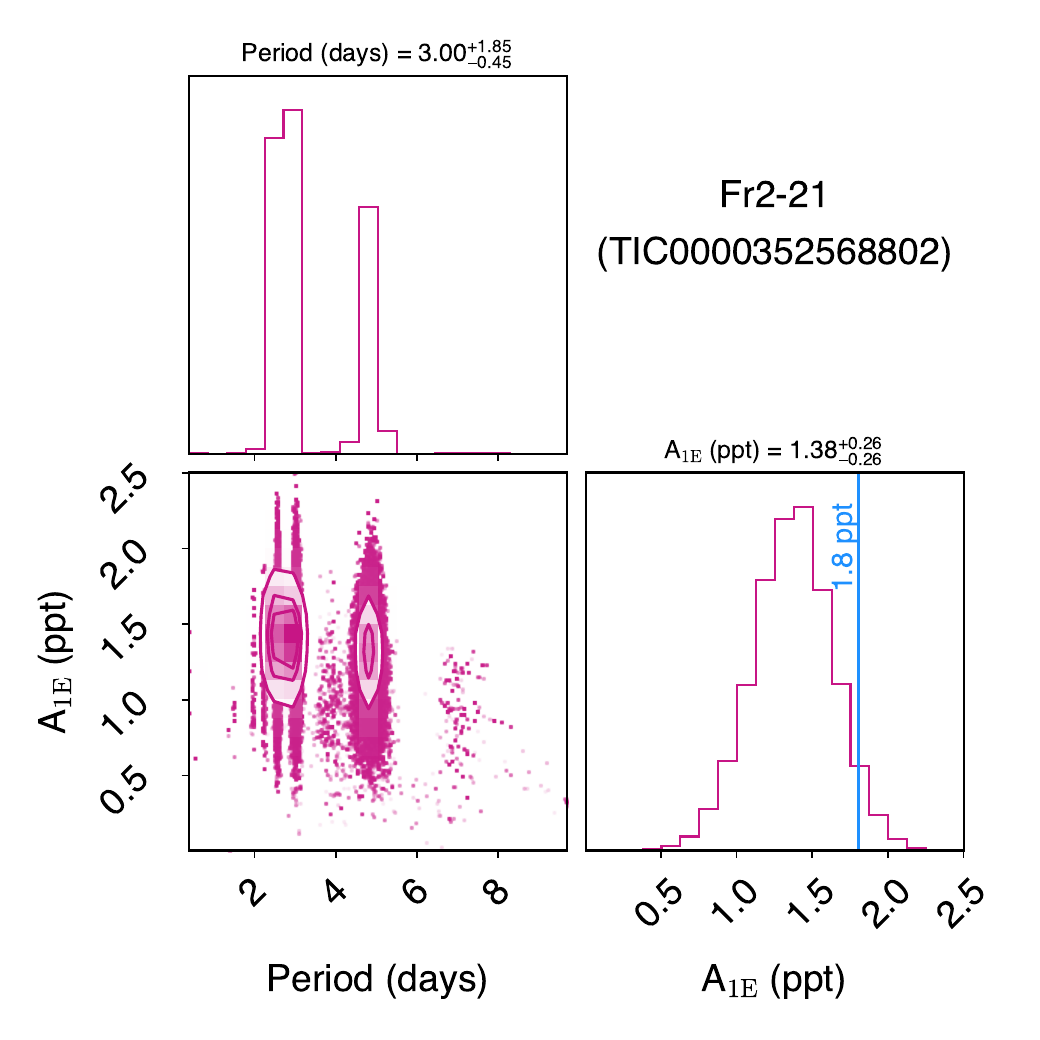}    
    \includegraphics[width=0.42\textwidth]{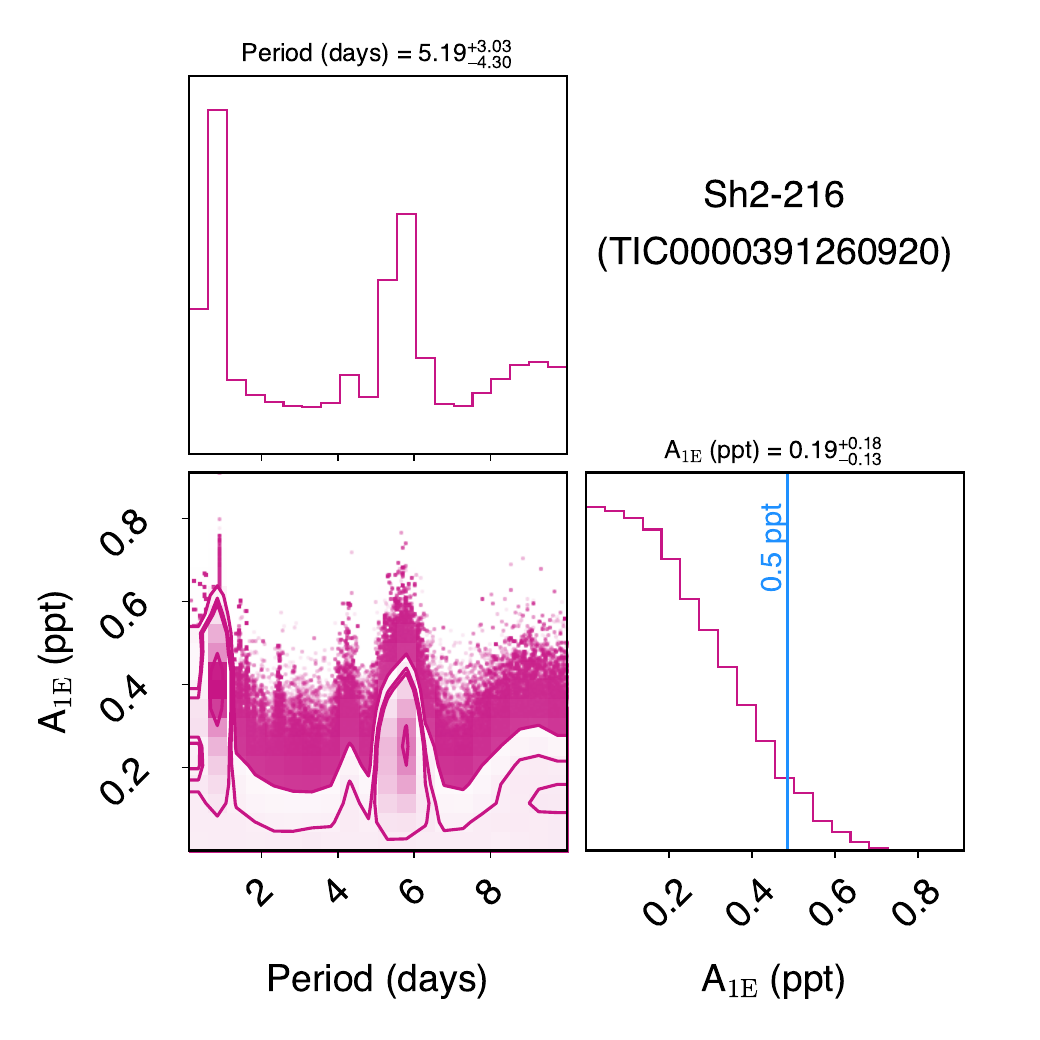}
      \includegraphics[width=0.42\textwidth]{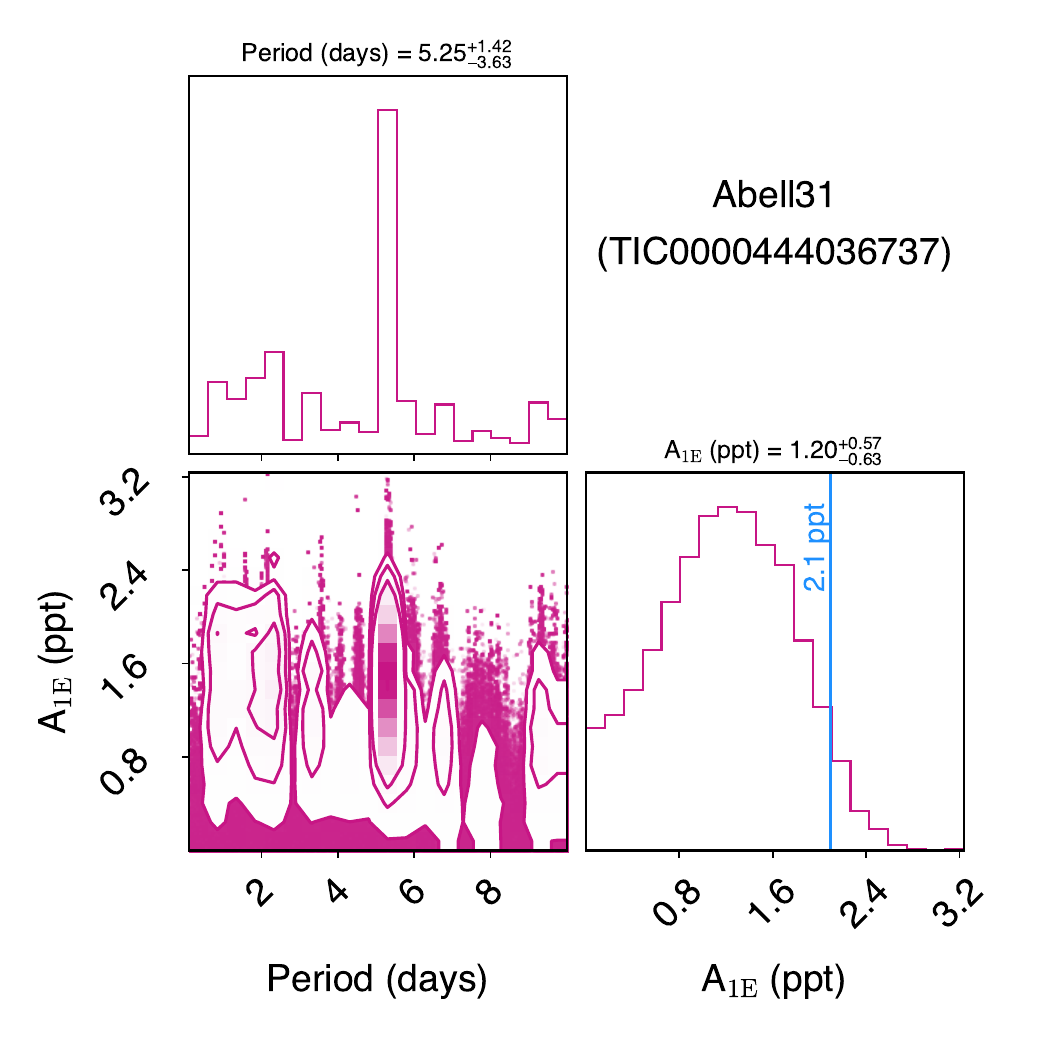}
       \includegraphics[width=0.42\textwidth]{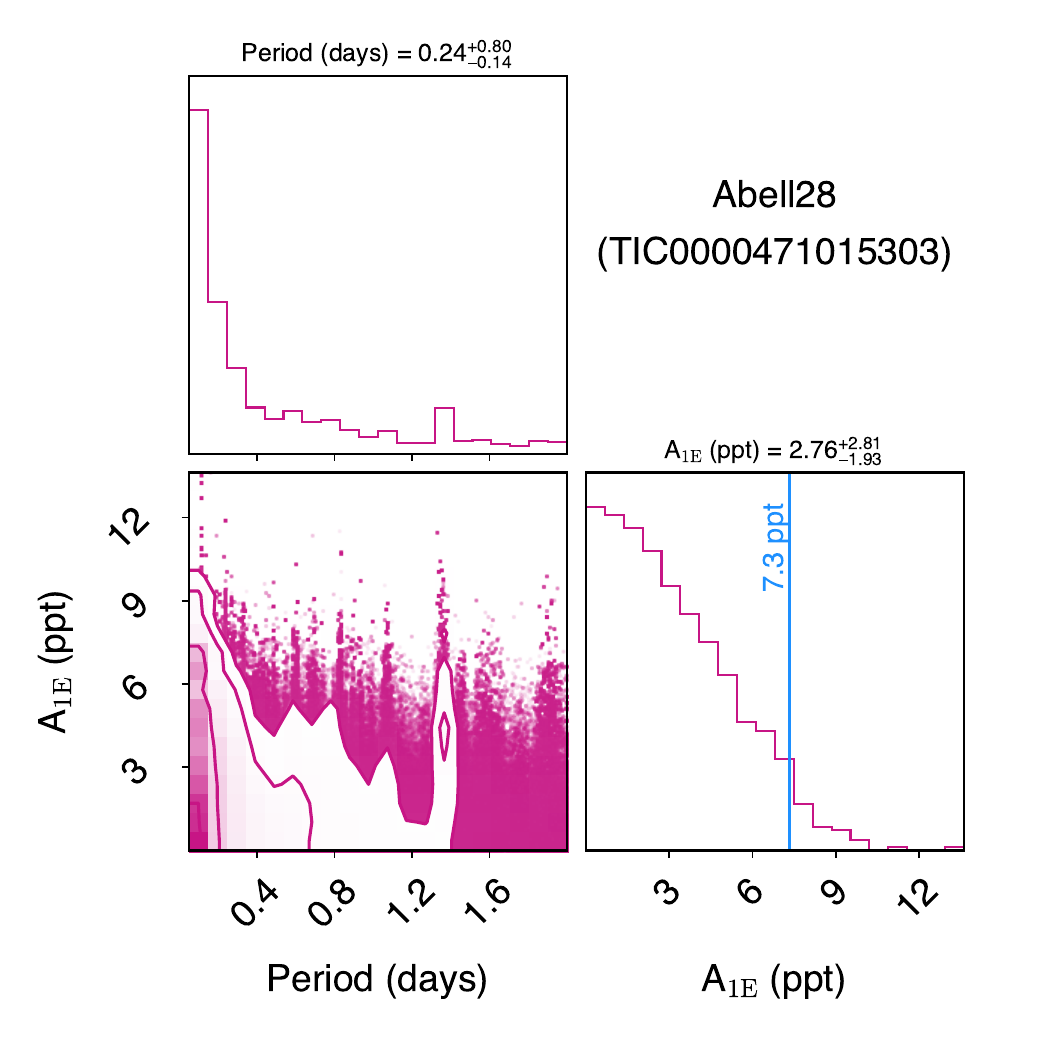}
    \includegraphics[width=0.42\textwidth]{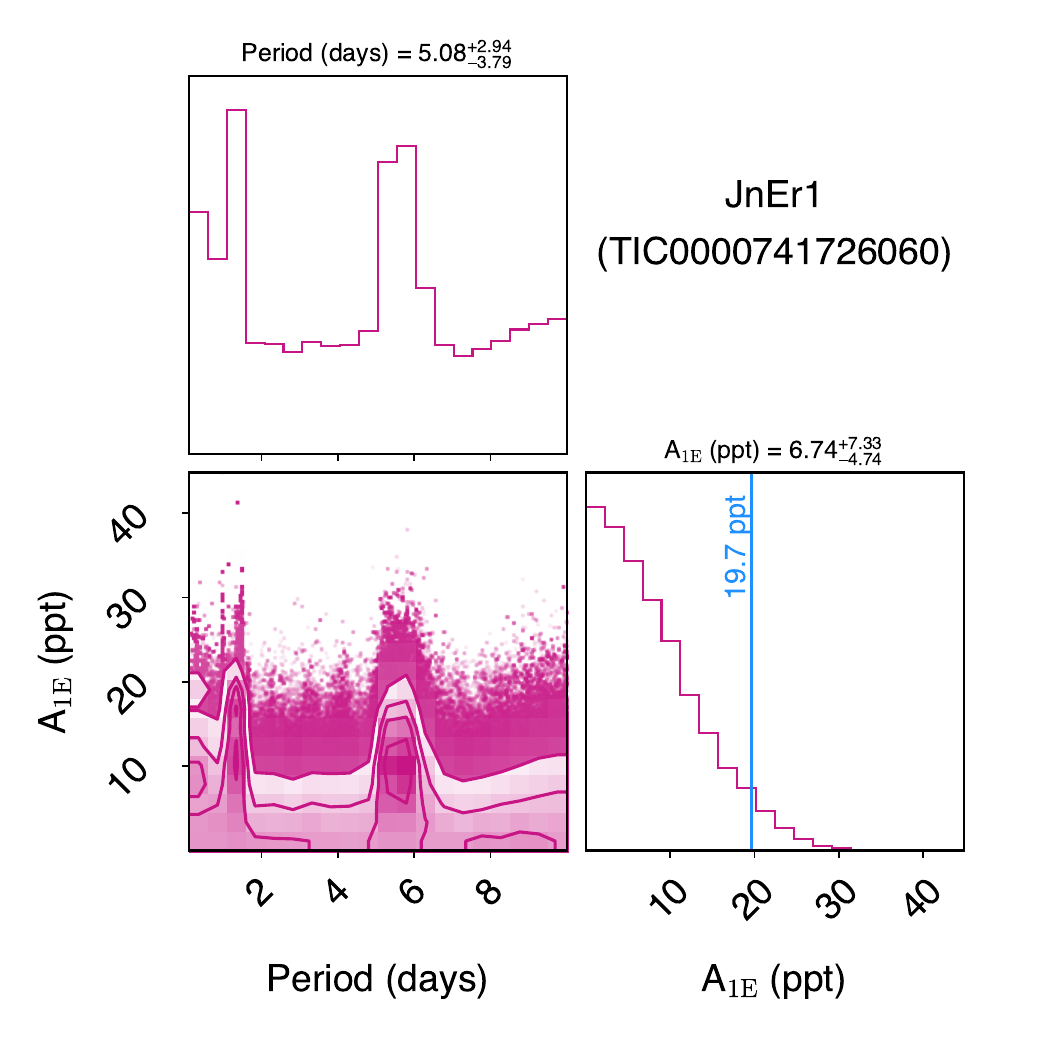}
       \includegraphics[width=0.42\textwidth]{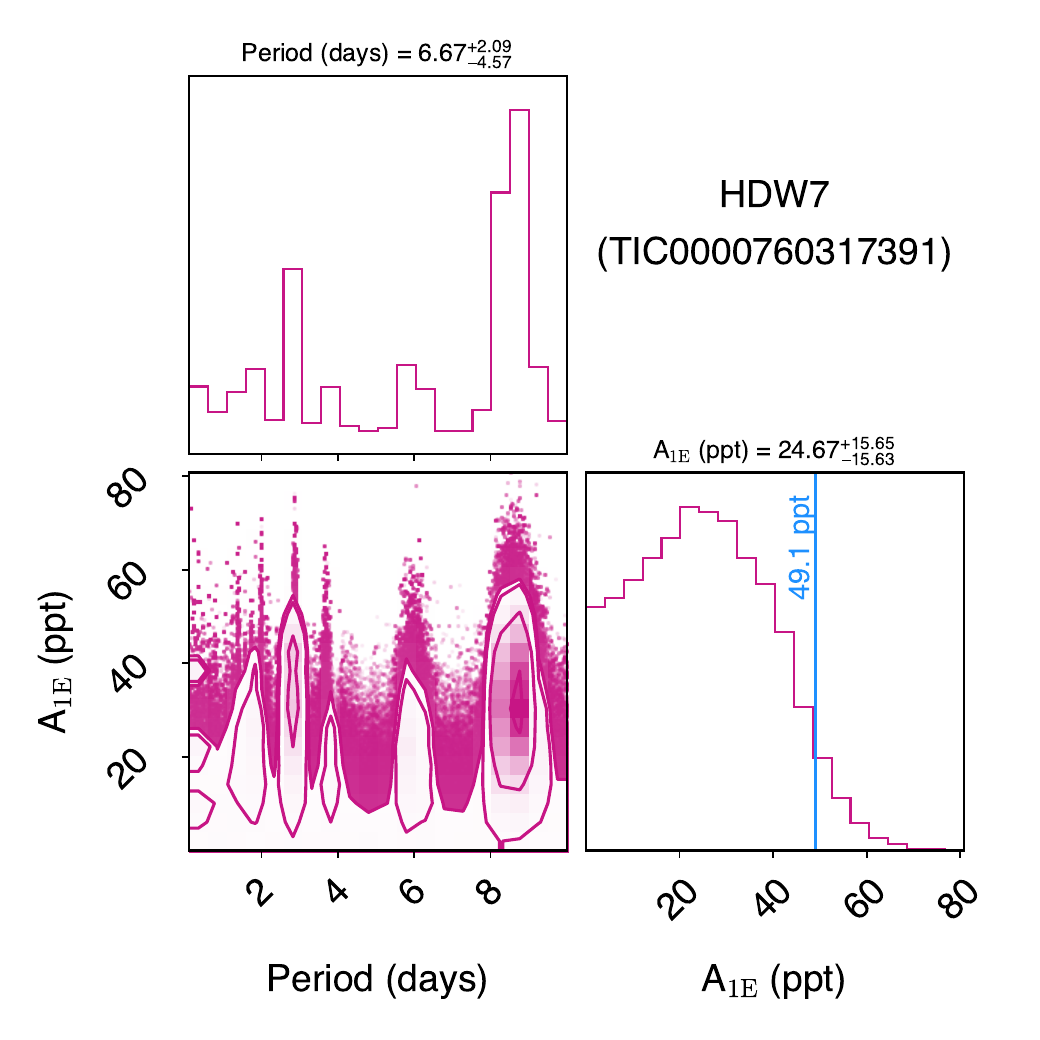}

\caption{continued.}
\label{fig:corner_plots_flats}
\end{figure*}

\addtocounter{figure}{-1}

  \begin{figure*}
  \centering   
    \includegraphics[width=0.42\textwidth]{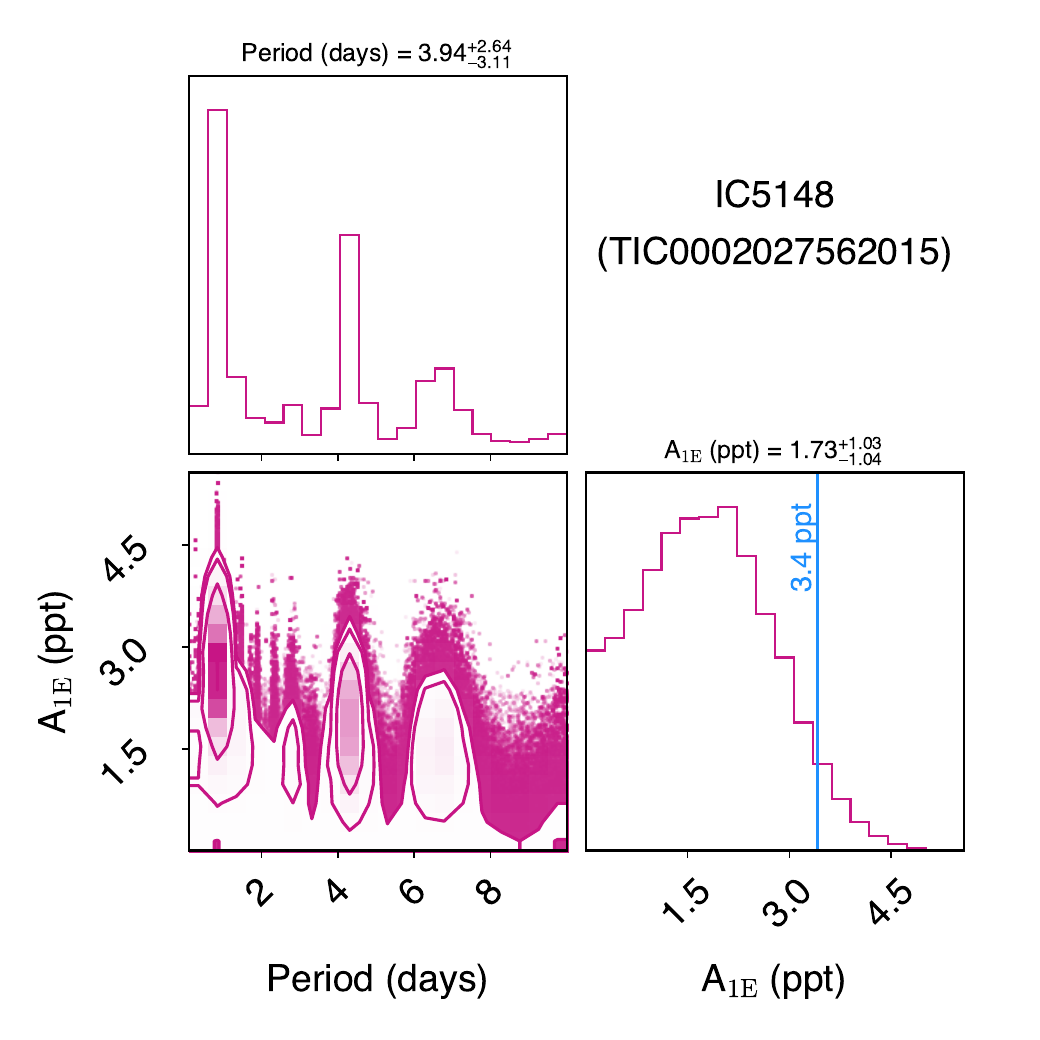}

\caption{continued.}
\label{fig:corner_plots_flats}
\end{figure*}

\end{appendix}

\end{document}